\documentclass[suppldata]{interact}   

\usepackage{epstopdf}
\usepackage[caption=false]{subfig}

\usepackage[numbers,sort&compress]{natbib}
\bibpunct[, ]{[}{]}{,}{n}{,}{,}
\makeatletter
\def\NAT@def@citea{\def\@citea{\NAT@separator}}
\makeatother

\theoremstyle{plain}

\theoremstyle{definition}

\theoremstyle{remark}

\usepackage[hyphens]{url}
\usepackage{float,epsfig,epstopdf}
\usepackage{graphicx}
\usepackage{amssymb}
\usepackage{amsmath}
\usepackage{bm}
\usepackage[colorlinks,citecolor=blue,linkcolor=blue,urlcolor=blue]{hyperref}
\usepackage{arydshln}
\usepackage{appendix}
\usepackage{pdflscape}
\usepackage{cancel}
\usepackage{multirow}

\begin{document}
	
			\title{A Gaussian copula joint model for longitudinal and time-to-event data with random effects}
	\author	{
		\name{Zili Zhang, Christiana Charalambous and Peter Foster}
		\affil{Department of Mathematics, University of Manchester, Manchester M13 9PL, UK}
	}
	\maketitle
{\bf Abstract:}~~Longitudinal and survival sub-models are two building blocks for joint modelling of longitudinal and time to event data. Extensive research indicates separate analysis of these two processes could result in biased outputs due to their associations. Conditional independence between measurements of biomarkers and event time process given latent classes or  random effects is a common approach for characterising the association between the two sub-models while taking the heterogeneity among the population into account. However, this assumption is tricky to validate because of the unobservable latent variables. Thus a Gaussian copula joint model with random effects is proposed to accommodate the scenarios where the conditional independence assumption is questionable. In our proposed model, the conventional joint model assuming conditional independence is a special case when the association parameters in the Gaussian copula shrink to zeros. Simulation studies and real data application are carried out to evaluate the performance of our proposed model. In addition, personalised dynamic predictions of survival probabilities are obtained based on the proposed model and comparisons are made to the predictions obtained under the conventional joint model.


\section{Introduction}
Longitudinal measurements of biomarkers at a sequence of informative (Dai and Pan, 2018\cite{dai18}) or uninformative time points and the time (censored or uncensored) until an event of interest, e.g. death or recurrence of a disease, occurs, are often jointly collected for each subject in clinical studies. Research interests could focus on the progression of the longitudinal process given covariates recorded at baseline,  prognosis of survival probabilities based on baseline covariates and history of biomarker measurements or characterising the relationship between the two processes. Exhaustive research (Verbeke and Molenberghs, 2000\cite{ver00} and  Kalbfleisch  and Prentice, 2002\cite{kal02}) has been done to analyse longitudinal and time to event data separately. However, when biomarkers are associated with patient's heath status, the event process would result in informative dropout in longitudinal measurements, which causes biased estimations if the analysis is based on the longitudinal process alone (Guo and Carlin, 2004\cite{guo04}). On the other hand, if the event process is of interest, measurement errors in the longitudinal process would lead to biases toward the null in the event process if not appropriately accounted for (Prentice, 1982\cite{pre82}). Ever since Faucett and Thomas (1996)\cite{fau96} and Wulfsohn and Tsiatis (1997)\cite{wul97}  modelled these two processes jointly by introducing dependency on a common set of latent random effects, there have been a variety of extensions on modelling this unobservable latent variable.  Henderson \textit{et al.} (2000)\cite{hen00} extended the conventional shared random-effects joint models by incorporating a bivariate Gaussian latent process shared by the two processes. In this way, the random effects represent a sustained trend while the stationary Gaussian process describes local deviations and is able to capture a more elaborate within-subject autocorrelation structure for the longitudinal process. The drawback of this approach is its high computational complexity, as it normally demands a Monte Carlo integration method for the steps requiring integration (e.g. E-step in the EM algorithm). Song, Davidian and Tsiatis (2002)\cite{son02} relaxed the normality assumption of the random effects by allowing them to have a smooth density. A latent class model for joint analysis of longitudinal biomarker and event process data was proposed by Lin \textit{et al.} (2002)\cite{lin02}. Liu \textit{et al.,} (2015)\cite{liu15}  came up with a latent class model with shared random effects, which is essentially a distinct shared random effects joint model of longitudinal and survival data within each latent class. Baghfalaki \textit{et al.} (2017)\cite{bag17} modelled a  heterogeneous random-effects distribution by substituting the usual normality assumption of the random effects with a finite mixture of multidimensional normal variables.  Incorporating functional covariates, which model the predictors as continuous curves over some domains such as time or space, into the joint model was considered by Li and Luo (2017)\cite{li17} and (2019)\cite{li19}  and Li \textit{et al.} (2021)\cite{li21}. For exhaustive overviews on joint modelling, the reader is referred to Tsiatis and Davidian (2004)\cite{tsi04},  Ibrahim \textit{et al.,} (2010)\cite{ibr10}, Papageorgiou \textit{et al.} (2019)\cite{pap19} and Alsefri \textit{et al.} (2020)\cite{als20}. Despite the diversity of the aforementioned joint models, they are all based on a common key assumption of conditional independence between the longitudinal and event sub-models given either latent random effects or classes or both of them. However, this assumption is difficult to evaluate in practice due to the fact that latent variables are unobservable. 

Emura \textit{et al.} (2017)\cite{emu17} and (2018)\cite{emu18} introduced a joint frailty-copula model, where the dependency between time to tumour progression (e.g., relapse of cancer) and informative terminal event (e.g., death) are introduced by both the Clayton copula and a common frailty term. The reasoning for this, as the authors point out, is that there could still exist residual dependence between the two sub-models after conditioning on a common frailty term and some insufficiently collected covariates across studies. The same concern could also exist in the field of joint modelling of longitudinal and time-to-event data. When there are only a few covariates measured across studies, conditional independence between the longitudinal and survival processes or within the longitudinal sub-model given these covariates and a simple random-effect structure may be questionable.  A score test was proposed by Jacqmin-Gadda \textit{et al.} (2010)\cite{jac10} to assess the conditional independence assumption for a latent class joint model, but it is only applicable for this type of joint model. Dutta \textit{et al.} (2021)\cite{dut21} modelled the log-transformed event time and longitudinal process, given latent subject-specific random effects, by a multivariate normal distribution, which allows the two sub-models to have one more layer of association by the covariance matrix of multivariate normal distribution. However, their assumption about the event process may be a little restrictive and lack flexibility.

In this paper, a Gaussian copula joint model with random effects is proposed to address these issues. Copula (Hofert \textit{et al,} 2018\cite{hof18}) is a useful tool to introduce non-linear correlation between marginals. In copulas, the dependency structure  is adjustable while the marginal distributions are kept the same. Rizopoulos \textit{et al.} (2008a)\cite{riz08a} and (2008b)\cite{riz08b} considered applying copulas to specify the joint distribution of random effects, instead of multivariate normality, for longitudinal and time-to-event processes, thus breaking the restriction of linear correlation between two sub-models and increasing its flexibility in considering different dependence structures by using various copula functions. Malehi \textit{et al.} (2015)\cite{mal15} adopted the same idea to model random effects but using them to join the longitudinal measurements and gap time between recurrent events. However, both of these models still fall under the framework of conditional independence. Joint modelling of all longitudinal measurements and event time formulated by a multivariate Gaussian copula was proposed by Ganjali and Baghfalaki (2015)\cite{gan15}. Zhang, Charalambous and Foster (2021)\cite{zha21} extended this work by also considering a multivariate $t$ copula, and compared the performance between the two copula joint models in the prediction of survival probabilities. Suresh \textit{et al.} (2021a)\cite{sur21a} and (2021b)\cite{sur21b} applied bivariate Gaussian copulas directly on event time and biomarker measured at a single time point and maximised a pseudo-likelihood to perform a dynamic prediction on event time. 
In the approaches of Ganjali and Baghfalaki (2015)\cite{gan15}, Suresh \textit{et al.} (2021a)\cite{sur21a} and (2021b)\cite{sur21b}, unlike the latent random effects joint model, the parameters in the  copulas introduce the non-linear dependency between the marginals (sub-models), thus avoid the unverifiable assumption of conditional independence. However, the lack of random effects in these marginal models undermine their capabilities to perform fitting and prediction at a subject level especially for the longitudinal process (Zhang, Charalambous and Foster, 2021\cite{zha21}).
  Our proposed model applies a Gaussian copula to characterise the joint distribution of longitudinal and event time processes after conditioning on random effects, thus allowing us to predict the longitudinal process at an individual level, and a conventional joint model is a special case when the correlation parameters in the Gaussian copula shrink to zero.  

The remainder of this paper is organised as follows. In Section 2, the modelling framework is briefly introduced for the proposed model. In Section 3, simulation studies are performed to compare the differences in parameter estimation and the corresponding impact on predictions between the conventional joint model and the proposed model. A real data application is considered in section 4. In Section 5, some limitations on this copula joint model with random effects are discussed and possible future work is proposed.

\section{Copula joint model framework}
Suppose there are $n$ subjects $(i=1,...,n)$ followed over time. For the $i$th subject, let $\bm y_{i}=\left\{y_{i1}=y_{i}(t_{i1}),...,y_{in_{i}}=y_{i}(t_{in_{i}})\right\}^{'}$ be the biomarker measured over time. An observed event time $T_{i}=\mbox{min}(C_{i},T_{i}^{*})$ is also recorded for this subject, where $C_{i}$ and $T_{i}^{*}$ denote the right censoring time and true event time, respectively. Let $\delta_{i}=I(T_{i}^{*}<C_{i})$ be the associated censoring indicator, which takes value 1 if the event is observed and 0 otherwise.  The recording time points and censoring process are assumed to be uninformative conditional on baseline covariates.

\subsection{Longitudinal process}
The progression of longitudinal process is assumed to follow a linear mixed model  (Laird and Ware, 1982\cite{lai82}):
\begin{eqnarray*}
	y_{ij}=\bm x_{ij}^{'}\bm\beta_{1}+\bm z_{ij}^{'}\bm b_{i}+\varepsilon_{ij},\mbox{  }i=1,...,n, \mbox{   }j=1,...,n_{i},
\end{eqnarray*}
where the $r$-dimensional random effects $\bm b_{i}\sim N_{r}(\bm0,\bm D)$ and the error term $\varepsilon_{ij}=\varepsilon_{i}(t_{ij})\sim N(0,\sigma^{2}).$ $\bm x_{ij}=\bm x_{i}(t_{ij})$  and $\bm z_{ij}=\bm z_{i}(t_{ij})$ are  $p\times1$ and $r\times1$ covariate vectors for fixed effect $\bm\beta_{1}$ and random effect $\bm b_{i}$ at time $t_{ij},$ respectively. There could be common covariates shared by $\bm x_{ij}$  and $\bm z_{ij}$.

\subsection{Survival process}
A relative risk hazard model (Cox, 1972\cite{cox72}) with frailty terms is considered for the time-to-event data given by:
\begin{eqnarray*}
	h_{i}(t)=h_{0}(t)\mbox{exp}(\bm w_{i}^{'}\bm\beta_{2}+\bm z_{i}(t)^{'}(\bm\alpha\circ\bm b_{i})),
\end{eqnarray*}
where  $\bm w_{i}$ denotes a $q$-dimensional vector of explanatory variables with corresponding regression coefficient vector $\bm\beta_{2}$ and the $r$-dimensional vector $\bm\alpha$ characterises the association between the longitudinal and survival processes through its components, e.g. the random intercept is allowed to have a different association parameter to the random slope ($\circ$ is the Hadamard product).  $\bm x_{ij}$  and $\bm w_{i}$ could share common covariates. Let the baseline hazard function be a piecewise-constant function with $K-1$ equally spaced internal knots,
\[h_{0}(t)=\sum_{k=1}^{K}\lambda_{k}I(v_{k-1}<t\leq v_{k}),\]
where $0=v_{0}<v_{1}<\cdots<v_{K}=\mbox{max}\{t_{i},\,i=1,...,n\}$ split the time scale into $K$ intervals with a different constant baseline hazard at each interval.

\subsection{Gaussian copula and joint model specification}
Copulas can separate the marginal distributions from the dependency structure in a multivariate distribution. A $d$-dimensional copula is a multivariate distribution of a $d$-dimensional random vector $\bm U$ whose $d$ components all have uniform distributions over $(0,1)$, such that:
\[C(\bm u)=C(u_{1},...,u_{d})=\mbox{Pr}(U_{1}\leq u_{1},...,U_{d}\leq u_{d}).\]
Given a random variable $X$ with CDF $F_{X}(x)$, $U=F_{X}(X)$ has an uniform distribution. Suppose a multivariate random vector $\bm X=(X_{1},...,X_{d})$ have a joint CDF $F_{\bm X}(\bm x)$ and continuously increasing marginals. Then the joint distribution of $\left(U_{1}=F_{X_{1}}(X_{1}),...,U_{d}=F_{X_{d}}(X_{d})\right)$ is a copula, say $C_{\bm X}$, specified as:
\begin{eqnarray*}
	\displaystyle
	C_{\bm X}(u_{1},...,u_{d})&=&\mbox{Pr}(F_{X_{1}}(X_{1})\leq u_{1},...,F_{X_{d}}(X_{d})\leq u_{d})\\
	\displaystyle
	&=&\mbox{Pr}(X_{1}\leq F^{-1}_{X_{1}}(u_{1}),...,X_{d}\leq F^{-1}_{X_{d}}(u_{d}))\\
	\displaystyle
	&=&F_{\bm X}(F^{-1}_{X_{1}}(u_{1}),...,F^{-1}_{X_{d}}(u_{d}))\\
	\displaystyle
	&=&F_{\bm X}(x_{1},...,x_{d}).
\end{eqnarray*}
The copula density can be derived as:
\begin{eqnarray*}
	\displaystyle
	c_{\bm X}(u_{1},...,u_{d})&=&\frac{\partial^{d}C_{\bm X}(u_{1},...,u_{d})}{\partial u_{1}\cdots\partial u_{d}}\\
	\displaystyle
	&=&\frac{f_{\bm X}\left(F_{X_{1}}^{-1}(u_{1}),...,F_{X_{d}}^{-1}(u_{d})\right)}{\prod_{i=1}^{d}f_{X_{i}}
		\left(F_{X_{i}}^{-1}(u_{i})\right)},
\end{eqnarray*}
and the relationship between the joint density of $\bm X$ and the copula density can be express as:
\begin{eqnarray*}
	\displaystyle
	f_{\bm X}(x_{1},...,x_{d})&=&\frac{\partial^{d}F_{\bm X}(x_{1},...,x_{d})}{\partial x_{1}\cdots\partial x_{d}}\\
	\displaystyle
	&=&c_{\bm X}\left(F_{X_{1}}(x_{1}),...,F_{X_{d}}(x_{d})\right)\prod_{i=1}^{d}f_{X_{d}}(x_{d}).
\end{eqnarray*}

The most commonly used copula is the Gaussian copula defined as:
\[C^{\bm R}(\bm u)=\bm\Phi_{d}\left(\Phi^{-1}(u_{1}),...,\Phi^{-1}(u_{d});\bm R\right),\]
where $\Phi$ is the standard univariate normal CDF and $\bm\Phi_{d}\left(\cdot;\bm R\right)$ denotes the CDF of a $d$-dimensional multivariate normal variable with mean vector $\bm0$ and covariance matrix $\bm R.$ In Gaussian copula, $\bm R$ is essentially the correlation (covariance) matrix of $\left(\Phi^{-1}(U_{1}),...,\Phi^{-1}(U_{d})\right).$  Therefore, the Gaussian copula is fully specified by a correlation matrix. 

We are ready to discard the conventional conditional independence assumption typically assumed in the latent random effects joint models, and use the Gaussian copula to characterise the conditional joint distribution of $(\bm y_{i},T_{i}^{*}|\bm b_{i})$ for subject $i,$ which is specified by:
\begin{eqnarray}
	F_{T_{i}^{*},\bm y_{i}}(t_{i},\bm y_{i}|\bm  b_{i})=\bm\Phi_{n_{i}+1}\left(\Phi^{-1}(F_{T_{i}^{*}}(t_{i}|\bm b_{i})),\Phi^{-1}(F_{y_{i1}}(y_{i1}|\bm b_{i})),...,\Phi^{-1}(F_{y_{in_{i}}}(y_{in_{i}}|\bm b_{i}));\bm R_{i}\right),\label{mgcrejmcdf}
\end{eqnarray}
where $\Phi^{-1}(F_{T_{i}^{*}}(t_{i}|\bm b_{i}))$ and $\Phi^{-1}(F_{y_{ij}}(y_{ij}|\bm b_{i})),$ $j=1,...,n_{i},$ can be derived according to the longitudinal and survival sub-models in Sections 2.1 and 2.2. The correlation between $\bm y_{i}$ and $T_{i}^{*}$ is introduced by both random effects $\bm b_{i}$ and the copula correlation matrix, i.e., 
\begin{eqnarray*}
	\displaystyle
	\bm R_{i}=\left\{
	\begin{array}{ll}
		\displaystyle
		R_{(t_{i})}& \bm R_{(t_{i})(\bm y_{i})}
		\\
		\displaystyle
		\\
		\bm R_{(\bm y_{i})(t_{i})}& \bm R_{(\bm y_{i})}
	\end{array}
	\right\},
\end{eqnarray*}
where $R_{(t_{i})}=\mbox{var}\left(\Phi^{-1}(F_{T_{i}^{*}}(t_{i}|\bm b_{i}))\right)=1$, \[\bm R_{\bm y_{i}}=\mbox{corr}\left(\left[\Phi^{-1}(F_{y_{i1}}(y_{i1}|\bm b_{i})),...,\Phi^{-1}(F_{y_{in_{i}}}(y_{in_{i}}|\bm b_{i}))\right]\right)\]
and
\[\bm R_{(t_{i})(\bm y_{i})}=\mbox{corr}\left(\Phi^{-1}(F_{T_{i}^{*}}(t_{i}|\bm b_{i})),\left[\Phi^{-1}(F_{y_{i1}}(y_{i1}|\bm b_{i})),...,\Phi^{-1}(F_{y_{in_{i}}}(y_{in_{i}}|\bm b_{i}))\right]\right).\]
Under this model specification, the conditional independence between $T_{i}^{*}$ and $\bm y_{i}$ will be a special case when $\bm R_{(t_{i})(\bm y_{i})}$ degenerates to a vector of zeros. Further assuming that $\bm R_{\bm y_{i}}$ is an identity matrix, implies conditional independence between successive longitudinal measurements within subjects, which is the most commonly seem assumption in the literature of joint modelling of longitudinal and time-to-event data.

Let $\bm\theta$ be the vector of parameters in the joint model and $\bm Z_{i|\bm b_{i}}=\left(Z_{t_{i}|\bm b_{i}},\bm Z_{\bm y_{i}|\bm b_{i}}^{'}\right)^{'}$, where $Z_{t_{i}|\bm b_{i}}=\Phi^{-1}\left(F_{T_{i}^{*}}(t_{i}|\bm b_{i})\right)$ is a scalar and $\displaystyle\bm Z_{\bm y_{i}|\bm b_{i}}=\frac{\bm y_{i}-\bm x_{i}\bm\beta_{1}-\bm z_{i}\bm b_{i}}{\sigma}$ is a vector with  $\bm x_{i}=\left\{\bm x_{i1},...,\bm x_{in_{i}}\right\}^{'}$ and $\bm z_{i}=\left\{\bm z_{i1},...,\bm z_{in_{i}}\right\}^{'}.$ Note vector $\bm Z_{i|\bm b_{i}}$ has mean $\bm0$ and covariance matrix $\bm R_{i}$ conditional on random effects $\bm b_{i}.$
Under the normality assumption of the longitudinal process, equation (\ref{mgcrejmcdf}) can be simplified as:
\begin{eqnarray*}
	F_{T_{i}^{*},\bm y_{i}}(t_{i},\bm y_{i}|\bm  b_{i})=\bm\Phi_{n_{i}+1}\left(\bm Z_{i|\bm b_{i}};\bm R_{i}\right).
\end{eqnarray*}
If the  event time is observed at $t_{i}$ with $\delta_{i}=1,$ the corresponding conditional joint pdf of $(T_{i}=t_{i}, \bm y_{i}|\bm b_{i})$ is given by:
\begin{eqnarray*}
	\displaystyle
	f_{T_{i},\bm y_{i}}(t_{i},\bm y_{i}|\bm  b_{i})
	=f_{T_{i}^{*},\bm y_{i}}(t_{i},\bm y_{i}|\bm  b_{i})=\sigma^{-n_{i}}\bm\phi_{n_{i}+1}\left(\bm Z_{i|\bm b_{i}};\bm R_{i}\right)\frac{f_{T_{i}^{*}}(t_{i}|\bm b_{i})}{\phi\left(Z_{t_{i}|\bm b_{i}}\right)},
\end{eqnarray*}
where $\phi$ is the standard univariate normal pdf and $\bm\phi_{n}\left(\mbox{ };\bm\mu,\bm\Sigma\right)$ denotes the pdf of a $n$ dimensional multivariate normal variable with mean vector $\bm\mu$ and covariance matrix $\bm\Sigma.$ When $\bm\mu=\bm0,$ it is further simplified as $\bm\phi_{n}\left(\mbox{ };\bm\Sigma\right).$ 
If this individual is censored at $t_{i}$ with $\delta_{i}=0,$ the conditional joint pdf of $(T_{i}>t_{i}, \bm y_{i}|\bm b_{i})$ is given by:
\begin{eqnarray*}
	\displaystyle
	f_{T_{i}^{*},\bm y_{i}}(T_{i}^{*}>t_{i},\bm y_{i}|\bm  b_{i})
	&=&\int_{t_{i}}^{\infty}\sigma^{-n_{i}}\bm\phi_{n_{i}+1}\left(\bm Z_{i|\bm b_{i}};\bm R_{i}\right)\frac{f_{T_{i}^{*}}(u|\bm b_{i})}{\phi\left(Z_{u|\bm b_{i}}\right)}\,du\\
	\displaystyle
	&=&\sigma^{-n_{i}}\bm\phi_{n_{i}}\left(\bm Z_{\bm y_{i}|\bm b_{i}};\bm R_{(\bm y_{i})}\right)\int_{t_{i}}^{\infty}\phi_{1}\left(Z_{u|\bm b_{i}};\mu_{i}^{t_{i}|\bm y_{i},\bm b_{i}},\left(\sigma_{i}^{t_{i}|\bm y_{i},\bm b_{i}}\right)^{2} \right)
	\frac{f_{T_{i}^{*}}(u|\bm b_{i})}{\phi\left(Z_{u|\bm b_{i}}\right)}\,du
	\\
	\displaystyle
	&=&\sigma^{-n_{i}}\bm\phi_{n_{i}}\left(\bm Z_{\bm y_{i}|\bm b_{i}};\bm R_{(\bm y_{i})}\right)\int_{Z_{t_{i}|\bm b_{i}}}^{\infty}\phi_{1}\left(z;\mu_{i}^{t_{i}|\bm y_{i},\bm b_{i}},\left(\sigma_{i}^{t_{i}|\bm y_{i},\bm b_{i}}\right)^{2}\right)\,dz
	\\
	&=&\sigma^{-n_{i}}\bm\phi_{n_{i}}\left(\bm Z_{\bm y_{i}|\bm b_{i}};\bm R_{(\bm y_{i})}\right)\Phi\left(-\frac{Z_{t_{i}|\bm b_{i}}-\mu_{i}^{t_{i}|\bm y_{i},\bm b_{i}}}{\sigma_{i}^{t_{i}|\bm y_{i},\bm b_{i}}}\right),
\end{eqnarray*}
where
\begin{eqnarray*}
	\mu_{i}^{t_{i}|\bm y_{i},\bm b_{i}}&=&\bm R_{(t_{i})(\bm y_{i})}\bm R_{(\bm y_{i})}^{-1}\bm Z_{\bm y_{i}|\bm b_{i}},\\
	\left(\sigma_{i}^{t_{i}|\bm y_{i},\bm b_{i}}\right)^{2}&=&1-\bm R_{(t_{i})(\bm y_{i})}\bm R_{(\bm y_{i})}^{-1}\bm R_{(\bm y_{i})(t_{i})}.
\end{eqnarray*}
The log-likelihood function is:
\begin{eqnarray}
	\displaystyle
\nonumber	l(\bm\theta)&=&\sum_{i}^{n}\mbox{log}\left\{\delta_{i}f_{T_{i}^{*},\bm y_{i}}(t_{i},\bm y_{i})+(1-\delta_{i})f_{T_{i}^{*},\bm y_{i}}(T_{i}^{*}>t_{i},\bm y_{i})\right\}\\
	&=&\sum_{i}^{n}\mbox{log}\int_{\bm b_{i}}\left\{\delta_{i}f_{T_{i}^{*},\bm y_{i}}(t_{i},\bm y_{i}|\bm b_{i})+(1-\delta_{i})f_{T_{i}^{*},\bm y_{i}}(T_{i}^{*}>t_{i},\bm y_{i}|\bm b_{i})\right\}f_{\bm b_{i}}(\bm b_{i})\,d\bm b_{i}.\label{jointmgcopranlik}
\end{eqnarray}

As the above integral does not have a closed form expression, a multivariate Gaussian quadrature technique (J\"{a}ckel, 2005\cite{jac05}) is applied to approximate the integration. However, the main density of $f_{\bm b_{i}}(\bm b_{i})$ is centred around $\bm0,$ if it is directly applied as a weighting kernel, the  locations of its quadrature points are evenly scatter around $\bm0,$ while $\displaystyle \delta_{i}f_{T_{i}^{*},\bm y_{i}}(t_{i},\bm y_{i}|\bm b_{i})+(1-\delta_{i})f_{T_{i}^{*},\bm y_{i}}(T_{i}^{*}>t_{i},\bm y_{i}|\bm b_{i})$ are more likely to be maximised around the subject-specific random effect, which is not usually around $\bm0.$ Therefore, applying $f_{\bm b_{i}}(\bm b_{i})$ as a common weighting kernel for all the subjects does not account for the difference in subject-specific random effects and can make the numerical approximation for (\ref{jointmgcopranlik}) inaccurate even with a large number of quadrature points.
For faster and more accurate calculation, a similar approach to the adaptive Gauss-Hermite rule in Rizopoulos (2012a)\cite{riz12a} is applied.  Expression (\ref{jointmgcopranlik}) is rearrange as:
\begin{eqnarray}
	\displaystyle
\nonumber	&&\mbox{log}\int_{\bm b_{i}}\left\{\delta_{i}f_{T_{i}^{*},\bm y_{i}}(t_{i},\bm y_{i}|\bm b_{i})+(1-\delta_{i})f_{T_{i}^{*},\bm y_{i}}(T_{i}^{*}>t_{i},\bm y_{i}|\bm b_{i})\right\}f_{\bm b_{i}}(\bm b_{i})\,d\bm b_{i}\\
\nonumber	&=&\mbox{log}\int_{\bm b_{i}}f_{\bm y_{i}}(\bm y_{i})\left\{\delta_{i}f_{T_{i}}(t_{i}|\bm y_{i},\bm b_{i})+(1-\delta_{i})f_{T_{i}^{*}}(T_{i}^{*}>t_{i}|\bm y_{i},\bm b_{i})\right\}f_{\bm b_{i}}(\bm b_{i}|\bm y_{i})\,d\bm b_{i}\\
\nonumber	&=&\mbox{log}f_{\bm y_{i}}(\bm y_{i})+\mbox{log}\int_{\bm b_{i}}\left\{\delta_{i}f_{T_{i}}(t_{i}|\bm y_{i},\bm b_{i})+(1-\delta_{i})f_{T_{i}^{*}}(T_{i}^{*}>t_{i}|\bm y_{i},\bm b_{i})\right\}f_{\bm b_{i}}(\bm b_{i}|\bm y_{i})\,d\bm b_{i}\\
\nonumber	&=&\mbox{log}\bm\phi_{n_{i}}\left(y_{i1}-\bm x_{i1}^{'}\bm \beta_{1},...,
	y_{in_{i}}-\bm x_{in_{i}}^{'}\bm \beta_{1};\bm V_{\bm y_{i}}\right)\\
	&+&\mbox{log}\int_{\bm b_{i}}\left\{\delta_{i}f_{T_{i}}(t_{i}|\bm y_{i},\bm b_{i})+(1-\delta_{i})f_{T_{i}^{*}}(T_{i}^{*}>t_{i}|\bm y_{i},\bm b_{i})\right\}f_{\bm b_{i}}(\bm b_{i}|\bm y_{i})\,d\bm b_{i},\label{jointmgcoprannewlik}
\end{eqnarray}
where $\bm V_{\bm y_{i}}$ is the variance-covariance matrix of longitudinal process after integrating out random effects $\bm b_{i},$ $f_{\bm b_{i}}(\bm b_{i}|\bm y_{i})$ can be derived by the joint normality of $\bm y_{i}$ and $\bm b_{i},$ 
\begin{eqnarray*}
	     f_{T_{i}}(t_{i}|\bm y_{i},\bm b_{i})=\phi_{1}\left(Z_{t_{i}|\bm b_{i}};\mu_{i}^{t_{i}|\bm y_{i},\bm b_{i}},\left(\sigma_{i}^{t_{i}|\bm y_{i},\bm b_{i}}\right)^{2}\right)\frac{f_{T_{i}^{*}}(t_{i}|\bm b_{i})}{\phi\left(Z_{t_{i}|\bm b_{i}}\right)}
	     \end{eqnarray*}
     and
	     \begin{eqnarray*}
	     f_{T_{i}^{*}}(T_{i}^{*}>t_{i}|\bm y_{i},\bm b_{i})=\Phi\left(-\frac{Z_{t_{i}|\bm b_{i}}-\mu_{i}^{t_{i}|\bm y_{i},\bm b_{i}}}{\sigma_{i}^{t_{i}|\bm y_{i},\bm b_{i}}}\right). 
	\end{eqnarray*}
The new weighting kernel $f_{\bm b_{i}}(\bm b_{i}|\bm y_{i}),$ after including the longitudinal information from this subject, is subject adaptive and more likely to be concentrated around its subject-specific random effects, thus the new locations of the quadrature points are expected to be closer to the main mass of $\delta_{i}f_{T_{i}}(t_{i}|\bm y_{i},\bm b_{i})+(1-\delta_{i})f_{T_{i}^{*}}(T_{i}^{*}>t_{i}|\bm y_{i},\bm b_{i})$ and higher accuracy of numerical approximation is expected to be achieved in  (\ref{jointmgcoprannewlik}) by fewer nodes compared to the original parameterisation in (\ref{jointmgcopranlik}). The maximisation of (\ref{jointmgcoprannewlik}) can be executed numerically by applying the Newton-type algorithm (Dennis, \textit{et al,} 1983\cite{den83}) or the approach of Nelder and Mead (1965)\cite{nel65}, which are implemented by the \verb|nlm| and \verb|optim| functions, respectively, in \verb|R| and the standard errors can be estimated from the inverse hessian matrix as a by-product of the two functions. The initial values for the longitudinal and survival sub-models can be obtained by fitting a regular joint model via  \verb|jointModel| or  \verb|joint| functions from \verb|JM| or \verb|joineR| packages, respectively, while the correlation parameters in the Gaussian copula can be initialised as 0s. In the simulation study of the next section,  parameter estimation  is conducted by the approach described here.

\section{Simulation studies}
We are interested in how the estimations of parameters are distorted by fitting a regular joint modelling assuming conditional independence if this assumption does not hold for the true dataset. In the data generating process, we consider different structures of correlation matrix $\bm R$ in the Gaussian copula while keeping the conditional distributions of the two sub-models fixed. 
\subsection{Parameter estimation}
In the following simulation studies, $N=500$ samples, each with a sample size of $n=200$ subjects, are generated from a Gaussian copula joint model with random effects. While the proposed joint model allows different association parameters for each component of random effects through vector $\bm\alpha,$ it is assumed to be a constant vector in the simulation studies for simplicity. 

The longitudinal process is taken to be:
\begin{eqnarray*}
	y_{ij}=\beta_{10}+\beta_{11}t_{ij}+\beta_{12}x_{i1}+\beta_{13}x_{i2}+\beta_{14}x_{i3}+\beta_{15}x_{i4}+b_{i0}+b_{i1}t_{ij}+\varepsilon_{ij},
\end{eqnarray*}

and the survival process is described by:
\begin{eqnarray*}
	h_{i}(t)=h_{0}(t)\mbox{exp}\left(\beta_{21}x_{i1}+\beta_{22}x_{i2}+\beta_{23}x_{i3}+\beta_{24}x_{i4}+\alpha(b_{i0}+b_{i1}t)\right),
\end{eqnarray*}
where $x_{i1}$ and $x_{i2}$ have probability 0.5 taking value 1 or 0, while $x_{i3}$, $x_{i4}$ and $x_{i5}$ are factors following a categorical distribution with $p=(0.3,0.5,0.2),$
mimicking variables like treatment, gender and age groups. Measurement times are scheduled at $t=0,1,2,...,10.$ An independent censoring process follows an exponential distribution with rate 0.011,  resulting in around $55\%$ dropout rate of longitudinal measurements and $52\%$ censoring rate of event time at $t=10.$ Although we assume a linear function for the longitudinal trajectory and a two dimensional random effects structure in this simulation study, the trend of the longitudinal process can be generalised to be non-linear and the dimension of random effects can be expended to more than random intercept and slope by using the same estimating procedure. Conditional on the random effects, which is also assumed for the simulation and discussion afterwards, two cases are considered for the correlation structure $\bm R_{(t_{i})(\bm y_{i})}$ between the longitudinal and time-to-event sub-models for data generation: 
\begin{itemize}
	\item \textbf{case 1}: an exchangeable structure with constant correlation $\rho_{ty}=0.4$; 
	\item  \textbf{case 2}: an AR1 structure with increasing correlation $\rho_{ty}^{|t-11|}$ and $\rho_{ty}=0.75,$ indicating later measurements have stronger correlations with the event time. 
\end{itemize}
In both cases, the correlation structure within the longitudinal process is assumed to be an AR1 with $\rho_{y}=0.5.$

For parameter estimations, assuming the marginals being correctly specified, four scenarios are considered: 
\begin{itemize}
	\item  \textit{scenario 1}: conditional independence between the two sub-models and conditional independence in the longitudinal process, i.e., $\bm R_{i}$ is misspecified as $\bm I_{n_{i}}.$
	\item  \textit{scenario 2}: the correlation structure $\bm R_{(\bm y_{i})}$ within the longitudinal process is correctly specified  as an AR1 but the correlation between the longitudinal and survival processes is ignored, i.e., $\bm R_{(t_{i})(\bm y_{i})}=\bm0;$  
	\item  \textit{scenario 3}: the correlation structure $\bm R_{(\bm y_{i})}$ within the longitudinal process is correctly specified as an AR1 but the correlation between the longitudinal and survival processes is misspecified as an AR1 structure for case 1 (an exchangeable structure for case 2); 
	\item  \textit{scenario 4}: the correlation structure $\bm R_{i}$ is correctly specified.
\end{itemize}

Scenarios 2 and 3, where the correlation structure of the longitudinal  process is correct but the extra correlation between the two sub-models is ignored or misspecified, are designed to verify the robustness of the parameter estimates under the misspecification of $\bm R_{(t_{i})(\bm y_{i})},$ given $\bm R_{(\bm y_{i})}$ is correct. Many authors, such as Henderson \textit{et al.} (2000)\cite{hen00}, Wang \textit{et al.} (2001)\cite{wan01} and Xu \textit{et al,} (2001)\cite{xu01},

\begin{landscape}
	\vspace*{\fill}
\begingroup
\setlength{\tabcolsep}{6pt} 
\renewcommand{\arraystretch}{1.18} 
\begin{table}[H]
	\tbl{\textbf{{\scriptsize Outputs of parameter estimations for \textbf{case 1} (constant $\rho_{ty}=0.4$ for $\bm R_{(t_{i})(\bm y_{i})}$) in \textit{scenario 1}: $\bm R_{i}$ is misspecified as $\bm I_{n_{i}},$  \textit{scenario 2}: $\bm R_{(\bm y_{i})}$ is correctly specified but $\bm R_{(t_{i})(\bm y_{i})}$ is misspecified as $\bm0,$ \textit{scenario 3}: $\bm R_{(\bm y_{i})}$ is correctly specified but $\bm R_{(t_{i})(\bm y_{i})}$ is misspecified as an AR1 structure and \textit{scenario 4}: $\bm R_{i}$ is correctly specified when there are at most 11 measurements.}}}
	{\begin{tabular}{lccccccccccccccccc}
			\toprule
			True value &$\beta_{10}$&$\beta_{11}$&$\beta_{12}$&$\beta_{13}$&$\beta_{14}$&$\beta_{15}$&$\beta_{21}$&$\beta_{22}$&$\beta_{23}$&$\beta_{24}$&$\sigma$&$D_{11}$&$D_{22}$&$D_{12}$&$\alpha$&$\rho_{ty}$  &$\rho_{y}$
			\\
			& 10         & -0.5       & 1                 & 0.5             &  0.5         & 1          &-2        &-1      & -1.5       &-2   &2 & 2       & 0.2       & -0.1         & -0.5      &     0.4        &     0.5 
			\\
			\hline
			Scenario 1 &                   &                   &                    &                   &                   &                   &                    &                    &                     &                   &
			\\
			\cdashline{1-18}
			Est.        &10.131 & -0.462 & 0.926 & 0.457&0.420 & 0.907 & -2.287 & -1.138 &-1.678&-2.259 &1.652&3.672 & 0.259  & -0.348 & -0.626&\textemdash& \textemdash
			\\
			SE         & 0.404 & 0.041   &  0.288&  0.287& 0.427 & 0.393& 0.324  & 0.290 & 0.385& 0.367 &0.035& 0.493 & 0.038 &  0.111   &  0.074  & \textemdash &  \textemdash
			\\
			SD        &0.400  & 0.044  & 0.298  & 0.294 &0.415  &0.389  &0.316    &0.287  &0.359 &0.344 &0.040&0.512  &0.035  & 0.108   & 0.071   & \textemdash&  \textemdash 
			\\
			RMSE   & 0.421 & 0.058  & 0.307  &  0.297 &0.422  &0.399  &0.427   & 0.318 & 0.401 & 0.430 &0.350&1.748  & 0.068 & 0.270  &  0.144  & \textemdash&  \textemdash
			\\
			CP	      & 0.948 & 0.840 & 0.932  & 0.934 & 0.952 & 0.946  &0.866  &0.936  & 0.938 &0.898&0.000 & 0.070 & 0.704 &0.396  & 0.606   &  \textemdash&  \textemdash
			\\
			ECP	     &0.942 & 0.864 & 0.940  & 0.940 & 0.952 & 0.942  & 0.856 & 0.928 & 0.922 & 0.870 &0.000 & 0.076 & 0.610 & 0.364 & 0.578    & \textemdash&  \textemdash
			\\
			\cdashline{1-18}
			Scenario 2   & &  &   &   &    &   &  &  &  &  &
			\\
			\cdashline{1-18}
			Est.     &10.192  & -0.473&0.900  &0.445   &0.402 &0.882&-2.447 &-1.211 &-1.755&-2.377&1.910 &2.386   & 0.207 &-0.152 &-0.777 & \textemdash& 0.453 
			\\
			SE       &0.406   & 0.042 &  0.290&  0.289& 0.429& 0.394 & 0.367& 0.318& 0.414& 0.399&0.067 &0.506   & 0.037 &  0.111&  0.114 &  \textemdash& 0.037
			\\
			SD       & 0.402  & 0.043 & 0.294&  0.291& 0.410 & 0.382 & 0.351& 0.316 & 0.388& 0.377&0.061  & 0.519  & 0.034& 0.109 & 0.102  &  \textemdash&0.034 
			\\
			RMSE  &0.445   & 0.051  &0.311  & 0.296 &0.421  & 0.399&0.568 & 0.380 & 0.464 & 0.532 &0.108& 0.646 & 0.034 & 0.120 & 0.295 &  \textemdash&0.058 
			\\
			CP	    & 0.934 & 0.882  & 0.934 & 0.934& 0.956 & 0.938 &0.778  &0.904 &0.920 & 0.852&0.752 & 0.870 & 0.958 & 0.946 & 0.318&  \textemdash &0.784 
			\\
			ECP	   & 0.932 & 0.892 & 0.938 & 0.936 &0.942  & 0.932 & 0.758 &0.904 & 0.896& 0.822&0.694& 0.872 & 0.942 & 0.936 &0.228 &  \textemdash & 0.726
			\\
			\cdashline{1-18}
			Scenario 3   & &  &   &   &    &   &  &  &  &  &
			\\
			\cdashline{1-18}
			Est.    &10.197  &-0.474 &0.895    &0.443   & 0.395 & 0.878 &-2.433& -1.210& -1.736&-2.363 &1.912&2.369   &0.207  &-0.151  &-0.770 & -0.030 &  0.454
			\\
			SE      & 0.405  & 0.043 & 0.290   &  0.289 &  0.428& 0.394 & 0.365 & 0.317 & 0.411 & 0.397 &0.068& 0.506  & 0.037 &  0.111 &  0.115 & 0.216   & 0.037
			\\
			SD      & 0.400  & 0.046 &  0.293 &  0.293 & 0.409 &  0.377& 0.356 & 0.317 & 0.370  & 0.366 &0.061& 0.516  & 0.034 &  0.107 &  0.103&0.241    &0.034
			\\
			RMSE & 0.446  & 0.053 & 0.311   & 0.298  & 0.422 & 0.396 & 0.560 & 0.380 & 0.439 & 0.515  &0.107& 0.634 & 0.034 & 0.119   & 0.288 & 0.493  & 0.057
			\\
			CP	    & 0.938  & 0.876 & 0.928  & 0.934  & 0.954 & 0.944 &0.788  &0.898 &0.936   & 0.866 &0.758& 0.880  & 0.960 & 0.950  & 0.348 & 0.446  &0.790 
			\\
			ECP	   & 0.930 & 0.908 & 0.932  & 0.938  & 0.936 &  0.934 & 0.776 & 0.898& 0.896 & 0.824 &0.706& 0.882 & 0.944 & 0.936  & 0.262 & 0.526   &  0.748
			\\
			\cdashline{1-18}
			Scenario 4   & &  &   &   &    &   &  &  &  &  &
			\\
			\cdashline{1-18}
			Est.    &10.027  &-0.496&0.984   &0.480    & 0.480 & 0.989  &-2.037& -1.001 & -1.527&-2.038 &2.016&1.791   &0.199   &-0.091  &-0.520 & 0.405 &  0.507  
			\\
			SE      & 0.398  & 0.043 &  0.286 &  0.285 &  0.422& 0.388  & 0.292 & 0.249 & 0.328 & 0.318  &0.092& 0.599 & 0.038 &  0.115  &  0.090 & 0.051  & 0.043
			\\
			SD      & 0.402  & 0.044 &  0.295 &  0.285& 0.417  &  0.386  & 0.288 & 0.253 & 0.313  & 0.308 &0.085& 0.596 & 0.034 &  0.109  &  0.090 &0.046  &0.042
			\\
			RMSE  & 0.403 & 0.044 & 0.295  &0.286  &0.417   & 0.386   & 0.291 & 0.253  & 0.313 & 0.310  & 0.086& 0.631 &0.034  &  0.110   & 0.092  & 0.046  & 0.042
			\\
			CP	    & 0.958 & 0.956 & 0.950  & 0.944 & 0.956 & 0.954   &0.948  &0.954  &0.962  & 0.954 &0.972 & 0.940 & 0.974 & 0.948  & 0.948  & 0.980  &0.956 
			\\
			ECP	   &0.958  & 0.960 & 0.952 & 0.944  & 0.956 & 0.952  & 0.948 & 0.954 & 0.944  & 0.942 &0.944& 0.938 & 0.948 & 0.940  & 0.948  & 0.956  & 0.952  
			\\
		\hline
	\end{tabular}}
\label{sim11excar}
\end{table}
\endgroup
		\vspace*{\fill}
\end{landscape}

\begin{landscape}
	\vspace*{\fill}
\begingroup
\setlength{\tabcolsep}{6pt} 
\renewcommand{\arraystretch}{1.18} 
\begin{table}[H]
	\tbl{\textbf{{\scriptsize Outputs of parameter estimations for \textbf{case 2} (continuous AR1 with $0.5^{|t-10|}$ for $\bm R_{(t_{i})(\bm y_{i})}$) in \textit{scenario 1}: $\bm R_{i}$ is misspecified as $\bm I_{n_{i}},$  \textit{scenario 2}: $\bm R_{(\bm y_{i})}$ is correctly specified but $\bm R_{(t_{i})(\bm y_{i})}$ is misspecified as $\bm0,$ \textit{scenario 3}: $\bm R_{(\bm y_{i})}$ is correctly specified but $\bm R_{(t_{i})(\bm y_{i})}$ is misspecified as an exchangeable structure and \textit{scenario 4}: $\bm R_{i}$ is correctly specified when there are at most 11 measurements.}}}
	{\begin{tabular}{lccccccccccccccccc}
			\toprule
		&$\beta_{10}$&$\beta_{11}$&$\beta_{12}$&$\beta_{13}$&$\beta_{14}$&$\beta_{15}$&$\beta_{21}$&$\beta_{22}$&$\beta_{23}$&$\beta_{24}$&$\sigma$&$D_{11}$&$D_{22}$&$D_{12}$&$\alpha$&$\rho_{ty}$  &$\rho_{y}$
		\\
		& 10      & -0.5     & 1         & 0.5        &  0.5       & 1         &-2        &-1        & -1.5        &-2       &2      & 2       & 0.2       & -0.1          & -0.5         &     0.5        &     0.5  
			\\
			\hline
			Scenario 1 &                   &                   &                    &                   &                   &                   &                    &                    &                     &                   &
			\\
			\cdashline{1-18}
			Est.       &9.930  & -0.463& 1.011   & 0.506  &0.519  & 0.975  & -1.930 & -0.956 &-1.459&-1.925 &1.655 &3.653 & 0.256  & -0.340 & -0.404&\textemdash  & \textemdash
			\\
			SE         & 0.408 & 0.044 &  0.289 &  0.288& 0.430 & 0.395  & 0.273  & 0.248  & 0.326& 0.307  &0.035 & 0.491 & 0.038 &  0.108   &  0.059 & \textemdash &  \textemdash
			\\
			SD         &0.392  &0.045  & 0.300  & 0.291 &0.455  &0.396   &0.274   &0.246   &0.316  &0.294  &0.038 &0.491  &0.039  & 0.110    & 0.060  & \textemdash &  \textemdash 
			\\
			RMSE    &0.398  &0.058  & 0.300  & 0.291  & 0.455& 0.396  & 0.282  & 0.250  &0.319  &0.303  &0.347 & 1.724 & 0.069 & 0.264   & 0.113  &\textemdash  &  \textemdash
			\\
			CP	      & 0.968  & 0.858 &0.948  & 0.948 & 0.936 & 0.944  &0.930  &0.962    & 0.954 &0.960  &0.000 & 0.066 & 0.696 &0.434   & 0.614   &  \textemdash &  \textemdash
			\\
			ECP	     & 0.958  &0.874  &0.954  &0.950  & 0.942 & 0.944  & 0.932 & 0.962   & 0.946 & 0.944 &0.000 & 0.066 & 0.704 & 0.444  & 0.632   & \textemdash  &  \textemdash 
			\\
			\cdashline{1-18}
			Scenario 2   & &  &   &   &    &   &  &  &  &  &
			\\
			\cdashline{1-18}
			Est.     &9.987   & -0.474&0.993  &0.496   &0.499  &0.954  &-2.126  &-1.045 &-1.560 &-2.077 &2.016  &1.748   & 0.194   &-0.090   &-0.576   & \textemdash   & 0.505
			\\
			SE       &0.400   & 0.044 &  0.285&  0.284 & 0.422 & 0.389 & 0.314  & 0.273 & 0.353 & 0.339  &0.089 &0.548  & 0.038   &  0.109   &  0.107   &  \textemdash  & 0.042
			\\
			SD       & 0.389  & 0.045 & 0.296 &  0.284 & 0.451  & 0.386 & 0.317   & 0.268 & 0.339 & 0.322 &0.091 & 0.552  & 0.038  & 0.114    & 0.103    &  \textemdash  &0.041
			\\
			RMSE  & 0.389  & 0.052 & 0.295 & 0.284  & 0.451  & 0.389 & 0.341  & 0.272  & 0.344 & 0.330 &0.092 & 0.607  & 0.039   & 0.112    & 0.134    & \textemdash   & 0.045
			\\
			CP	    & 0.964  & 0.910  & 0.948  & 0.954  & 0.936 & 0.948  &0.938  &0.956   &0.960 & 0.960 &0.938 & 0.928  & 0.940   & 0.944  & 0.888    &  \textemdash   &0.924 
			\\
			ECP	   & 0.960  & 0.912  & 0.956  & 0.954  & 0.946 & 0.948  & 0.940 & 0.952  & 0.948 & 0.942&0.940 & 0.928  & 0.942   & 0.946  & 0.890    &  \textemdash   &0.938 
			\\
			\cdashline{1-18}
			Scenario 3   & &  &   &   &    &   &  &  &  &  &
			\\
			\cdashline{1-18}
			Est.    &9.984   &-0.474  &0.996   &0.498   & 0.496 & 0.957 &-2.118  & -1.046& -1.554&-2.069 &2.018 & 1.737   &0.194    &-0.089&-0.568 & 0.008 &  0.506
			\\
			SE      & 0.402  & 0.044  &  0.286 &  0.284 &  0.423& 0.389 & 0.324  & 0.275 & 0.356 & 0.345  &0.095 & 0.590  & 0.038  &  0.112 &  0.118 & 0.117   & 0.045 
			\\
			SD      & 0.388  & 0.045  &  0.297 &  0.284& 0.451  &  0.386& 0.343  & 0.270 & 0.355 & 0.343  &0.096 & 0.582  & 0.039  &  0.113 &  0.128&0.129   &0.047 
			\\
			RMSE & 0.388  & 0.052  & 0.296  & 0.283 &  0.450 & 0.388 & 0.362 & 0.274 & 0.359  & 0.349  &0.097 & 0.638  & 0.039  & 0.113  & 0.145  &0.509  &0.047 
			\\
			CP	    & 0.966  & 0.916  & 0.946  & 0.958 & 0.934  & 0.950 &0.932  &0.946  &0.952   & 0.950  &0.942 & 0.924  & 0.934  & 0.940  & 0.900 & 0.012 &0.932
			\\
			ECP	   & 0.960  & 0.916  & 0.952  & 0.958 & 0.942  & 0.948 & 0.946 & 0.944 & 0.950  & 0.950  &0.942 & 0.924  & 0.936  & 0.944  & 0.914 & 0.028 &0.944
			\\
			\cdashline{1-18}
			Scenario 4   & &  &   &   &    &   &  &  &  &  &
			\\
			\cdashline{1-18}
			Est.    &9.982   &-0.503 &1.021    &0.507  & 0.526  & 0.992   &-2.027& -1.004 & -1.516 &-2.010 &2.009 &1.814    &0.198  &-0.093  &-0.509 & 0.511 &  0.501 
			\\
			SE      & 0.403  & 0.046 &  0.287 &  0.286 &  0.425& 0.391   & 0.300 & 0.256  & 0.333  & 0.323 &0.089 & 0.560  & 0.038 &  0.110  &  0.104 & 0.177 & 0.043  
			\\
			SD      & 0.383 & 0.049  &  0.296 &  0.278& 0.455  &  0.385  & 0.302  & 0.258  & 0.329 & 0.309 &0.091 & 0.559 & 0.039 &  0.111   &  0.110  &0.226 &0.045
			\\
			RMSE & 0.383 & 0.049  & 0.296  & 0.277  & 0.455  & 0.385  & 0.303  & 0.258  & 0.329 & 0.309 & 0.091 &0.589  & 0.039 &  0.111   & 0.111   & 0.226 &0.045   
			\\
			CP	    & 0.968 & 0.944 & 0.936  & 0.964  & 0.940  & 0.964  &0.958   &0.956   &0.960  & 0.972 &0.942 & 0.940 & 0.936 & 0.942  & 0.944  & 0.926 &0.928  
			\\
			ECP	   & 0.968 & 0.954 & 0.946  & 0.950  & 0.950  & 0.960  & 0.960  & 0.956  & 0.960 & 0.958 &0.948 & 0.940 & 0.942 & 0.944  & 0.944  & 0.958 &0.944
			\\
			\hline
	\end{tabular}}
\label{sim11carcar}
\end{table}
\endgroup
		\vspace*{\fill}
\end{landscape}

\noindent  have  discussed applying an elaborate mean-zero stochastic process to characterise serial correlation within longitudinal measurements, which enhances the flexibility of the linear mixed model to accommodate the different correlation structures in the longitudinal data. However, increasing the flexibility of the longitudinal process by adding additional random effects or covariates may result in overfitting problems. For example, adding extra random effects to scenario 1, where the marginals have already been correctly modelled, makes the sub-models overcomplicated.  Even though the great flexibility of these models may allow them to capture the true structure of $\bm R_{(\bm y_{i})},$ the extra dependency introduced by $\bm R_{(t_{i})(\bm y_{i})}$ is still not considered in these models since they assume conditional independence between the two sub-models. According to the simulation results of scenarios 2 and 3, we are able to evaluate whether fitting the joint models, which are capable of capturing the true correlation structure within the longitudinal process, is a sufficient action when there is in fact extra correlation between the two sub-models.

Each subject in the data set is originally generated with up to 11 measurements. To study the impact of the number of longitudinal measurements on parameter estimation, we consider keeping only the measurements at $t= 0,2,4,6,8,10$ and $t= 0,2,6,10$ to let each subject have up to 6 and 4 measurements, respectively. The outputs summarised from $N=500$ replications of parameters estimates for the two cases under all four scenarios with 11 longitudinal measurements are displayed in Tables 1 and 2, while the results for 4 and 6 longitudinal measurements are displayed in Tables A.1 to A.4 in Appendix A. The model-based standard errors from the inverse Hessian matrix and the empirical standard deviations are represented by SE and SD, respectively, while CP and ECP stand for the coverage probabilities from the model-based and empirical 95\% normality confidence intervals. The root mean square error of parameter $\theta$ is defined as $RMSE(\theta)=\sqrt{\frac{1}{N}\sum_{i=1}^{N}\left(\hat{\theta}_{i}-\theta\right)^{2}},$ where  $\hat{\theta}_{i}$ is the parameter estimate for the $i$th sample.

According to the outputs from Tables \ref{sim11excar} and \ref{sim11carcar},  for case 1, the impacts of misspecification of $\bm R$ 
  are mainly on the slope term $\beta_{11}$ of the longitudinal process, the regression parameters $\bm\beta_{2}$ of the survival process, the components of covariance matrix $\bm D$ of the random effects, the variance $\sigma^{2}$ of the error terms and the association parameter $\alpha$ of the two processes. In scenario 1, the parameters $\bm\beta_{2}$, $\bm D$  and $\alpha^{2}$ are overestimated, while there is an underestimation of parameter $\sigma$.  In scenario 2, after correctly considering the correlation structure within the longitudinal process, there are smaller discrepancies between the estimates of $\bm D,$ $\sigma^{2}$ and $\rho_{y}$ and their true values, although $\alpha$ is more overestimated compared to scenario 1. Scenario 3 is essentially a replication of scenario 2, since the estimate of $\rho_{ty}$ for the continuous AR1 correlation structure between the two sub-models is close to 0 and can be ignored. This can be further validated by the prediction study carried out
 in the next section.

For case 2, unlike case 1, the biases in the regression parameters  $\bm\beta_{2}$ of the survival process are always very small. Similarly, the estimates of $\sigma^{2}$ and $\rho_{y}$ also exhibit quite small biases in scenarios 2 and 3, except in scenario 1, where  $\rho_{y}$ is fixed to be 0 and $\sigma^{2}$ is clearly underestimated.  In scenario 1, the components of $\bm D$ are overestimated while $\alpha$ is underestimated. Interestingly, in scenarios 2 and 3, where the correlation structure within the longitudinal process is correctly specified, the biases in the components of $\bm D$ (underestimated) and $\alpha$ (overestimated) exhibit opposite trends compared with scenario 1. The performance of scenarios 2 and 3 is very close in case 2, which is also the case in case 1, since the estimate of $\rho_{ty}$ for the constant correlation structure between the two sub-models is again close to 0. These biases in parameter estimates for scenarios 2 and 3 in both cases 1 and 2 indicate it is not sufficient to correctly specify the correlation structure  $\bm R_{(\bm y_{i})}$ in the longitudinal process alone.

For both cases, the biases for the regression parameter $\bm\beta_{1}$ of the longitudinal process are generally very small under the misspecification of $\bm R$, although moderate underestimation can be observed on slope  $\beta_{11},$ and this is because subjects with lower longitudinal values are more likely to dropout in the actual data generating process than the fitted models assume.  As a result of the correct specification of the model, in scenario 4 for both cases, all estimates are very close to their true values, with coverage probabilities close to the 95\% nominal level,  although there is some moderate but insignificant underestimation of the variance component of the random intercept $D_{11}.$ It is noticeable that its bias and RMSE decrease (compared with results in Tables 6 to 9 in Appendix B) when there are more longitudinal measurements per subject.

Contrasting the results in Tables 1 and 2 with those in Tables A.1 to A.4 in Appendix A, we notice that there are some tendencies in the biases for some of the parameters estimates as the length of the longitudinal measurements per subject changes. For example, in scenario 1 of case 1, the estimates of $\bm D$ and $\sigma^{2}$ deviate further away from their true values as the length of longitudinal observations increases, while the bias in the estimation of $\alpha,$ conversely, exhibits a reduction. In scenarios 2 and 3 of case 1, the deviations of the estimates of $\bm D,$ $\sigma^{2}$ and $\rho_{y}$ from their true values decrease with more longitudinal measurements per subject, but there is an opposite trend for the bias of $\alpha.$ For case 2, the biases of the estimates of $\bm D,$ $\sigma^{2}$ and $\alpha$ are increasing with more longitudinal measurements in scenario 1, on the other hand, these parameters exhibit a reduction in biases under scenarios 2 and 3.
In Appendix B, we give a heuristic explanation about why there are biases in some of the parameter estimates under the conventional joint model through the EM algorithm and the connection between the bias in $\bm D$ and the length of the longitudinal observations recorded.

In both cases, the AIC and likelihood ratio test can be use to select the best models. For example, when there are up to 11 longitudinal measurements in each subject, in case 1, scenario 4 outperforms other scenarios  in 500 out of 500 Monte Carlo samples by both criteria, while scenario 3  outperforms scenario 2 in only  72 out of 500 Monte Carlo samples by AIC. In case 2, scenario 4 outperforms scenario 1, scenario 2 and scenario 3  in 500, 384 and 424 out of 500 Monte Carlo samples by AIC, while scenario 3 outperforms scenario 2 in only in 33 out of 500 Monte Carlo samples by AIC. This means scenario 3, given having  one more parameter, does not provide better fitting than scenario 2 and the correlation introduced by $\rho_{ty}$ between the two sub-models is stronger in case 1 than case 2, which is also reflected in the prediction study presented in the next section.

A simulation study where both $\rho_{ty}$ and $\rho_{y}$ shrink to zero is also conducted. We generate $N=500$ Monte Carlo samples of size $n=200$ under the same conditionals as in Tables \ref{sim11excar} and \ref{sim11carcar}, then fit the dataset under the four scenarios. It turns out the likelihood ratio test rejects the null model of conditional independence at a significant level of 5\% in about 5\% of the Monte Carlo samples when compared to the more complicated models.

\subsection{Dynamic prediction}
The prediction of the survival probability for a specific subject with some baseline covariates and longitudinal measurements is of main interest after a joint model is fitted. Rizopoulos (2011)\cite{riz11} proposed an approach for performing a dynamic prediction on survival probabilities for a specific subject which can be updated as new longitudinal information becomes available. A similar predicting procedure is provided for the proposed model here.  The only difference compared with Rizopoulos (2011)\cite{riz11} is that the posterior distribution of $T_{i}^{*}$ in (\ref{dypredmgcjre}) still depends on the longitudinal process given $\bm b_{i}$, since the assumption of conditional independence does not hold here.

Suppose a joint model is fitted based on a random sample of $n$ subjects $\mathcal{\bm D}_{n}=\left\{T_{i},\delta_{i},\bm y_{i};i=1,...,n\right\}.$ Predictive survival probabilities for a new subject $i$ which has a set of longitudinal measurements $\mathcal{\bm Y}_{i}(t)=\left\{y_{i}(s);0\leq s<t\right\}$ up to $t$ and a vector of baseline covariates $\bm w_{i}$

\begin{figure}[H]
	\centering
	\includegraphics[width=\textwidth]{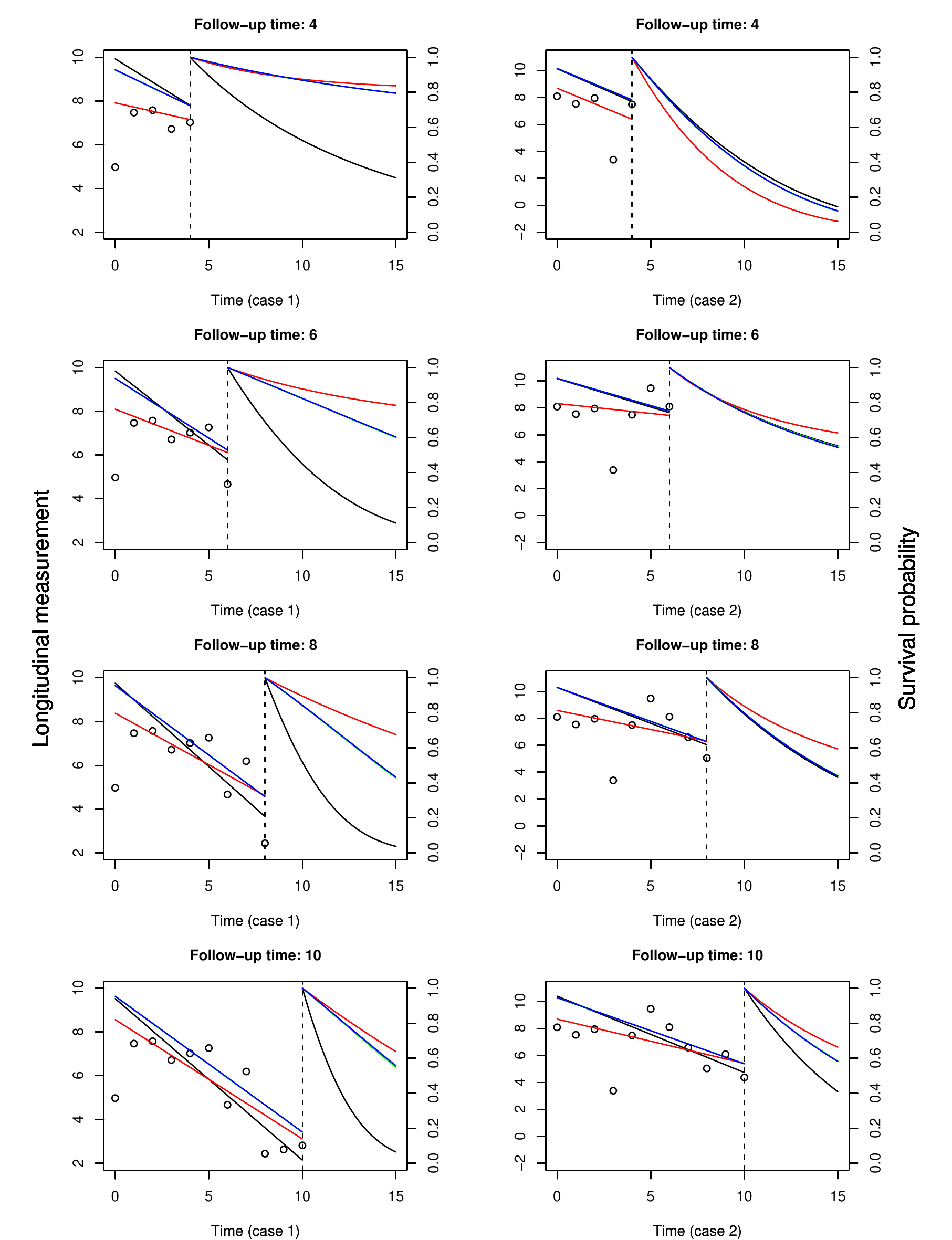}\\
	\caption{\textbf{{\scriptsize Dynamic  prediction of survival probabilities and fitted longitudinal trajectories for subjects simulated from cases 1 (left) and 2 (right). The red, green, blue and black lines represent scenarios 1, 2, 3 and 4, respectively. The cases and scenarios correspond to the outputs in Tables 1 and 2.}}}
	\label{bigdynacase1case2}
\end{figure}

\begin{figure}[H]
	\centering
	\includegraphics[width=\textwidth]{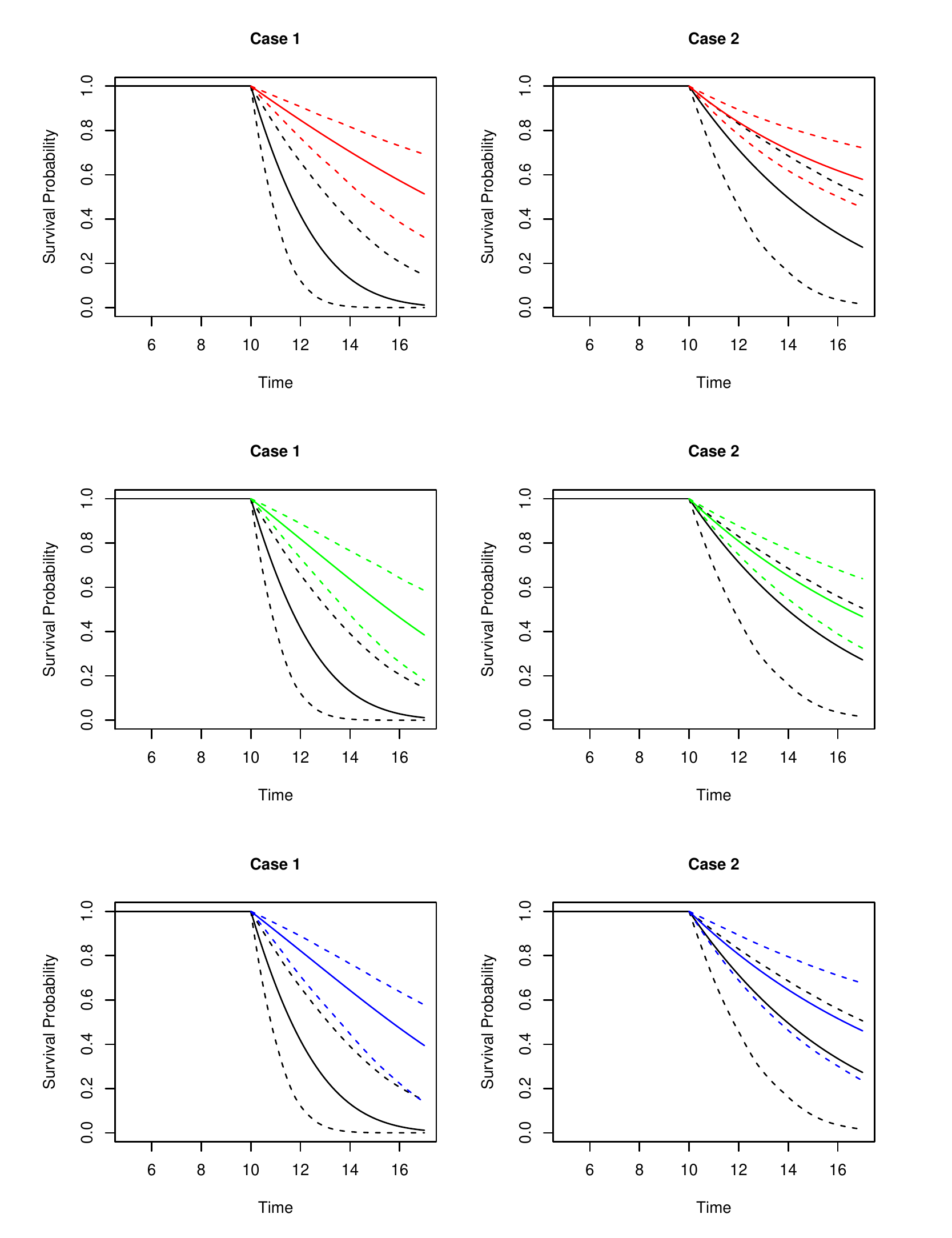}\\
	\caption{\textbf{{\scriptsize Predicted survival probabilities and the corresponding 95\% confidence intervals for subjects simulated from cases 1 and 2 at $\bm t\textbf{=10}.$ The red, green, blue and black lines represent scenarios 1, 2, 3 and 4, respectively. The cases and scenarios correspond to the outputs in Tables 1 and 2.}}}
	\label{bigcfsurcase1case2}
\end{figure}

\noindent are given by:
\begin{eqnarray}
	\nonumber	\pi_{i}(u|t)&=&P(T_{i}^{*}>u|T_{i}^{*}>t,\mathcal{\bm Y}_{i}(t),\bm w_{i},\mathcal{\bm D}_{n};\bm\theta)=P(T_{i}^{*}>u|T_{i}^{*}>t,\mathcal{\bm Y}_{i}(t),\bm w_{i};\bm\theta)\\ 
	\nonumber	&=&\int_{R^{q}}P(T_{i}^{*}>u|T_{i}^{*}>t,\mathcal{\bm Y}_{i}(t),\bm w_{i},\bm b_{i};\bm\theta)f_{\bm b_{i}}(\bm b_{i}|T_{i}^{*}>t,\mathcal{\bm Y}_{i}(t),\bm w_{i};\bm\theta)d\bm b_{i}\\  
	&=&\int_{R^{q}}\frac{S_{T_{i}^{*}}\left(u|\mathcal{\bm Y}_{i}(t),\bm w_{i},\bm b_{i};\bm\theta\right)}{S_{T_{i}^{*}}\left(t|\mathcal{\bm Y}_{i}(t),\bm w_{i},\bm b_{i};\bm\theta\right)}f_{\bm b_{i}}(\bm b_{i}|T_{i}^{*}>t,\mathcal{\bm Y}_{i}(t),\bm w_{i};\bm\theta)d\bm b_{i}.\label{dypredmgcjre}
\end{eqnarray}
This expression can be approximated by its first-order estimate (Rizopoulos, 2011\cite{riz11}) using the empirical Bayes estimate for $\bm b_{i}:$
\begin{eqnarray*}
	\hat{\pi}_{i}(u|t)\approx\frac{S_{T_{i}^{*}}\left(u|\mathcal{\bm Y}_{i}(t),\bm w_{i},\hat{\bm b}_{i};\hat{\bm\theta}\right)}{S_{T_{i}^{*}}\left(t|\mathcal{\bm Y}_{i}(t),\bm w_{i},\hat{\bm b}_{i};\hat{\bm\theta}\right)},
\end{eqnarray*} 
where $\hat{\bm\theta}$ denotes the maximum likelihood estimators and $\hat{\bm b}_{i}$ denotes the mode of the conditional 
distribution $f_{\bm b_{i}}(\bm b_{i}|T_{i}^{*}>t,\mathcal{\bm Y}_{i}(t),\bm w_{i};\bm\theta).$

It can be shown that:
\begin{eqnarray*}
	S_{T_{i}^{*}}\left(t|\mathcal{\bm Y}_{i}(t),\bm w_{i},\hat{\bm b}_{i};\hat{\bm\theta}\right)&=&1-\Phi\left(\frac{\Phi^{-1}\left(F_{T_{i}^{*}}(t|\hat{\bm b}_{i},\bm w_{i};\hat{\bm\theta})\right)-\hat{\mu}_{i}^{t|\bm y_{i},\hat{\bm b}_{i}}}{\hat{\sigma}_{i}^{t|\bm y_{i},\hat{\bm b}_{i}}}\right)\\  \nonumber
	&=&\Phi\left(\frac{\Phi^{-1}\left(S_{T_{i}^{*}}(t|\hat{\bm b}_{i},\bm w_{i};\hat{\bm\theta})\right)+\hat{\mu}_{i}^{t|\bm y_{i},\hat{\bm b}_{i}}}{\hat{\sigma}_{i}^{t|\bm y_{i},\hat{\bm b}_{i}}}\right),
\end{eqnarray*}
and
\begin{eqnarray*}f_{\bm b_{i}}(\bm b_{i}|T_{i}^{*}>t,\mathcal{\bm Y}_{i}(t),\bm w_{i};\bm\theta)
	&=&\frac{f_{\bm b_{i},T_{i}^{*}}(\bm b_{i},T_{i}^{*}>t|\mathcal{\bm Y}_{i}(t),\bm w_{i};\bm\theta)}{f_{T_{i}^{*}}(T_{i}^{*}>t|\mathcal{\bm Y}_{i}(t),\bm w_{i};\bm\theta)}\\
	&=&\frac{f_{T_{i}^{*}}(T_{i}^{*}>t|\bm b_{i},\mathcal{\bm Y}_{i}(t),\bm w_{i};\bm\theta)f_{\bm b_{i}}(\bm b_{i}|\mathcal{\bm Y}_{i}(t),\bm w_{i};\bm\theta)}{\int_{-\infty}^{\infty}f_{T_{i}^{*}}(T_{i}^{*}>t|\bm b_{i},\mathcal{\bm Y}_{i}(t),\bm w_{i};\bm\theta)f_{\bm b_{i}}(\bm b_{i}|\mathcal{\bm Y}_{i}(t),\bm w_{i};\bm\theta)\,d\bm b_{i}}\\
	&=&\frac{S_{T_{i}^{*}}(t|\bm b_{i},\mathcal{\bm Y}_{i}(t),\bm w_{i};\bm\theta)f_{\bm b_{i}}(\bm b_{i}|\mathcal{\bm Y}_{i}(t),\bm w_{i};\bm\theta)}{\int_{-\infty}^{\infty}S_{T_{i}^{*}}(t|\bm b_{i},\mathcal{\bm Y}_{i}(t),\bm w_{i};\bm\theta)f_{\bm b_{i}}(\bm b_{i}|\mathcal{\bm Y}_{i}(t),\bm w_{i};\bm\theta)\,d\bm b_{i}}.
\end{eqnarray*}

The predictive survival probability at time $u>t$ can be derived as:
\begin{eqnarray*}
	\displaystyle
	\hat{\pi}_{i}(u|t)&\approx&\frac{S_{T_{i}^{*}}\left(u|\mathcal{\bm Y}_{i}(t),\bm w_{i},\hat{\bm b}_{i};\hat{\bm\theta}\right)}{S_{T_{i}^{*}}\left(t|\mathcal{\bm Y}_{i}(t),\bm w_{i},\hat{\bm b}_{i};\hat{\bm\theta}\right)}\\  \nonumber
	&=&\frac{\Phi\left(\frac{\Phi^{-1}\left(S_{T_{i}^{*}}(u|\hat{\bm b}_{i},\bm w_{i};\hat{\bm\theta})\right)+\hat{\mu}_{i}^{u|\bm y_{i},\hat{\bm b}_{i}}}{\hat{\sigma}_{i}^{u|\bm y_{i},\hat{\bm b}_{i}}}\right)}{\Phi\left(\frac{\Phi^{-1}\left(S_{T_{i}^{*}}(t|\hat{\bm b}_{i},\bm w_{i};\hat{\bm\theta})\right)+\hat{\mu}_{i}^{t|\bm y_{i},\hat{\bm b}_{i}}}{\hat{\sigma}_{i}^{t|\bm y_{i},\hat{\bm b}_{i}}}\right)},  \nonumber
\end{eqnarray*}
 which reduces to $\displaystyle\frac{S_{T_{i}^{*}}(u|\hat{\bm b}_{i},\bm w_{i};\hat{\bm\theta})}{S_{T_{i}^{*}}(t|\hat{\bm b}_{i},\bm w_{i};\hat{\bm\theta})}=\displaystyle\mbox{exp}\left\{-\int_{t}^{u}h_{i}(s|\hat{\bm b}_{i},\bm w_{i};\hat{\bm\theta})ds\right\}$ given $\bm R_{(t)(\bm y_{i})}=\bm0$  and $\bm R_{(u)(\bm y_{i})}=\bm0.$ 
The survival probabilities are updated once new measurements of longitudinal values are introduced.
Figure \ref{bigdynacase1case2} presents dynamic predictions for two new subjects simulated based on cases 1 and 2 in Section 3.1, which have true event times at $t=12.670$ and 15.424, respectively. The parameters used for predicting in the four scenarios are from the outputs in Tables \ref{sim11excar} and \ref{sim11carcar}.

We also adopt the Monte Carlo simulation scheme introduced in Rizopoulos (2012b)\cite{riz12b} to construct confidence intervals for predictions of survival probabilities:
\begin{enumerate}
	\item Draw $\hat{\bm\theta}^{(l)}\sim N\left\{\hat{\bm\theta}, \hat{\mbox{var}}\left(\hat{\bm\theta}\right)\right\}$;
	
	\item Draw $\hat{\bm b}_{i}^{l}\sim f\left(\bm b_{i}|T_{i}^{*}>t,\mathcal{\bm Y}_{i}(t),\bm\theta^{(l)}\right)$;
	
	\item Compute 	\begin{eqnarray*}
		\displaystyle
		\hat{\pi}_{i}^{(l)}(u|t)&\approx&\frac{S_{T_{i}^{*}}\left(u|\mathcal{\bm Y}_{i}(t),\bm w_{i},\hat{\bm b}_{i}^{(l)};\hat{\bm\theta}^{(l)}\right)}{S_{T_{i}^{*}}\left(t|\mathcal{\bm Y}_{i}(t),\bm w_{i},\hat{\bm b}_{i}^{(l)};\hat{\bm\theta}^{(l)}\right)}\\  \nonumber
		&=&\frac{\Phi\left(\frac{\Phi^{-1}\left(S_{T_{i}^{*}}(u|\hat{\bm b}_{i}^{(l)},\bm w_{i};\hat{\bm\theta}^{(l)})\right)+\hat{\mu}_{i}^{u|\bm y_{i},\hat{\bm b}_{i}^{(l)}}}{\hat{\sigma}_{i}^{u|\bm y_{i},\hat{\bm b}_{i}}}\right)}{\Phi\left(\frac{\Phi^{-1}\left(S_{T_{i}^{*}}(t|\hat{\bm b}_{i}^{(l)},\bm w_{i};\hat{\bm\theta}^{(l)})\right)+\hat{\mu}_{i}^{t|\bm y_{i},\hat{\bm b}_{i}^{(l)}}}{\hat{\sigma}_{i}^{t|\bm y_{i},\hat{\bm b}_{i}^{(l)}}}\right)},  \nonumber
	\end{eqnarray*}
\end{enumerate}

Step 1-3 are repeated $L$ times. The confidence interval can be derived from the sample percentiles of the realizations $\left\{\hat{\pi}_{i}^{(l)}(u|t),l=1,...,L\right\}.$
The corresponding 95\% confidence intervals are constructed for both subjects after $t=10$. The plots are presented in Figure \ref{bigcfsurcase1case2}. We notice that the upper intervals for scenario 1 are under the lower intervals of the other scenarios for this simulated  subject in case 1, indicating quite a huge impact on prediction of survival probabilities caused by the misspecification of $\bm R_{i}.$

Intuitively, the proposed joint model  provides most accurate prediction of survival probabilities while the misspecified models  overestimate the predicted survival probabilities, especially in the standard joint model in scenario 1, considering the true event times of these two subjects are at $t=12.7$ and $15.4.$ In fact, the predicted mean survival times for cases 1 and 2 are calculated for different scenarios given the subjects have survived up to $t=10.$ For case 1, the predicted mean survival times are 18.566, 16.174, 16.281, 12.062 for scenarios 1, 2, 3 and 4, respectively. For case 2,  they are 18.807, 18.643 and 15.184 for scenarios 2, 3 and 4, respectively, while the survival curve is above 0 when $t$ approaches $\infty$ for scenario 1 and this is because the positive random slope (negative for the other scenarios) makes the hazard function decrease to 0 quickly (the event is not expected to happen after a certain time point). The predicted expected event times 12.1 and 15.2 in scenario 4 are close to the true event times of 12.7 and 15.4  for both cases. Meanwhile, the accuracies of predictions are slightly improved by correctly taking the correlation within the longitudinal process into consideration as shown in scenarios 2 and 3, although the differences between these two scenarios  are indistinguishable.

	To calibrate the overall discriminative performance among the four models, consider a pair of randomly selected subjects at risk by $t$, denoted as $i_{1}$ and $i_{2}$, from the population. Suppose subject $i_{1}$ experiences the event between the interval of $(t,u]$ whereas subject $i_{2}$ does not.  A good predictive model is supposed to assign higher survival probabilities to subject $i_{2}$ than subject $i_{1},$ i.e.,  $\pi_{i_{2}}(u|t)>\pi_{i_{1}}(u|t),$ and the area under the  receiver operating characteristics curve (AUC) defined as:
	\[\mbox{AUC}(u|t)=P\left\{\pi_{i_{2}}(u|t)>\pi_{i_{1}}(u|t) | T_{i_{2}}^{*}>u, t<T_{i_{1}}^{*}\leq u\right\}\]
	is a formal tool to assess the discriminative capability of the model.  
		Meanwhile, the overall predictive accuracy of the survival probability is calibrate by prediction error (PE).  If a subject, say $i,$ is event free up to time $u,$ an accurate predicting model is expected to provide  $\pi_{i}(u|t)$ close to 1 and close to 0 otherwise. The PE, or the Brier score, defined as:
	\[\mbox{PE}(u|t)=E\left[\left\{I(T_{i}^{*}>u)-\pi_{i}(u|t)\right\}^{2}| T_{i}^{*}>t\right]\]
	can be used to evaluate the predicting accuracy of the survival model. A good model is expected to have a high AUC value close to 1 and a low PE value close to 0. To account for censoring in the population,
	the weighted AUC and PE estimators from Andrinopoulou  \textit{et al.} (2018)\cite{and18} are adopted and their values are calculated based on $N=100$ new Monte Carlo samples each of size $n=200,$ generated from cases 1 and 2.

\begin{figure}[H]
	\centering
	\begin{minipage}{0.325\textwidth}
		\includegraphics[width=\linewidth]{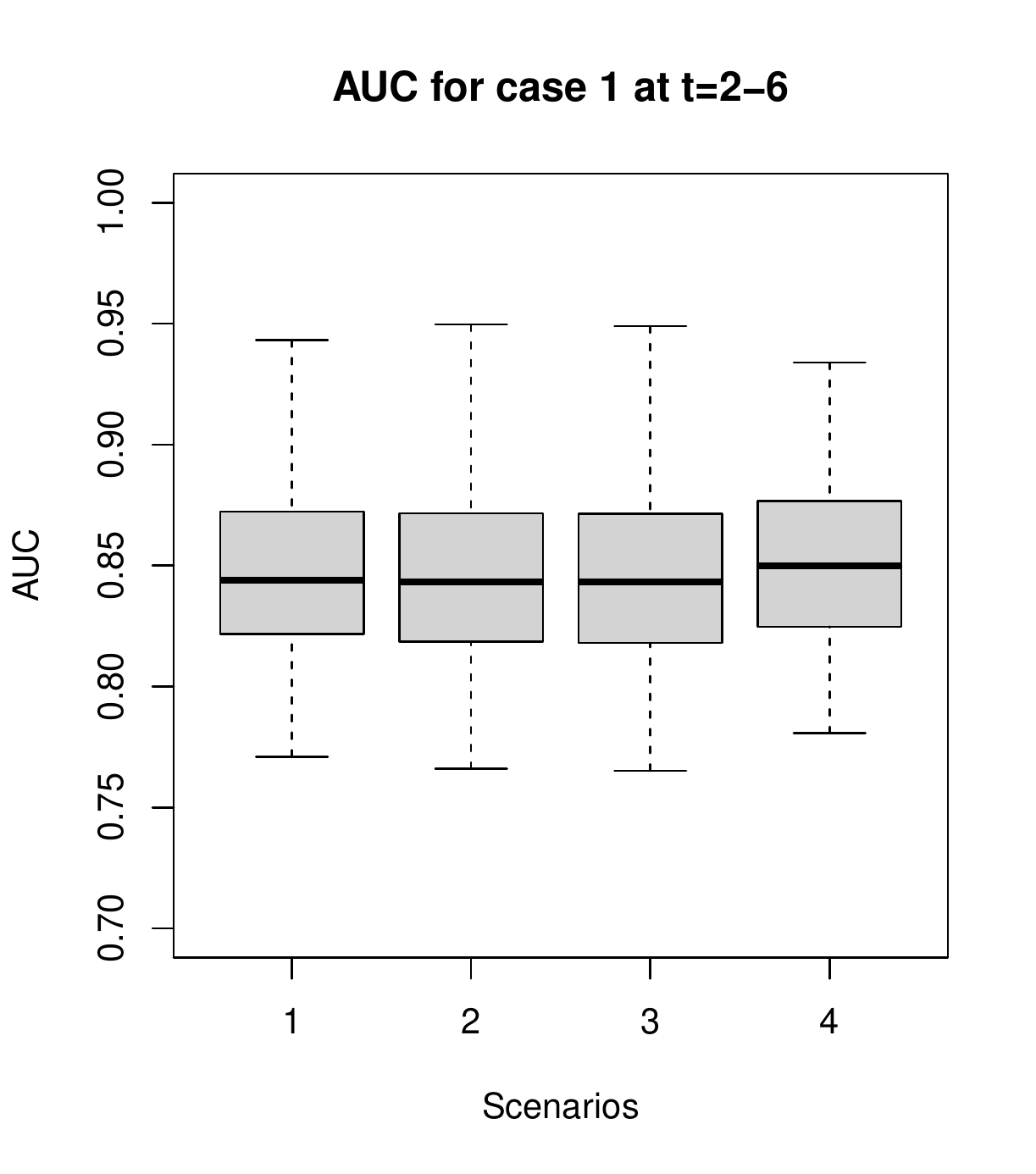}
	\end{minipage}
	\begin{minipage}{0.325\textwidth}
	\includegraphics[width=\linewidth]{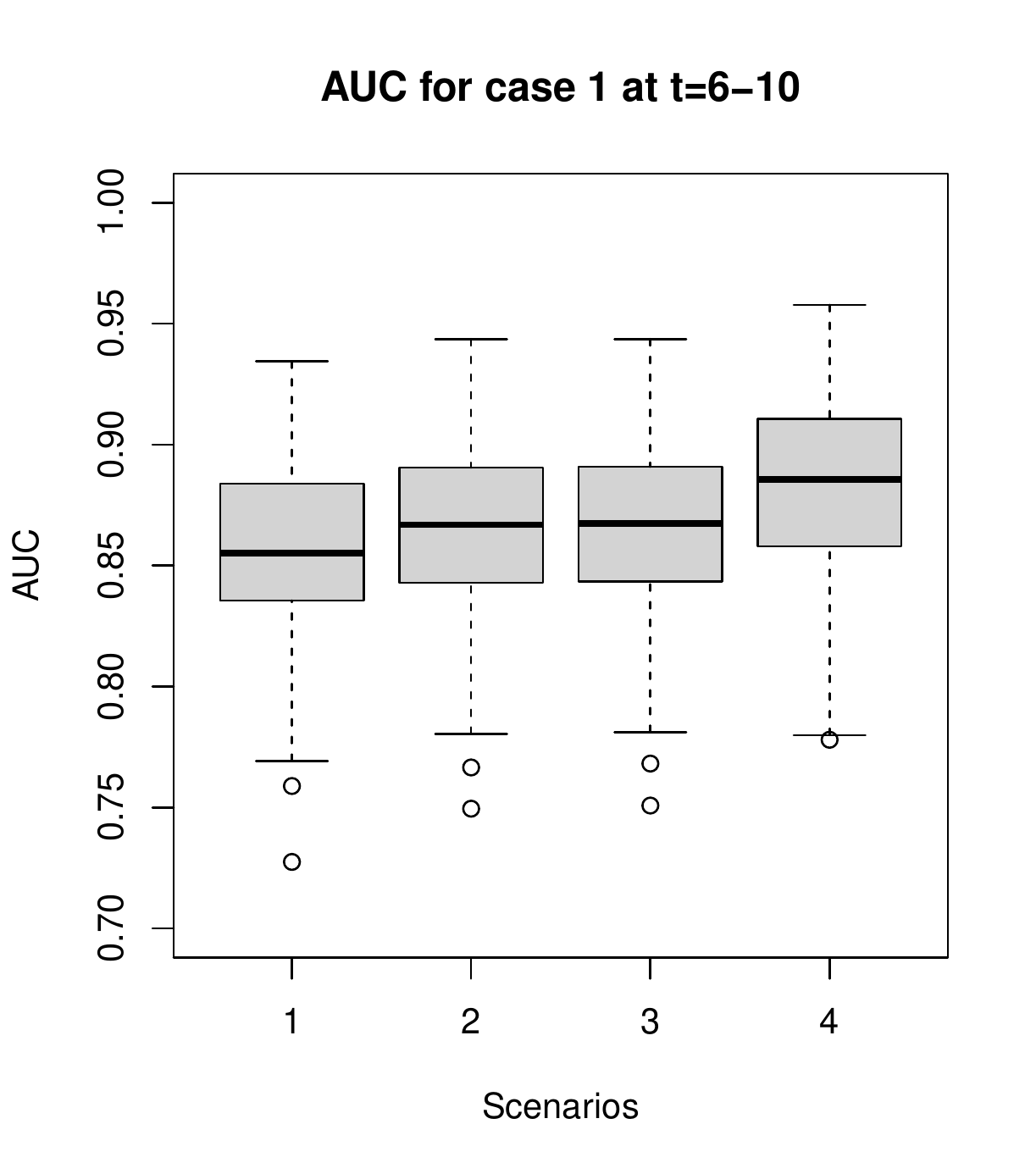}
\end{minipage}
	\begin{minipage}{0.325\textwidth}
	\includegraphics[width=\linewidth]{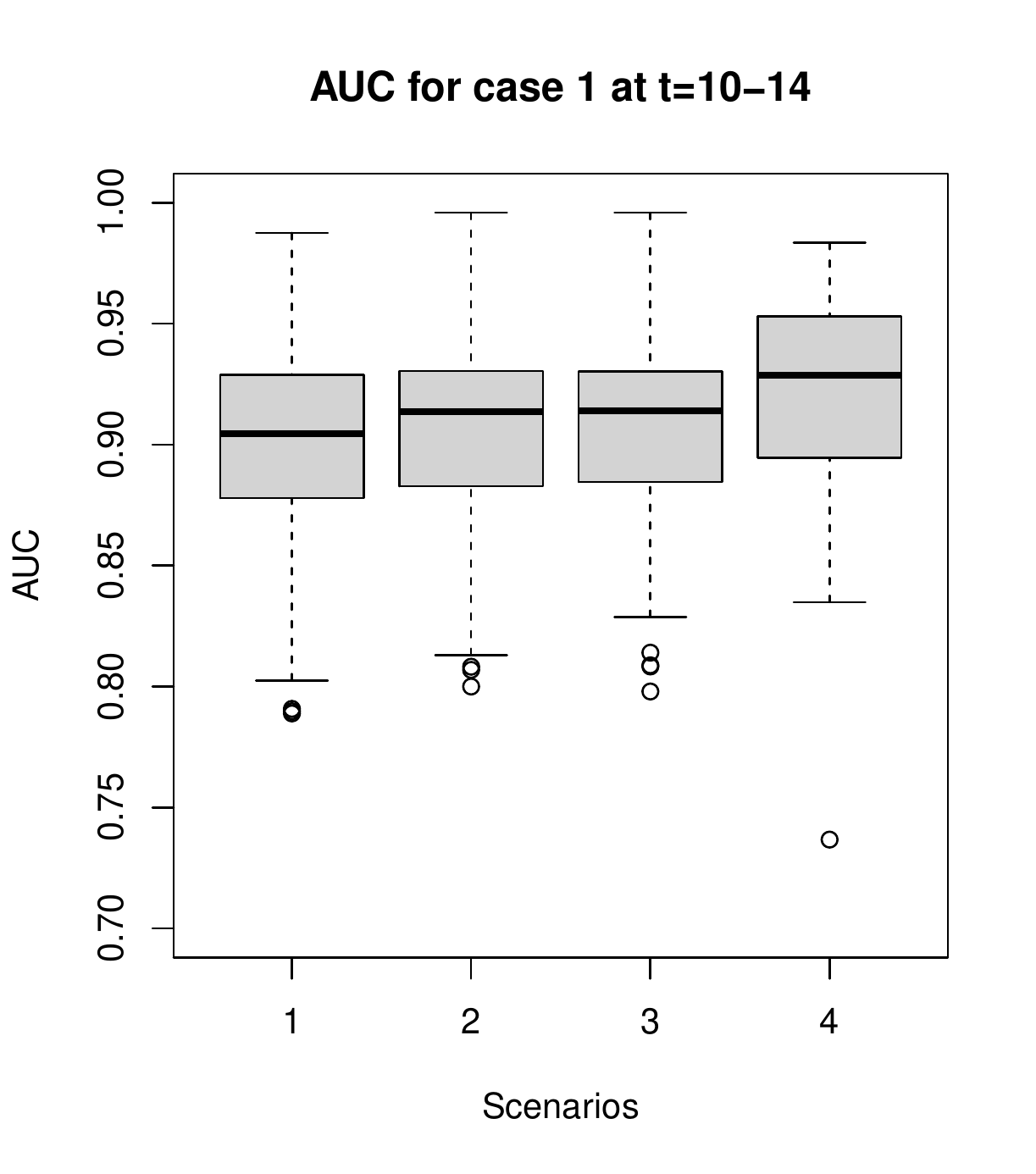}
\end{minipage}
	\begin{minipage}{0.325\textwidth}
		\includegraphics[width=\linewidth]{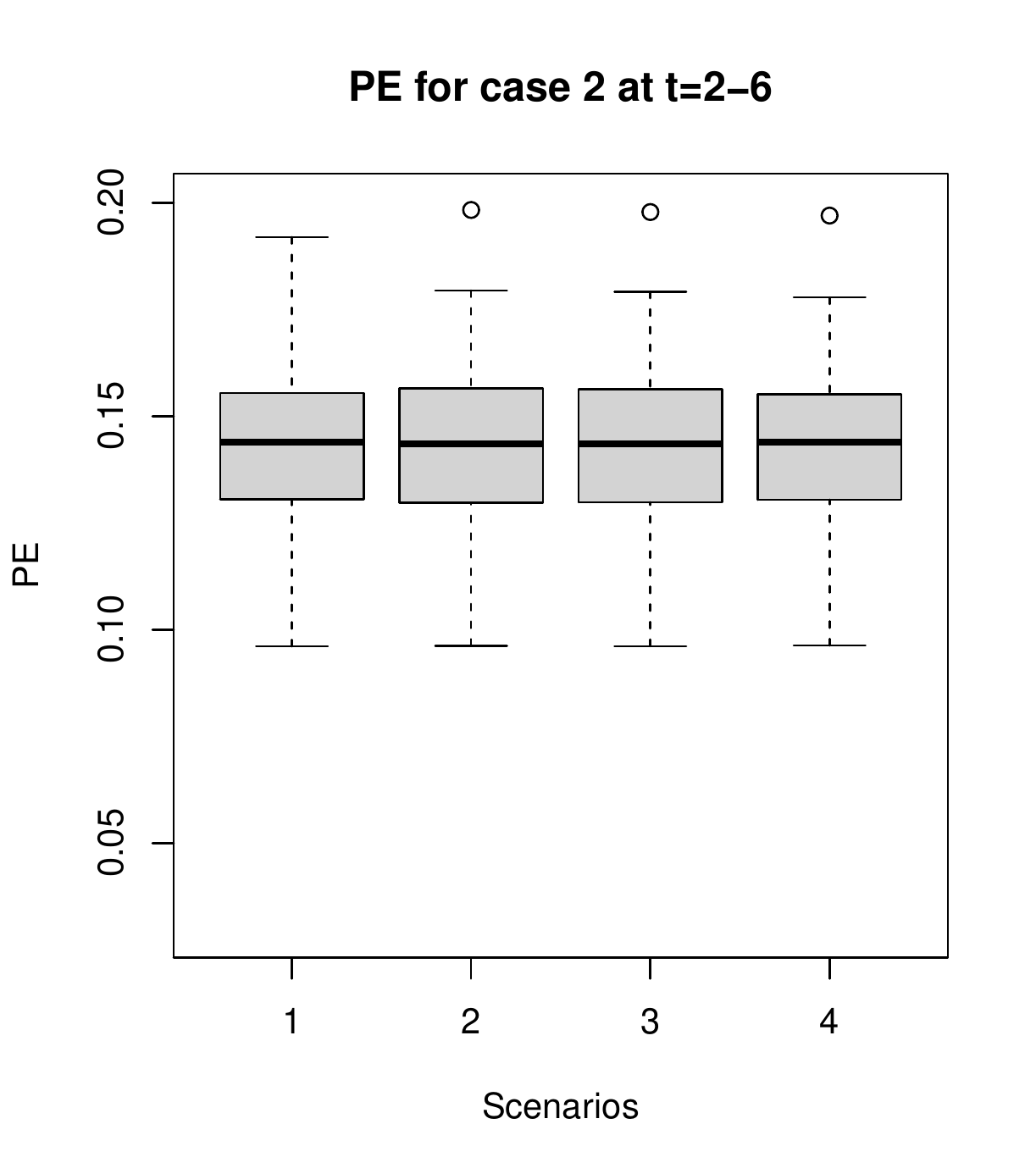}
	\end{minipage}
	\begin{minipage}{0.325\textwidth}
		\includegraphics[width=\linewidth]{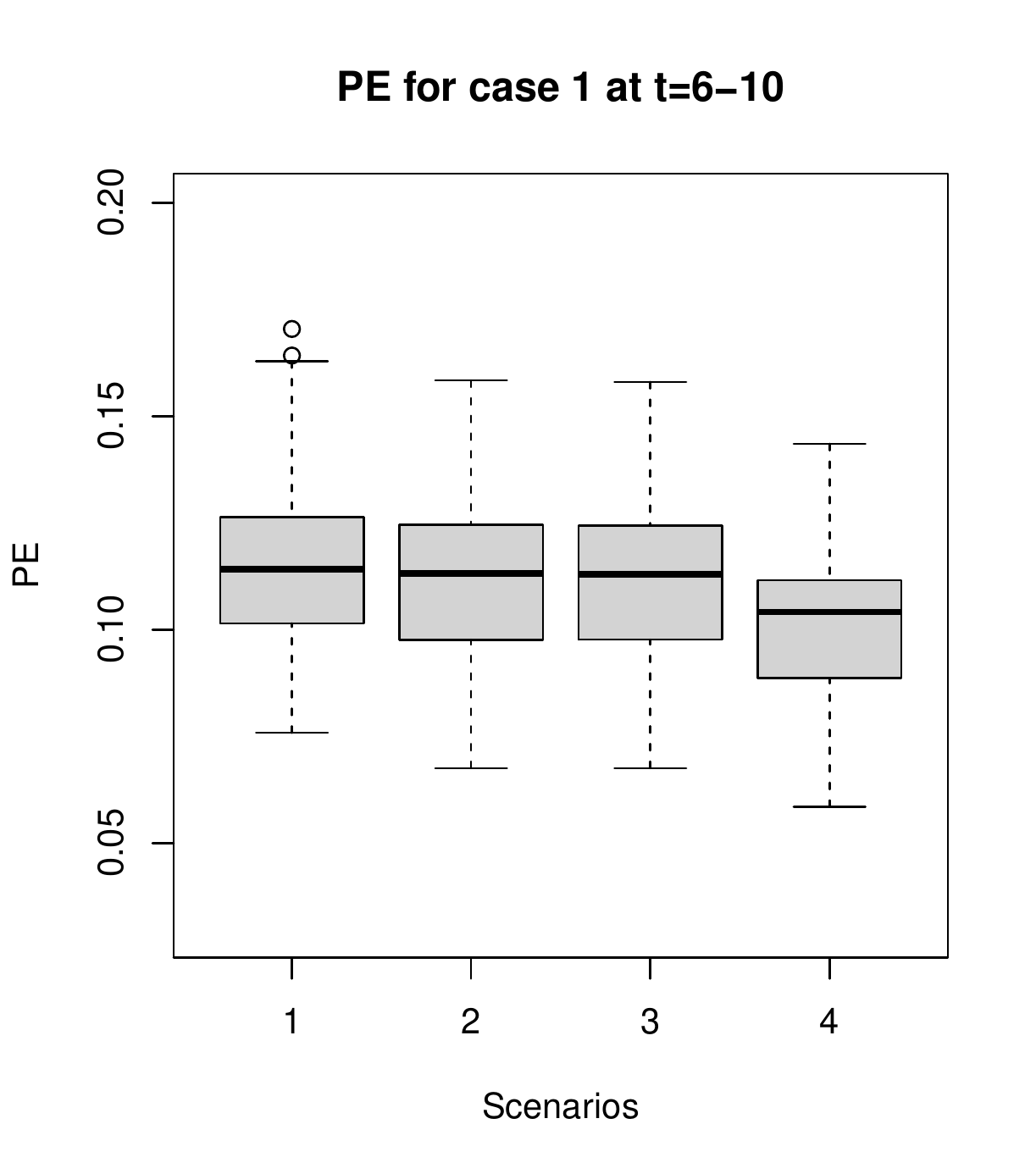}
	\end{minipage}
	\begin{minipage}{0.325\textwidth}
		\includegraphics[width=\linewidth]{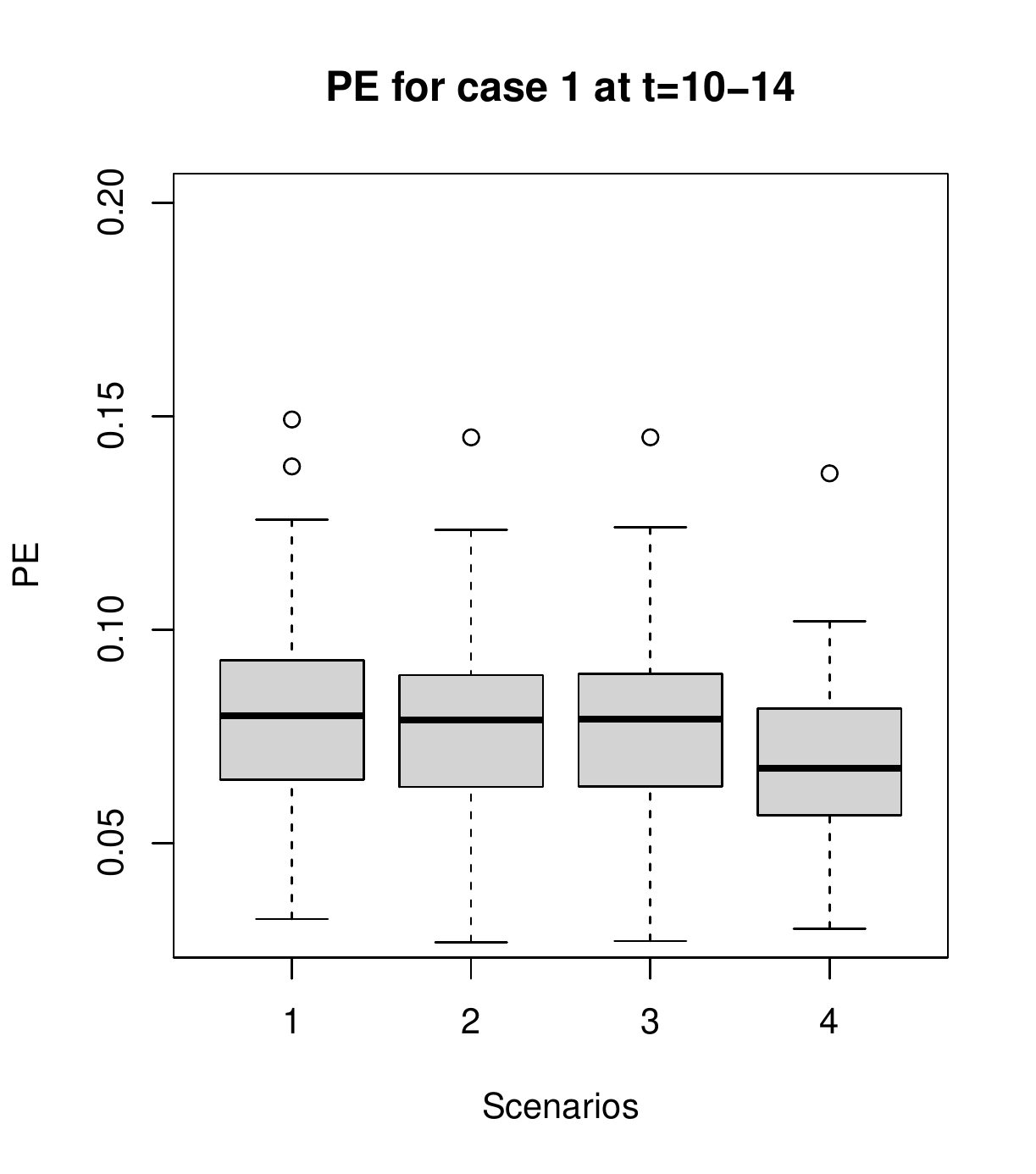}
	\end{minipage}
\caption{\textbf{{\scriptsize Box plots of AUC and PE at $t=$2, 6 and 10 with $\Delta t=4$  under 4 scenarios calculated from 100 Monte Carlo samples each with sample size $n=200$ simulated from cases 1. The cases and scenarios correspond to the outputs in Tables 1.}}}
	\label{dynaAUCPEcase1}
\end{figure}

\begin{figure}[H]
	\centering
	\begin{minipage}{0.325\textwidth}
		\includegraphics[width=\linewidth]{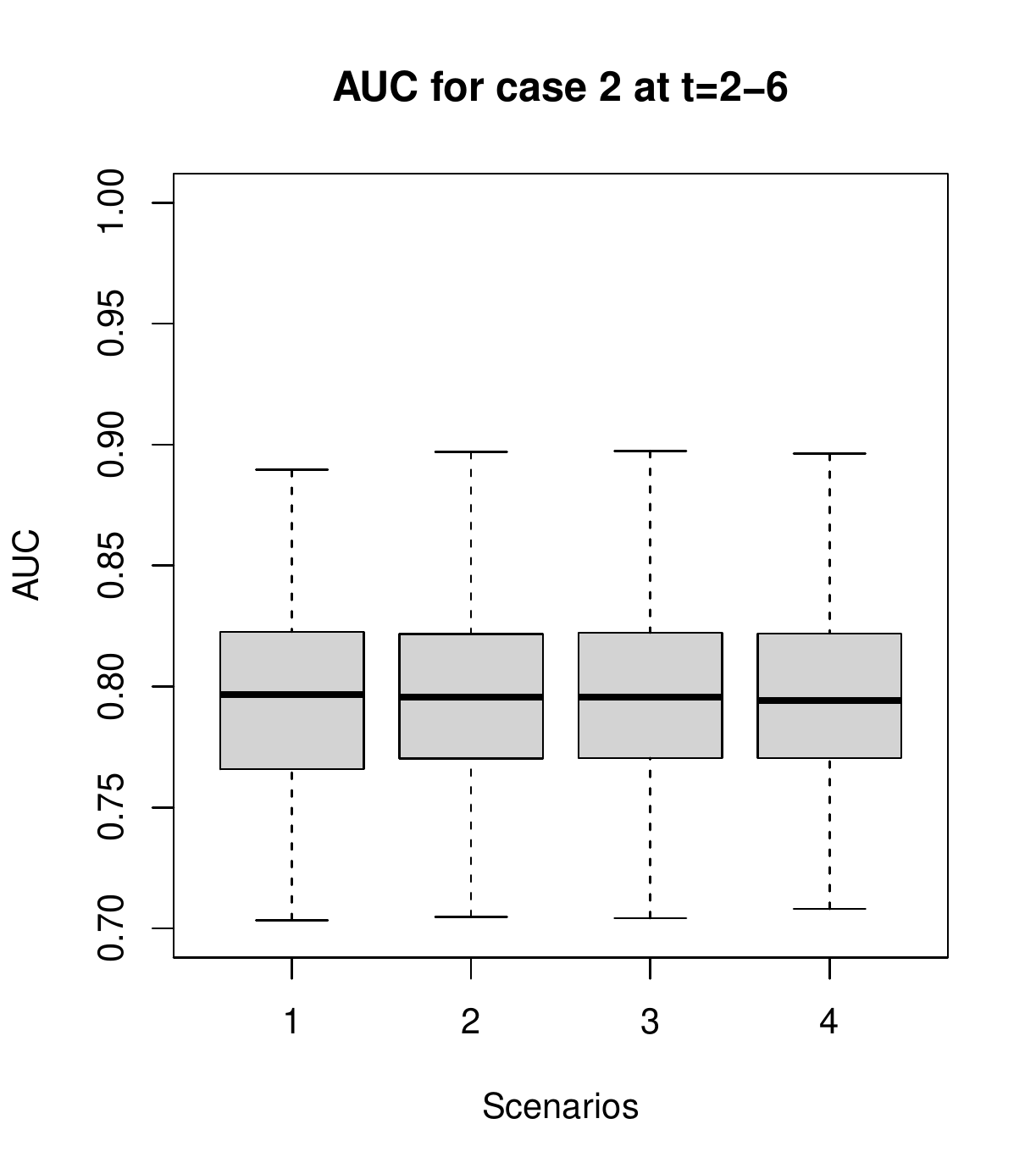}
	\end{minipage}
	\begin{minipage}{0.325\textwidth}
		\includegraphics[width=\linewidth]{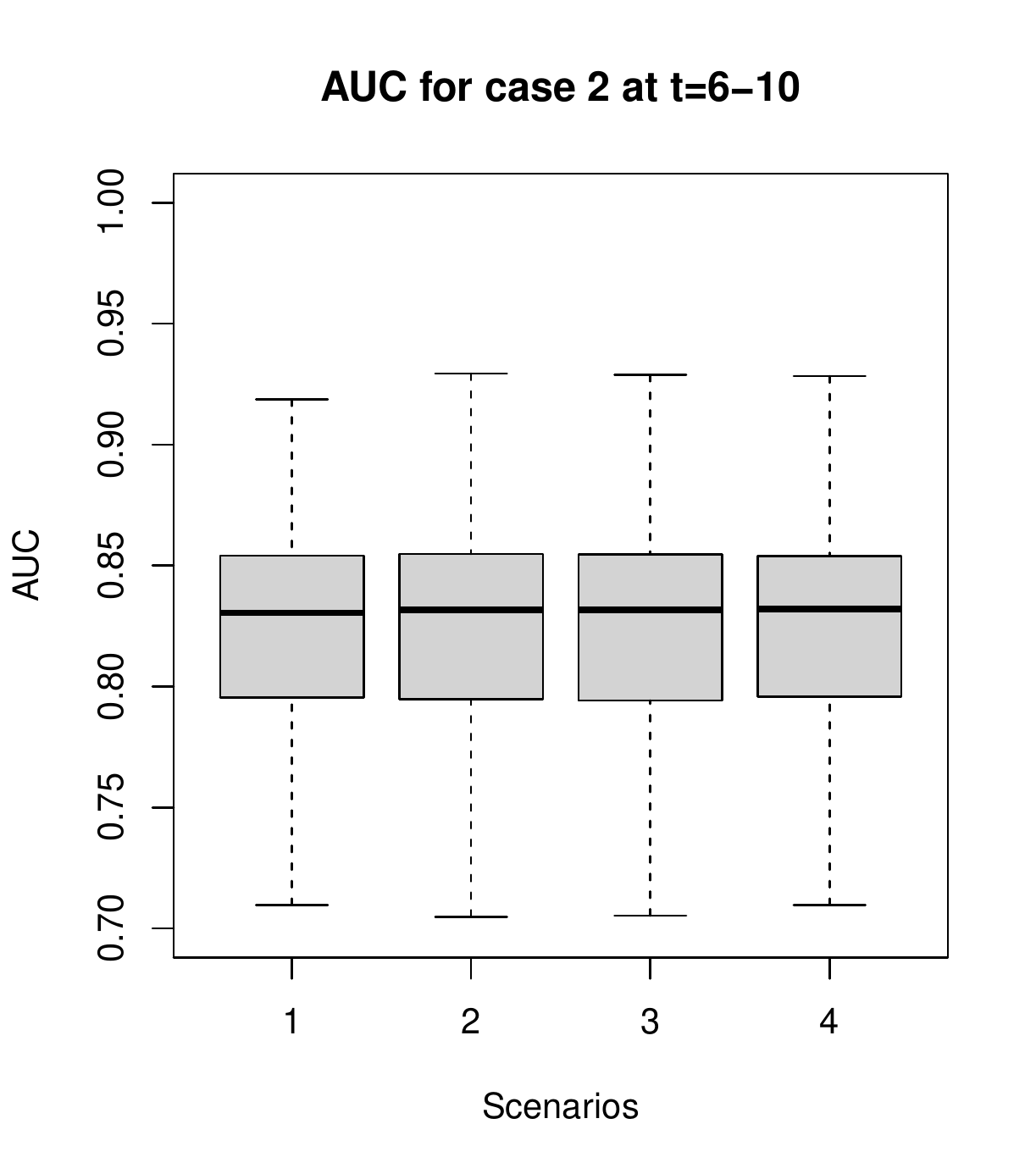}
	\end{minipage}
	\begin{minipage}{0.325\textwidth}
		\includegraphics[width=\linewidth]{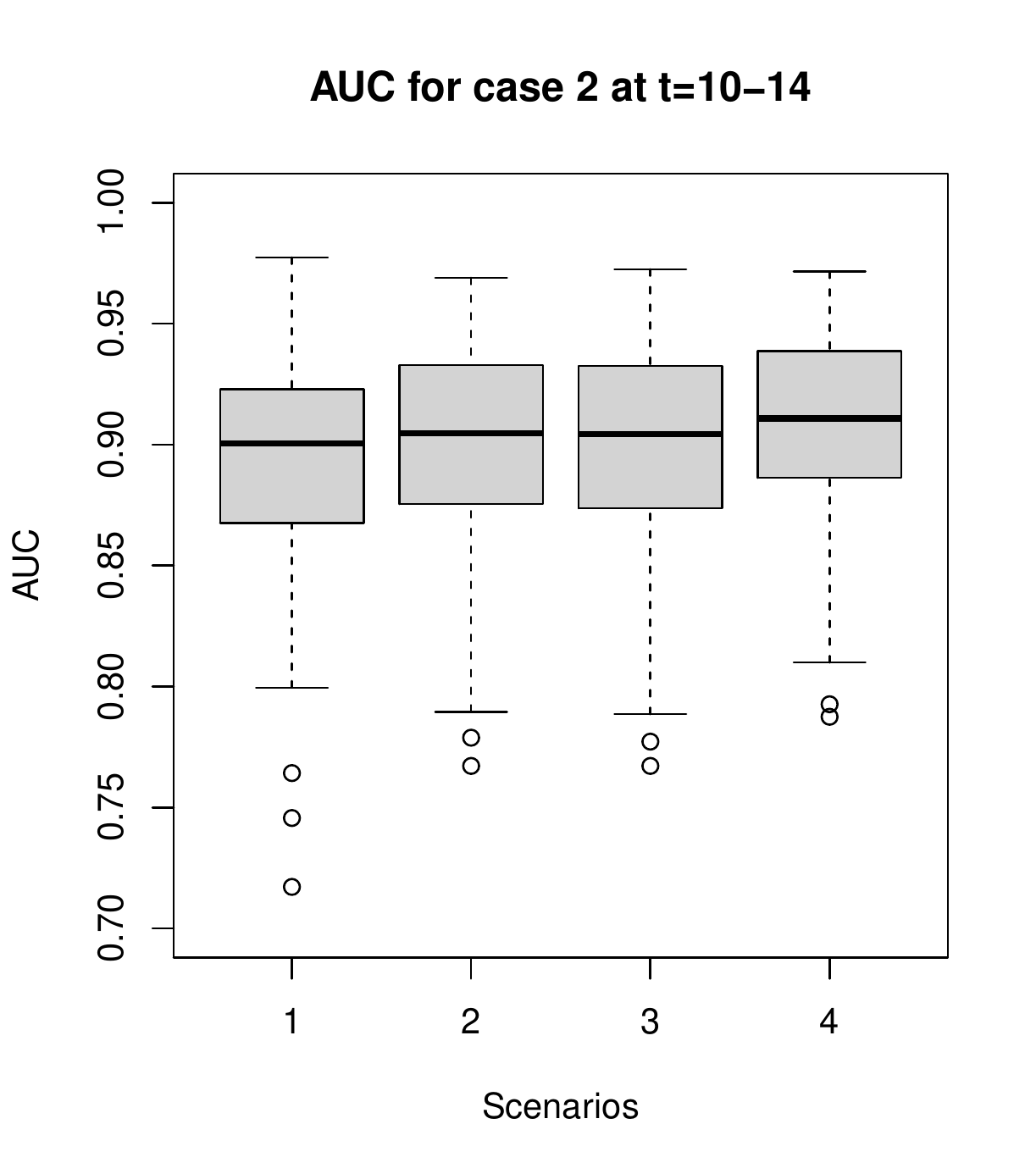}
	\end{minipage}
	\begin{minipage}{0.325\textwidth}
		\includegraphics[width=\linewidth]{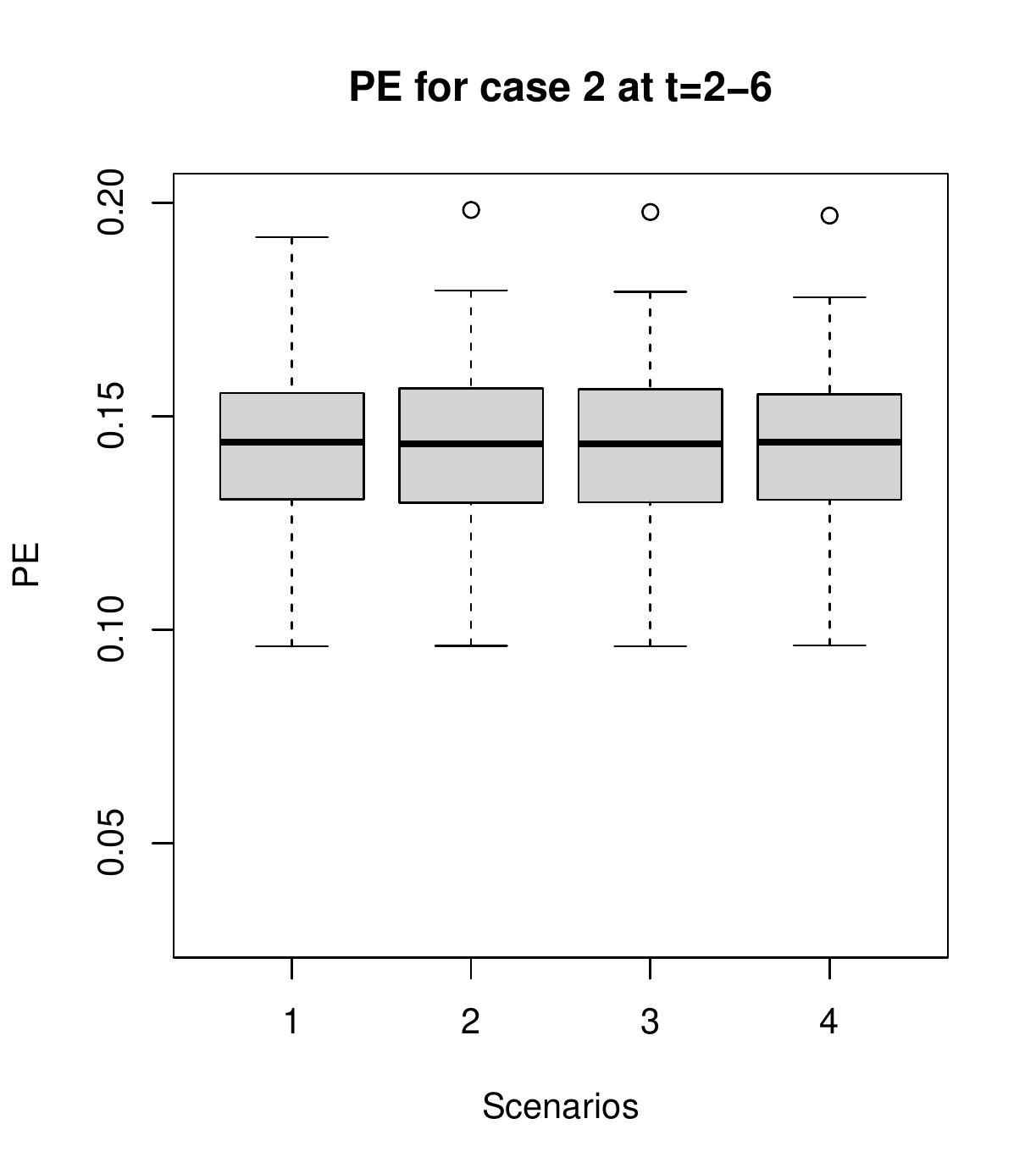}
	\end{minipage}
	\begin{minipage}{0.325\textwidth}
		\includegraphics[width=\linewidth]{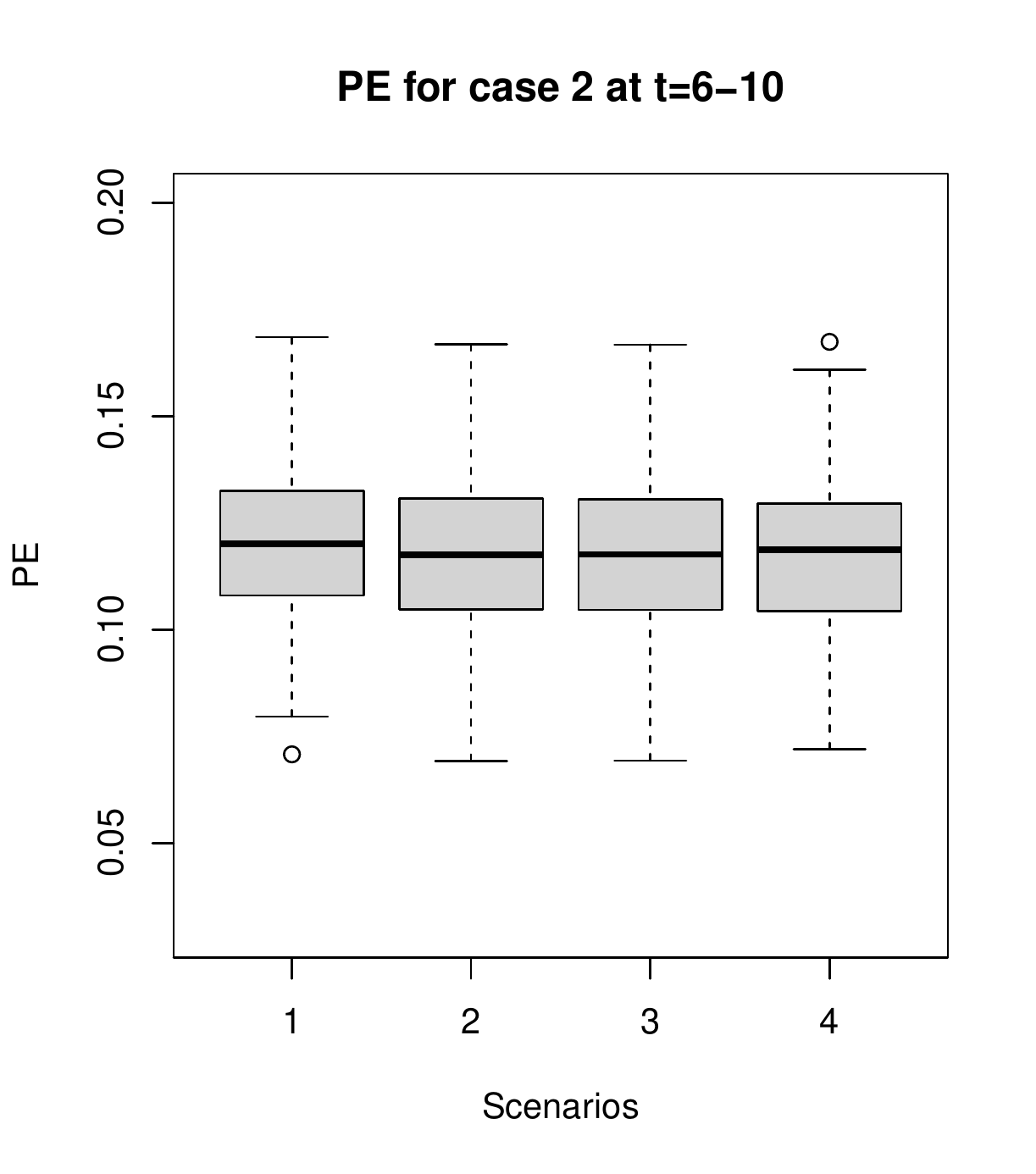}
	\end{minipage}
	\begin{minipage}{0.325\textwidth}
		\includegraphics[width=\linewidth]{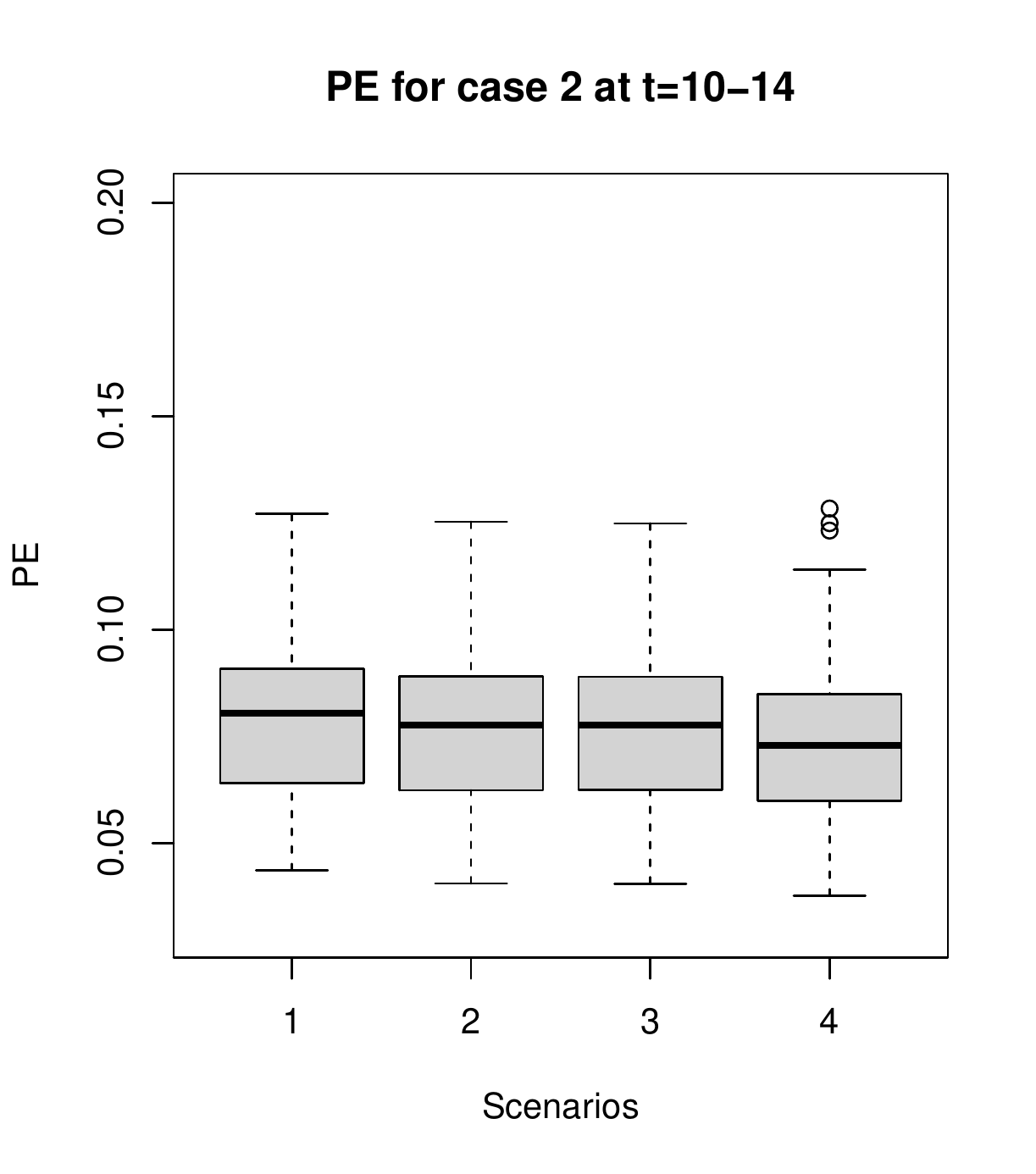}
	\end{minipage}
	\caption{\textbf{{\scriptsize Box plots of AUC and PE at $t=$2, 6 and 10 with $\Delta t=4$  under 4 scenarios calculated from 100 Monte Carlo samples each with sample size $n=200$ simulated from cases 2. The cases and scenarios correspond to the outputs in Tables 2.}}}
	\label{dynaAUCPEcase2}
\end{figure}

	Figures \ref{dynaAUCPEcase1} and \ref{dynaAUCPEcase2} display the box plots of dynamic AUC and PE values at $t=$2, 4 and 10 with $\Delta t=u-t=4$ for the 100 Monte Carlo samples. For both cases, the gaps of the discriminative abilities and predictive accuracies among the four models increase with $t,$ this may be because there is extra information, besides the shared or correlated random effects,  carried by $\rho_{ty}$ from the longitudinal process to help with improving the prediction of survival  probabilities as $t$ increases. To be more specific, for the subjects from case 1, the correlation between the two sub-models is  constant with $\rho_{ty}=0.4.$ At $t=2,$ the four models have similar levels of discrimination and accuracy. At this stage, the information introduced by $\rho_{ty}$ is not sufficient to distinguish among the four models. At $t=6,$ as more information from the longitudinal process is brought in by $\rho_{ty},$ scenario 4 has the best performance among the four models while scenario 2 and 3 perform similarly and worse than scenario 4 but both slightly better than scenario 1. This pattern is most obvious at $t=10,$ where $\rho_{ty}$ has the most effect. Surprisingly, the performance of scenario 1 is not very far behind than others and this may be due to the severe biases in the association parameter $\alpha$ and variance components of $\bm D$ in scenario 1, which are in fact trying to incorporate the extra dependency from the two sub-models. Similar but less obvious phenomena can also be observed for subjects in case 2. The four models still perform closely up to $t=6,$ since the correlation between the two sub-models is still weak, in fact $\rho_{ty}=0.5^4$, at this time and more obvious distinction can be observed at $t=10.$ These findings match those from Figure \ref{bigdynacase1case2}, where for case 1 we notice large discrepancies in individual predictions at an earlier stage compared to the differences observed in case 2.  The fact that $\rho_{ty}$ introduces more dependency  in case 1 than case 2 is also verified by the generally higher AUC and lower PE values in case 1  compared to their counterparts at the same time points for case 2.

These results demonstrate the necessity of taking both $\bm R_{(t_{i})(\bm y_{i})}$ and $\bm R_{(\bm y_{i})}$ into account correctly.

\section{Application to AIDS data}
The AIDS data (Goldman \textit{et al.,} 1996\cite{gol96}) is built in some \verb|R| packages such as \verb|JM| (Rizopoulos, 2010\cite{riz10}) or \verb|joineR| (Philipson, {\it et al,} 2017\cite{phi17}). The dataset comprises of the square root of CD4 cell counts per cubic millimeter ($mm^{3}$
) in blood for 467 subjects with advanced human immunodeficiency virus infection at study entry and later on at 2, 6, 12, and 18 months. The CD4 cell count is a important indicator for the progression from HIV infection to acquired immune
deficiency syndrome (AIDS).  Higher CD4 levels indicate a stronger immune system, making it less likely for a subject to progress to AIDS diagnosis or death. Baseline covariates such as gender (female=0 ,male=1), previous opportunistic infection (prevOI) with noAIDS=0 and AIDS=1, AZT (intolerance=0, failure=1) and randomly assigned drug by zalcitabine (ddC=0) or  didanosine (ddI=1)  are also recorded. The main aim of this study was to compare the efficacy and safety of the two drugs ddI and ddC. 
By the end of the study censoring was at about 59.7\% (188 patients had died) and only 1405 were recorded (out of the 2335 planned measurements), leading to 39.8\% missing responses.
A Gaussian copula joint model with random effects is applied to the real data with the longitudinal and survival process specified as follows:
\[y_{ij}=\beta_{10}+\beta_{11}t_{ij}+\beta_{12}t_{ij}Drug_{i}+\beta_{13}Gender_{i}+
\beta_{14}PrevOI_{i}+\beta_{15}Stratum_{i}+b_{i0}+b_{i1}t+\varepsilon_{ij}\]
and
\[h_{i}(t)=h_{0}(t)\mbox{exp}\left\{\beta_{21}Drug_{i}+\beta_{22}Gender_{i}+\beta_{23}PrevOI_{i}
+\beta_{24}Stratum_{i}+\alpha\left(b_{i0}+b_{i1}t\right)\right\},\]
where $\varepsilon_{ij}\sim N(0,\sigma^{2})$ and $h_{0}(t)$
is a piecewise-constant baseline function with eight knots equally spaced.

Preliminary analysis using the \verb|lme| function in \verb|R| for the longitudinal process alone suggests an AR1 structure for $\bm\varepsilon_{i}$ provides the best fit. Thus in scenarios 2, 3 and 4 below, which take into account the correlation in the longitudinal process, we assume an AR1 correlation structure.

	Three scenarios are considered for the correlation matrix of the Gaussian copula. Given the random effects, 
\begin{itemize}
	\item  \textit{scenario 1}: there are no correlations between the two processes as well as within the longitudinal process (conditional independence); 
	\item  \textit{scenario 2}: no correlation is assumed between the survival and longitudinal processes while the correlation within the longitudinal process is assumed to be an AR1 (longitudinal correlated);
	\item  \textit{scenario 3}: there is an increasing correlation structure between the survival and longitudinal processes, which is $\displaystyle\left(\rho_{ty}^{16},\rho_{ty}^{8},\rho_{ty}^{4},\rho_{ty}^{2},\rho_{ty}\right),$ since it is natural to assume that survival time is likely to be more related with later longitudinal measurements and an AR1 correlation structure within the longitudinal process (correlated in both). 
	\end{itemize}

In fact, a constant correlation between the two sub-models and an AR1 correlation structure within the longitudinal process is also fitted but it provides worse fitting than scenario 3 in terms of AIC (results not shown here). The estimation results are summarised in Table \ref{aidsest}.  Under scenarios 1 and 2, the estimates for the slope term for time, $\beta_{11},$ and the error standard deviation, $\sigma,$ are smaller than those in scenario 3 while the estimate for the association parameter $\alpha$ is larger. The reasoning for this was discussed in Section 3.1. While there doesn't seem to be much evidence of a correlation within the longitudinal process, the results indicate there is significant correlation between the two processes. The correlation between the two random effects is positive for scenario 3 while it is negative for scenarios 1 and 2. In general, the regression parameter estimates are similar among the three scenarios. Overall, the joint model with correlation in the longitudinal process (scenario 2) does not perform significantly better than the standard joint model (scenario 1) according to all the information criteria.  On the other hand, the proposed model provides much better fitting than its two counterparts as indicated by a log-likelihood ratio test and AIC criteria.

\begingroup
\setlength{\tabcolsep}{6pt} 
\renewcommand{\arraystretch}{1.2} 
\begin{table}[H]
	\tbl{\textbf{{\scriptsize Parameter estimations based on AIDS data for all three scenarios.}}}
	{\begin{tabular}{ccccccccc}
			\toprule
			&\multicolumn{2} {c} {Scenario 1}&&\multicolumn{2} {c} {Scenario 2}&&\multicolumn{2} {c} {Scenario 3}
			\\
			&\multicolumn{2} {c}{(Con. Independent)}&&\multicolumn{2} {c}{(Longitudinal Corr.)}&&\multicolumn{2} {c}{(Both Corr.)}
			\\
			\cline{2-9}	
			& Est. & SE &&  Est.  &  SE     &&Est.  &SE
			\\
			\hline
			$\beta_{10}$                                           & 10.611 &  0.799 && 10.810   & 0.847  && 10.647  & 0.808  
			\\
			$\beta_{11}$                                            & -0.186 &  0.021 && -0.184   & 0.021  && -0.211  & 0.022  
			\\
			$\beta_{12}$                                            & 0.017  & 0.030  && 0.017     & 0.030  && 0.002   & 0.030  
			\\
			$\beta_{13}$                                            & -0.249 & 0.755  && -0.439 & 0.794  && -0.223  & 0.764  
			\\
			$\beta_{14}$                                            & -4.695 & 0.493  && -4.778 & 0.505  && -4.760  & 0.498  
			\\
			$\beta_{15}$                                            & -0.284 & 0.474  && -0.242 & 0.482  && -0.294  & 0.479  
			\\
			$\beta_{21}$                                            & 0.301   & 0.167   && 0.283   & 0.169 && 0.264   & 0.160  
			\\
			$\beta_{22}$                                            & -0.308 & 0.452  && -0.261  & 0.493 && -0.228 & 0.410  
			\\
			$\beta_{23}$                                            & 1.744   & 0.395  && 1.763    & 0.427 && 1.731   & 0.355  
			\\
			$\beta_{24}$                                            & 0.171    & 0.201   && 0.161   & 0.202&& 0.138  & 0.190  
			\\
			$D_{11}$                                                  & 15.934  & 1.170   && 16.193  & 1.318 &&15.224  &1.325  
			\\
			$D_{22}$                                                 & 0.032   & 0.006  && 0.027  & 0.008 &&0.023   & 0.009 
			\\
			$D_{12}$                                                 & -0.093 & 0.062  && -0.082  & 0.069 && 0.011   &0.073   
			\\
			$\rho_{ty}$                                             &\textemdash&\textemdash&&\textemdash&\textemdash  && 0.763   &0.102   
			\\
			$\rho_{y}$                                              &\textemdash&\textemdash&& 0.103     &  0.093 && 0.186   & 0.113  
			\\
			$\sigma$                                                & 1.738   & 0.048  && 1.826     & 0.105  && 1.933   & 0.149  
			\\
			$\alpha$                                                 & -0.246 & 0.046  && -0.248  & 0.049  && -0.238  & 0.045 
			\\
			Loglik &\multicolumn{2} {c}{-4255.669}  &&\multicolumn{2} {c}{-4255.041} &&\multicolumn{2} {c}{-4248.388}
			\\
			AIC    &\multicolumn{2} {c}{8555.338}    &&\multicolumn{2} {c}{8556.082} &&\multicolumn{2} {c}{8544.776}  
			\\
			\hline
	\end{tabular}}
\label{aidsest}
\end{table}
\endgroup

\begin{figure}[H]
	\centering
	\includegraphics[width=\textwidth]{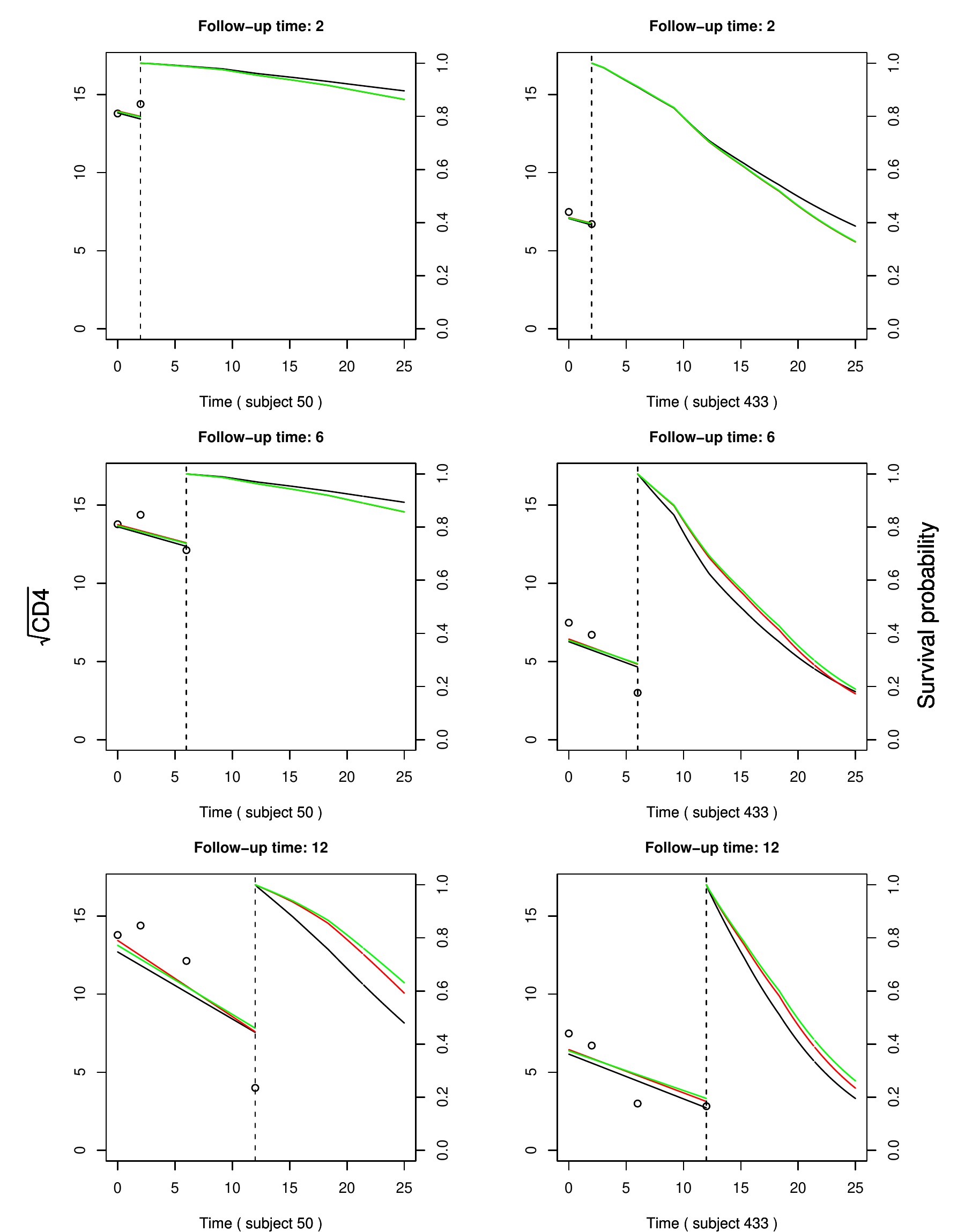}\\
	\caption{\textbf{{\scriptsize Dynamic prediction of survival probabilities and fitted longitudinal trajectories for subjects 50 and 433 in AIDS dataset. The red, green, black lines represent scenarios 1, 2 and 3, respectively. The three scenarios correspond to the outputs in Table 3.}}}
	\label{cd4surpredsub50sub433}
\end{figure}

	Figure \ref{cd4surpredsub50sub433} depicts the dynamic predictions of subjects 50 and 433, which have censoring and true event times at $t=14.4$ and $t=14.13,$ respectively, from the AIDS dataset. The parameters applied for predicting in the three scenarios are from the outputs in Table \ref{aidsest}. 
    For subject 50, the  predictive survival probabilities for the three scenarios are quite close until $t=12,$ where a larger decline in predicted survival probabilities can be observed in scenario 3 compared with the other two scenarios. 
	We notice that there is a large deviation from the fitted lines to the observed longitudinal measurement at $t=12$ and this dip from the fitted longitudinal values to the longitudinal observation is taken into account in our proposed model by correlation parameter $\rho_{ty},$ while it is ignored in scenarios 1 and 2. On the other hand, the discrepancies in the predicted survival  probabilities among all scenarios are not significant for subject 433 across time and this can be explained by the fact that there is no large deviation between the fitted trajectories and the observed longitudinal measurements, which are closely randomly scattered around the fitted lines. Also, the fitted trajectories  are fairly close among the three scenarios across time.
	  Nevertheless, the predicted event times given subject 433 survived up to $t=12$ are 20.9, 21.5 and 20.0 for scenario 1, 2 and 3, which means the proposed model still  provides relatively the most accurate predictions on survival probabilities among the three candidates. We note that the predicted event time of 20.0 is

	\begingroup
\setlength{\tabcolsep}{6pt} 
\renewcommand{\arraystretch}{1.18} 
\begin{table}[H]
	\tbl{\textbf{{\scriptsize AUC for three scenarios at $t=$2, 6, 12 and 18 with $\Delta t=$2 and 4.}}}
	{\begin{tabular}{lcccccccc}
			\toprule
			&$t=2-4$&$t=2-6$&$t=6-8$& $t=6-10$ & $t=12-14$ & $t=12-16$  & $t=18-20$ & $t=18-22$   
			\\
			\hline
			AUC($t+\Delta t|t$)
			\\
			\cdashline{1-9}
			Scenario 1           &  0.744  & 0.746   & 0.720 & 0.699  & 0.741 & 0.758  & 0.764 & 0.760
			\\            
			Scenario 2          &  0.744  &   0.746  & 0.720 & 0.699  & 0.739 & 0.757 & 0.770 & 0.769
			\\            
			Scenario 3          &  0.752  &  0.750  &  0.705 & 0.699  & 0.743 & 0.766 & 0.796 & 0.815
			\\
			\cdashline{1-9}
			PE($t+\Delta t|t$)
			\\
			\cdashline{1-9} 
			Scenario 1           &  0.052  & 0.085   & 0.051  & 0.099  & 0.071 & 0.120  & 0.101 & 0.145
			\\            
			Scenario 2          &  0.052  &  0.085   & 0.051 & 0.099  & 0.071 & 0.120  & 0.100 & 0.144
			\\            
			Scenario 3          &  0.053  &  0.085   & 0.052 & 0.099  & 0.070 & 0.118  & 0.095 & 0.132
			\\
			\hline
	\end{tabular}}
	\label{dynaAUCBSaids}
\end{table}
\endgroup

\begingroup
\setlength{\tabcolsep}{6pt} 
\renewcommand{\arraystretch}{1.2} 
\begin{table}[H]
	\tbl{\textbf{{\scriptsize Outputs of parameter estimations for \textit{scenario 1}: conditional independent for two sub-models as well as within the longitudinal process, \textit{scenario 2}: conditional independent for two sub-models but the correlation within the longitudinal process is assumed to be an AR1 and \textit{scenario 3}: the correlation between two sub-models is $\left(\rho_{ty}^{16},\rho_{ty}^{8},\rho_{ty}^{4},\rho_{ty}^{2},\rho_{ty}\right)$ while the correlation within the longitudinal process is assumed to be an AR. The true parameters values are set to be the estimates in scenario 3 from Table 3 for generating $N=500$ dataset mimicking the AIDS data.}}}
	{\begin{tabular}{lccccccccccc}
			\toprule
			True value &$\beta_{10}$&$\beta_{11}$&$\beta_{12}$&$\beta_{13}$&$\beta_{14}$&$\beta_{15}$&$\beta_{21}$&$\beta_{22}$&$\beta_{23}$&$\beta_{24}$&$\sigma$
			\\
			                  & 10.647        & -0.211        & 0.002         & -0.223        &  -4.760        & -0.294        &0.264          &-0.228         & 1.731           &0.138            &1.933
			\\
			\hline
			Scenario 1  &                   &                   &                    &                   &                   &                    &                    &                   &                    &                    &
			\\
			\cdashline{2-12}
			Est.          &10.572         & -0.176        & 0.006          & -0.245        &-4.682         &-0.264         & 0.274         & -0.235        &1.814             &0.130           &1.743
			\\
			SE            & 0.680          & 0.020        &  0.029         &  0.619         & 0.434         & 0.407           & 0.154         & 0.295          & 0.244           & 0.182         &0.046
			\\
			RMSE       & 0.683           &0.041         &  0.029         &  0.618         &0.440           &0.408           & 0.154         &0.295           &0.258            &0.182         & 0.195
			\\
			CP            &0.944           &  0.554       &  0.944        & 0.942          & 0.948          &0.940           & 0.952         & 0.946          &  0.934          & 0.944        & 0.018
			\\
			\cdashline{1-12}
			Scenario 2  &                    &                   &                   &                    &                     &                    &                   &                    &                     &                  &
			\\
			\cdashline{2-12}
			Est.         &10.572        & -0.176          &0.006          &-0.246          &-4.681           &-0.263          &0.275          &-0.236          &1.820           &0.131           &1.875
			\\
			SE            &0.678          & 0.020          &  0.029        &  0.617         & 0.434            & 0.407           & 0.155         & 0.297          & 0.245         & 0.183         &0.095
			\\
			RMSE       &0.682          & 0.040          & 0.029          &  0.617        & 0.441             &0.407           & 0.155         & 0.297          & 0.260          & 0.183         &0.111
			\\
			CP            &0.946          &0.556          & 0.952          & 0.946         & 0.944            &  0.942         & 0.950         & 0.950         &  0.928        & 0.946          &0.922
			\\
			\cdashline{1-12}
			Scenario 3 &                   &                   &                     &                   &                       &                     &                   &                   &                    &                    &
			\\
			\cdashline{2-12}
			Est.         &10.645         &-0.211        &0.002            &-0.224         & -4.758            & -0.260         &0.267          & -0.228         & 1.749          &0.134           & 1.945
			\\
			SE          & 0.617           & 0.019        &  0.027           &  0.570         &  0.397            & 0.378           & 0.139         & 0.257           & 0.213          & 0.163          & 0.097
			\\
			RMSE     & 0.617           &0.019         &  0.027           &  0.569         & 0.396             &  0.379          & 0.139        & 0.257           & 0.213           & 0.162          & 0.097
			\\
			CP          &0.956           &0.948        &0.944             & 0.954          &0.950              & 0.942           &0.956         &0.948           &0.952            & 0.958         &0.940
			\\
			\midrule
			True value &$D_{11}$&$D_{22}$&$D_{12}$&$\alpha$&$\rho_{ty}$  &$\rho_{y}$&&&&&
			\\
		            	& 15.224       & 0.023      & 0.011            & -0.238         &     0.763   &     0.186         &                &                &               &                &
			\\
			\cline{1-7}
			Scenario 1  &                   &                   &                    &                   &                   &                   &                    &                   &                     &                   &
			\\
			\cdashline{2-7}
			Est.      &15.686      & 0.031       & -0.096          & -0.266          &\textemdash& \textemdash &  &  && &
			\\
			SE        & 1.131       & 0.006       &  0.071           &  0.023           & \textemdash &  \textemdash &  &   & &  &
			\\
			RMSE   & 1.221       & 0.010        &0.129            & 0.037            & \textemdash&  \textemdash &  & && &
			\\
			CP       &0.922       & 0.764        & 0.678            &0.778            & \textemdash&  \textemdash &&&&&
			\\
			\cdashline{1-7}
			Scenario 2   & &  &   &   &    &   &  &  &  &  &
			\\
			\cdashline{2-7}
			Est.      &15.164      & 0.024       &-0.058           &-0.274           & \textemdash& 0.143             && &&&
			\\
			SE        &1.134       & 0.007        & 0.074           &  0.024          & \textemdash& 0.078           & &  & & &
			\\
			RMSE   & 1.134      & 0.007        & 0.101            & 0.043           & \textemdash&0.089            && && &
			\\
			CP       &0.936       & 0.944       & 0.838           &0.706            & \textemdash&0.910            &&&&&
			\\
			\cdashline{1-7}
			Scenario 3   & &  &   &   &    &   &  &  &  &  &
			\\
			\cdashline{2-7}
			Est.      &15.151    &0.022         &0.016             &-0.241           & 0.762          &  0.191          && & & &
			\\
			SE        & 1.150    & 0.006        &  0.062            &  0.022         & 0.027          & 0.070          &  &  &  &  &
			\\
			RMSE   & 1.151     & 0.006        &  0.062            &  0.022         &0.027           &0.070           &&&&&
			\\
			CP       &0.946     & 0.958        &  0.956            &0.938           &0.930          &0.948            &&&&&
			\\
			\cline{1-7}
	\end{tabular}}
\end{table}
\endgroup

\noindent  not as close to the true event time of 14.13 as seen in previous results. As pointed out in Dafni and Tsiatis (1998)\cite{daf98},  the CD4 count is not a successful surrogate for disease progression and a combination of other virological  and immunological markers might be required to achieve better predictive accuracy.

	The overall assessment of performance for the three scenarios is also performed by leave-one-out cross-validation. The predicted on survival probability for a subject is computed on parameters estimated from the remaining population excluding that subject. Table \ref{dynaAUCBSaids} summarises the AUC and BS values based on the predictive survival probabilities for these 467 subjects. Generally speaking, scenario 3 has the best discriminative capability among the three models and this is most obvious at $t=12$ and 18 since the correlation structure is quite weak before $t=12.$ This pattern is similar to the results demonstrated by case 2 from the simulation study.

 Another simulation study is conducted with the true parameters values set to be the estimates in scenario 3 from Table \ref{aidsest}. There are $N=500$ Monte Carlo samples each with sample size $n=467.$ 
 The covariates are simulated according  Bernoulli distributions with empirical probabilities calculated from the AIDS data. An independent censoring process is sampled from a $U(12,21.4),$ since the maximum observed time are 21.4 and censoring times are generally after $t=12$ in the AIDS data. The censoring rate is adjusted to be  about 59.7\% by the end of the study. The missing rates of longitudinal measurements at each scheduled time point are also adjusted according to the AIDS data. Table 5 summarises the estimation results  in all the three scenarios. In scenarios 1 and 2, the slope term of time, $\beta_{11},$ and the standard error, $\sigma,$ of the error terms are underestimated while the association term, $\alpha,$ is overestimated as expected.  The estimate of $\rho_{y}$ is not significant while for $\rho_{ty}$ it is highly significant. The estimate of $D_{12}$ is positive only in scenario 3. The slope $\beta_{11}$ and association parameter $\alpha$  affect the interpretation of the model most and the effects of misspecification on them in this simulation study are generally similar to the results in Table \ref{aidsest}.

 In terms of the likelihood ratio test at a 95\% significance level, scenario 2 outperforms scenario 1 only in 192 out of the 500 Monte Carlo samples while scenario 3 outperforms scenario 1 in 438 out of the 500 Monte Carlo samples. The mean of log-likelihoods over the 500 Monte Carlo samples are -4255.377 , -4253.617 and -4247.285  for scenarios 1, 2 and 3, respectively, which are very close to the log-likelihoods in Table 3. Overall, the proposed model performs much better than a standard joint model for data having extra correlation after conditioning on random effects. The performance is not significantly improved if only the extra correlation within longitudinal measurements is considered.

\section{Discussion}
Our proposed Gaussian copula model with random effects, in which the conditional distribution associated with the longitudinal and survival processes given the random effects is specified by a Gaussian copula, relaxes the most common but rarely verified assumption of conditional independence given latent random effects, assumed in the conventional joint model.  Compared with the conventional joint model, the proposed model is able to utilise the remaining information, if any, in the error terms of the longitudinal measurements and provides more accurate parameter estimations as well as survival prediction. 
Compared with the marginal copula joint model in Zhang, Charalambous and Foster (2021)\cite{zha21} and Ganjali and Baghfalaki (2015)\cite{gan15}, our model allows  predictions at precise subject-specific level for both sub-models. 
The two simulation studies indicate noticeable biases are introduced in parameter $\beta_{11}$ describing the tendency of the longitudinal process and association parameter $\alpha$ between the two sub-models if the extra correlation introduced by copula is ignored or misspecified. Generally, compared with the regression parameters $\bm\beta_{1}$ in the longitudinal process, the regression parameters $\bm\beta_{2}$ of the survival process are more likely to be affected by the misspecification of the correlation matrix $\bm R$ in the Gaussian copula as suggested by the first simulation study. A further dynamic prediction study suggests there are quite significant differences in predictive survival probabilities in the four scenarios considered, especially for case 1 with moderate exchangeable correlation dependency between the two sub-models. Overall, among the four candidate models, scenario 4 has the best discriminative capability and the lowest predictive error. In addition, parameter estimation and predictions of survival probabilities of scenario 2 are slightly improved compared to scenario 1. On the other hand, the differences or improvements from scenario 2 to scenario 3 are negligible and they are both inferior to scenario 4. All the above suggest the necessity for specifying a correct correlation matrix $\bm R$ for modelling the longitudinal and time-to-event processes even after proposing the correct marginal structures conditional on the random effects.

A real data application suggests there are indeed extra correlations, especially between the longitudinal and survival process, after conditioning on all the covariates and random effects (random intercept and slope) for the AIDS data. The proposed model provides significantly better fitting and prediction than its competitors.

As discussed by Diggle \textit{et al.} (2008)\cite{dig08}, it is not straightforward to have a scientific interpretation of the correlation parameters in $\bm R.$ Choosing a proper structure for  $\bm R$ is also problematic, since it is the correlation matrix for the marginals after conditioning on unobservable random effects and can not be estimated empirically from the raw data. Besides, the optimal structure for $\bm R$ may depend on the structure of both the random and fixed effects considered in the model and the computation price is higher compared to the conventional joint model. An unstructured (symmetric and positive definite) $\bm R$ might be a reasonable way to start, but it comes with even higher computational cost, especially when there is a large number of longitudinal measurements;  in this case, the optimal correlation structure may be selected according to AIC criteria or likelihood ratio test among the different options. Furthermore, the positive-definite restriction on $\bm R$ makes our approach more difficulty to be applied on the unbalanced datasets. Suresh \textit{et al.} (2021a)\cite{sur21a} and (2021b)\cite{sur21b} overcame this issue of finding an appropriate correlation structure between the two sub-models since their approach only requires the specification of two dimensional copulas. However, their approach falls under the marginal model regime as in Zhang, Charalambous and Foster (2021)\cite{zha21} and therefore may be improved by the addition of random effects. 

In future work, it could be interesting to develop a similar score test for verifying the conditional independence assumption given random effects as proposed in Jacqmin-Gadda \textit{et al.} (2010)\cite{jac10}. Different types of copula such as the multivariate $t$ copula can be applied if there are tail dependencies in the data. We are also interested in incorporating functional predictors, responses or even a functional correlation structure of $\bm R$ to make the proposed model more general and versatile.

\section*{Appendix A}
Tables (\ref{sim6excar}) to  (\ref{sim4carcar})  summarised outputs from $N=500$ replications of parameters estimates for the two cases under all four scenarios with 4 and 6 longitudinal measurements from Section 3.1. The discussions of the results are given in Section 3.1.

\begin{landscape}
	\vspace*{\fill}
	\begingroup
	\setlength{\tabcolsep}{6pt} 
	\renewcommand{\arraystretch}{1.18}
	\begin{table}[H]
		 \renewcommand\thetable{A.1}
		\tbl{\textbf{{\scriptsize Outputs of parameter estimations for \textbf{case 1} (constant $\rho_{ty}=0.4$ for $\bm R_{(t_{i})(\bm y_{i})}$) in \textit{scenario 1}: $\bm R_{i}$ is misspecified as $\bm I_{n_{i}},$  \textit{scenario 2}: $\bm R_{(\bm y_{i})}$ is correctly specified but $\bm R_{(t_{i})(\bm y_{i})}$ is misspecified as $\bm0,$ \textit{scenario 3}: $\bm R_{(\bm y_{i})}$ is correctly specified but $\bm R_{(t_{i})(\bm y_{i})}$ is misspecified as an AR1 structure and \textit{scenario 4}: $\bm R_{i}$ is correctly specified when there are at most 4 measurements.}}}
	{\begin{tabular}{lccccccccccccccccc}
				\toprule
				True value &$\beta_{10}$&$\beta_{11}$&$\beta_{12}$&$\beta_{13}$&$\beta_{14}$&$\beta_{15}$&$\beta_{21}$&$\beta_{22}$&$\beta_{23}$&$\beta_{24}$&$\sigma$ &$D_{11}$&$D_{22}$&$D_{12}$&$\alpha$&$\rho_{ty}$  &$\rho_{y}$
				\\
				& 10        & -0.5     & 1         & 0.5       &  0.5       & 1        &-2      &-1          & -1.5       &-2      &2   & 2        & 0.2         & -0.1           & -0.5             &     0.4        &     0.5
				\\
				\hline
				Scenario 1 &                   &                   &                    &                   &                   &                   &                    &                    &                     &                   &
				\\
				\cdashline{1-18}
				Est.        &10.142 & -0.470 & 0.927 & 0.457&0.427 & 0.909 & -2.347 & -1.160 &-1.709&-2.299&1.789 &2.887 & 0.214  & -0.196 & -0.700&\textemdash& \textemdash 
				\\
				SE         & 0.418 & 0.045 &  0.300 &  0.298& 0.443 & 0.407  & 0.358  & 0.313 & 0.409& 0.393 &0.077 &0.523 & 0.039 &  0.115  &  0.108& \textemdash &  \textemdash
				\\
				SD         &0.420  &0.047  & 0.308  & 0.303 &0.433 &0.403   &0.342  &0.307  &0.377 &0.358 &0.074   &0.542  &0.036  & 0.112  & 0.098 & \textemdash&  \textemdash
				\\
				RMSE    & 0.443 &0.055 & 0.316   & 0.306 & 0.439 & 0.413 & 0.488 &  0.346 & 0.430& 0.467&0.224  &1.040 & 0.039 &  0.147 & 0.223 & \textemdash&  \textemdash  
				\\
				CP	      & 0.938 & 0.894 &0.934  & 0.934 & 0.958 & 0.938   &0.850 &0.936  & 0.928 &0.896&0.222  & 0.634 & 0.960 &0.876  & 0.580 &  \textemdash&  \textemdash
				\\
				ECP	      & 0.942 & 0.904 &0.940  & 0.942 & 0.954 & 0.938   &0.814 &0.932  & 0.906 &0.842&0.196  &  0.644& 0.944& 0.866 & 0.484 & \textemdash&  \textemdash 
				\\
				\cdashline{1-18}
				Scenario 2   & &  &   &   &    &   &  &  &  &  &
				\\
				\cdashline{1-18}
				Est.     &10.233  & -0.478&0.887 &0.397   &0.407&0.895&-2.337 &-1.125 &-1.695&-2.352&1.805 &2.763 & 0.208 &-0.181 &-0.709 & \textemdash& 0.188 
				\\
				SE       &0.416   & 0.045 &  0.300&  0.298& 0.443& 0.406 & 0.363& 0.315& 0.410& 0.397 &0.091 &0.567  & 0.039 &  0.118&  0.114&  \textemdash& 0.323
				\\
				SD       & 0.401  & 0.043 & 0.300&  0.297& 0.438& 0.361 & 0.315 & 0.298& 0.377  & 0.361 &0.085& 0.562  & 0.036& 0.110 & 0.102&  \textemdash&0.171  
				\\
				RMSE   & 0.445 & 0.055 &0.316 &0.306  & 0.439 & 0.413 &0.501 & 0.353 & 0.434 & 0.476 &0.201& 0.937  & 0.038 & 0.138 & 0.237& \textemdash  & 0.379
				\\
				CP	    & 0.938 & 0.898 & 0.932 & 0.934 & 0.960 & 0.942  &0.840  &0.932 &0.924 & 0.890&0.440& 0.722 & 0.968 & 0.904 & 0.560&  \textemdash&0.982 
				\\
				ECP	   & 0.940 & 0.904 & 0.944 & 0.944 & 0.956 & 0.938  & 0.820 &0.930 &0.904 & 0.844& 0.384& 0.722 & 0.948& 0.892 & 0.468 & \textemdash & 0.528
				\\
				\cdashline{1-18}
				Scenario 3   & &  &   &   &    &   &  &  &  &  &
				\\
				\cdashline{1-18}
				Est.    &10.125 &-0.468  &0.920  &0.468 & 0.432& 0.922 &-2.338& -1.162& -1.688&-2.290 &1.817 &2.742  &0.211  &-0.182  &-0.701 & -0.042 &  0.164
				\\
				SE      & 0.417  & 0.050 &  0.300 & 0.299 &  0.443& 0.408 & 0.359 & 0.313 & 0.406 & 0.392 &0.090 & 0.565  & 0.040 &  0.118&  0.113 & 0.363  & 0.384
				\\
				SD      & 0.426  & 0.050 &  0.304 &  0.303& 0.448&  0.412& 0.349& 0.306 & 0.374  & 0.361&0.080& 0.562  & 0.036 &  0.111 &  0.095&0.282 &0.182 
				\\
				RMSE  & 0.444 & 0.060 & 0.314  & 0.305  & 0.453 &0.419 &0.486 & 0.346 & 0.418  &0.463 & 0.200& 0.931  & 0.037 & 0.138  &  0.223 & 0.524 & 0.382 
				\\
				CP	    & 0.942  & 0.900 & 0.936  & 0.934 & 0.954& 0.932 &0.846 &0.930 &0.940 & 0.900&0.456 & 0.746  & 0.966 & 0.908  & 0.584 & 0.802&0.990
				\\
				ECP	   & 0.948 & 0.904 & 0.940 & 0.940 & 0.956 & 0.936 & 0.826 & 0.926 &0.924 & 0.858& 0.374& 0.742 & 0.948 & 0.890   & 0.426 & 0.634 & 0.530
				\\
				\cdashline{1-18}
				Scenario 4   & &  &   &   &    &   &  &  &  &  &
				\\
				\cdashline{1-18}
				Est.    &10.022  &-0.499  &0.988  &0.489 & 0.486& 0.989 &-2.015& -1.005& -1.526&-2.028&2.023&1.748 &0.198  &-0.087  &-0.512 & 0.423&  0.460
				\\
				SE      & 0.411    & 0.048   &  0.298 &  0.296 &  0.438& 0.403  & 0.300 & 0.256 & 0.335 & 0.325 &0.155& 0.815  & 0.042 &  0.132&  0.104 & 0.069 & 0.192
				\\
				SD      & 0.428  & 0.048 &  0.312 &  0.309& 0.451&  0.402& 0.307  & 0.269& 0.330& 0.322&0.165& 0.842 & 0.039 &  0.129 &  0.108&0.067  & 0.187 
				\\
				RMSE  & 0.428 & 0.048 &  0.312  &0.309  &0.451 & 0.401 &0.307 & 0.269 & 0.330 & 0.323 &0.166& 0.878 & 0.039 & 0.129 & 0.109 & 0.071 & 0.191
				\\
				CP	   & 0.940   & 0.944 & 0.920  & 0.926  & 0.948& 0.956 &0.934 &0.944 &0.952 & 0.960&0.928& 0.916  & 0.978 & 0.966  & 0.928 & 0.956&0.928
				\\
				ECP	  & 0.952   & 0.942 & 0.944  & 0.940    & 0.950& 0.956 &0.942 &0.958 &0.950 & 0.956&0.946& 0.934  & 0.964 & 0.962  & 0.934 & 0.948&0.928
				\\
				\hline
		\end{tabular}}
\label{sim6excar}
	\end{table}
	\endgroup
	\vspace*{\fill}
\end{landscape}

\begin{landscape}
	\vspace*{\fill}
	\begingroup
	\setlength{\tabcolsep}{6pt} 
	\renewcommand{\arraystretch}{1.18} 
	\begin{table}[H]
		 \renewcommand\thetable{A.2}
		\tbl{\textbf{{\scriptsize Outputs of parameter estimations for \textbf{case 1} (constant $\rho_{ty}=0.4$ for $\bm R_{(t_{i})(\bm y_{i})}$) in \textit{scenario 1}: $\bm R_{i}$ is misspecified as $\bm I_{n_{i}},$  \textit{scenario 2}: $\bm R_{(\bm y_{i})}$ is correctly specified but $\bm R_{(t_{i})(\bm y_{i})}$ is misspecified as $\bm0,$ \textit{scenario 3}: $\bm R_{(\bm y_{i})}$ is correctly specified but $\bm R_{(t_{i})(\bm y_{i})}$ is misspecified as an AR1 structure and \textit{scenario 4}: $\bm R_{i}$ is correctly specified when there are at most 6 measurements.}}}
	{\begin{tabular}{lccccccccccccccccc}
				\toprule
				True value &$\beta_{10}$&$\beta_{11}$&$\beta_{12}$&$\beta_{13}$&$\beta_{14}$&$\beta_{15}$&$\beta_{21}$&$\beta_{22}$&$\beta_{23}$&$\beta_{24}$&$\sigma$&$D_{11}$&$D_{22}$&$D_{12}$&$\alpha$&$\rho_{ty}$  &$\rho_{y}$
				\\
				& 10          & -0.5       & 1            & 0.5         &  0.5     & 1         &-2          &-1           & -1.5          &-2         &2& 2        & 0.2         & -0.1       & -0.5        &     0.4        &     0.5  
				\\
				\hline
				Scenario 1 &                   &                   &                    &                   &                   &                   &                    &                    &                     &                   &
				\\
				\cdashline{1-18}
				Est.        &10.165   & -0.469&0.915   &0.451   &0.416   &0.895   &-2.360 &-1.171  &-1.715 &-2.314 &1.774&2.964 & 0.225  & -0.224 & -0.695&\textemdash& \textemdash 
				\\
				SE         & 0.412     & 0.043 &  0.293 &  0.292& 0.435 & 0.400  & 0.349  & 0.307 & 0.403& 0.387 &0.057& 0.502 & 0.038 &  0.112  &  0.096 & \textemdash &  \textemdash
				\\
				SD        & 0.410     & 0.044 &  0.301 &  0.297& 0.427  & 0.390  & 0.333   &0.300  &0.372 &0.356 &0.058&0.521  &0.035  & 0.111    & 0.089 & \textemdash&  \textemdash
				\\
				RMSE   & 0.442   & 0.054 &0.312  &0.301   & 0.434  & 0.404  &0.490  & 0.345   & 0.430 & 0.475 &0.233& 1.096  & 0.043 & 0.166 & 0.214 & \textemdash&  \textemdash 
				\\
				CP	      & 0.934   & 0.868 &  0.930 & 0.930  & 0.952 & 0.942  &0.836    &0.928  & 0.932&0.872& 0.022& 0.542 & 0.920 & 0.806 & 0.490 &  \textemdash&  \textemdash
				\\
				ECP	     & 0.934 & 0.888   & 0.932 & 0.934  & 0.950 & 0.934   & 0.810   & 0.916 & 0.910 & 0.846&0.026& 0.578 & 0.886 & 0.802 & 0.422 & \textemdash&  \textemdash
				\\
				\cdashline{1-18}
				Scenario 2   & &  &   &   &    &   &  &  &  &  &
				\\
				\cdashline{1-18}
				Est.     &10.180  &-0.472&0.907  &0.448   &0.412 &0.888&-2.407 &-1.193 &-1.735 &-2.348&1.871&2.539 & 0.210 &-0.168 &-0.746  & \textemdash& 0.375 
				\\
				SE       &0.414   & 0.043 &  0.295&  0.294& 0.437& 0.402 & 0.363& 0.316& 0.411 & 0.397&0.079 &0.537  & 0.039 &  0.115&  0.113  &  \textemdash& 0.092 
				\\
				SD       & 0.412  & 0.044 & 0.301&  0.297 &0.426& 0.389 & 0.344 & 0.310& 0.380 & 0.369&0.073& 0.550  & 0.035& 0.113 & 0.100  &  \textemdash&0.099  
				\\
				RMSE  & 0.449 & 0.053  &0.314 &0.301   &0.435 & 0.404 & 0.532& 0.365& 0.447 & 0.507&0.148& 0.769 & 0.036 & 0.132 & 0.266 & \textemdash& 0.159
				\\
				CP	     & 0.938 & 0.878 & 0.928 & 0.934 & 0.950& 0.938  &0.818  &0.918 &0.928 & 0.860&0.630& 0.822 & 0.966 & 0.924 & 0.436&  \textemdash&0.812
				\\
				ECP	    & 0.938 & 0.886 & 0.930 & 0.936 & 0.946 & 0.938& 0.790& 0.908 & 0.910 &0.838& 0.570& 0.830 & 0.942 & 0.918 & 0.342 & \textemdash & 0.840 
				\\
				\cdashline{1-18}
				Scenario 3   & &  &   &   &    &   &  &  &  &  &
				\\
				\cdashline{1-18}
				Est.    &10.178 &-0.469  &0.897  &0.455 & 0.401   & 0.884 &-2.394& -1.188& -1.715&-2.328 &1.872&2.535  &0.211    &-0.172  &-0.741 & -0.056  &  0.377
				\\
				SE      & 0.413  & 0.045 & 0.295 &  0.294 &  0.437& 0.402 & 0.362 & 0.315 & 0.409 & 0.395 &0.080& 0.540  & 0.039 &  0.116  &  0.114  & 0.306  & 0.093
				\\
				SD      & 0.410  & 0.048 &  0.297 & 0.294& 0.433  &  0.384 & 0.349 & 0.313& 0.367  & 0.358&0.075& 0.556  & 0.034 &  0.111  &  0.097& 0.260  & 0.102 
				\\
				RMSE  & 0.447 & 0.057 & 0.314  &0.298  &0.444  & 0.401  & 0.526 & 0.365 & 0.425 & 0.485 &0.148& 0.771  & 0.036  & 0.133 & 0.260 & 0.525  & 0.160
				\\
				CP	    & 0.934  & 0.870 & 0.926  & 0.932& 0.948 & 0.946  &0.812  &0.912  &0.944  & 0.882&0.638& 0.822  & 0.966 & 0.926  & 0.452 & 0.680  &0.832 
				\\
				ECP	   & 0.932 & 0.906 & 0.928  & 0.932 & 0.948 & 0.934 & 0.798 & 0.908 & 0.920 & 0.852&0.606& 0.828 & 0.938 & 0.914  & 0.320  & 0.588  & 0.864 
				\\
				\cdashline{1-18}
				Scenario 4   & &  &   &   &    &   &  &  &  &  &
				\\
				\cdashline{1-18}
				Est.    &10.023  &-0.498  &0.983  &0.494    & 0.476& 0.984   &-2.030& -1.011& -1.522&-2.030&2.024&1.742   &0.198  &-0.087  &-0.519 & 0.412& 0.503  
				\\
				SE      & 0.406  & 0.045 &  0.292 &  0.290  &  0.431 & 0.396  & 0.295 & 0.252 & 0.331 & 0.321 &0.118& 0.684 & 0.040 &  0.122&  0.096 & 0.059 & 0.080  
				\\
				SD      & 0.423  & 0.045 &  0.304 &  0.293 & 0.454 &  0.400  & 0.302 & 0.260 & 0.325& 0.315 &0.119& 0.706 & 0.035 &  0.119 &  0.097 &0.053  &0.086 
				\\
				RMSE  & 0.424 & 0.045 & 0.304  & 0.293  & 0.455  & 0.400  & 0.303 & 0.260 & 0.325 & 0.316 &0.122& 0.751 & 0.035  &  0.120 & 0.099  & 0.054 & 0.086
				\\
				CP	    & 0.938 & 0.958 & 0.942  & 0.934   & 0.946  & 0.954   &0.930 &0.940 &0.960  & 0.958&0.934& 0.914  & 0.980 & 0.950  & 0.932 & 0.964&0.958  
				\\
				ECP	   & 0.956 & 0.956 & 0.952  & 0.934   & 0.946 &  0.958  & 0.934& 0.952 & 0.948 &0.954 &0.934 & 0.922 & 0.944 & 0.948  & 0.932 & 0.942 & 0.972 
				\\
				\hline
		\end{tabular}}
	\label{sim4excar}
	\end{table}
	\endgroup
	\vspace*{\fill}
\end{landscape}

\begin{landscape}
	\vspace*{\fill}
	\begingroup
	\setlength{\tabcolsep}{6pt} 
	\renewcommand{\arraystretch}{1.18} 
	\begin{table}[H]
		 \renewcommand\thetable{A.3}
		\tbl{\textbf{{\scriptsize Outputs of parameter estimations for \textbf{case 2} (continuous AR1 with $0.5^{|t-10|}$ for $\bm R_{(t_{i})(\bm y_{i})}$) in \textit{scenario 1}: $\bm R_{i}$ is misspecified as $\bm I_{n_{i}},$  \textit{scenario 2}: $\bm R_{(\bm y_{i})}$ is correctly specified but $\bm R_{(t_{i})(\bm y_{i})}$ is misspecified as $\bm0,$ \textit{scenario 3}: $\bm R_{(\bm y_{i})}$ is correctly specified but $\bm R_{(t_{i})(\bm y_{i})}$ is misspecified as an exchangeable structure and \textit{scenario 4}: $\bm R_{i}$ is correctly specified when there are at most 4 measurements.}}}
		{\begin{tabular}{lccccccccccccccccc}
				\toprule
				True value &$\beta_{10}$&$\beta_{11}$&$\beta_{12}$&$\beta_{13}$&$\beta_{14}$&$\beta_{15}$&$\beta_{21}$&$\beta_{22}$&$\beta_{23}$&$\beta_{24}$&$\sigma$&$D_{11}$&$D_{22}$&$D_{12}$&$\alpha$&$\rho_{ty}$  &$\rho_{y}$
				\\
				& 10      & -0.5     & 1         & 0.5        &  0.5       & 1         &-2        &-1        & -1.5        &-2       &2      & 2       & 0.2       & -0.1          & -0.5         &     0.5        &     0.5  
				\\
				\hline
				Scenario 1 &                   &                   &                    &                   &                   &                   &                    &                    &                     &                   &
				\\
				\cdashline{1-18}
				Est.        &9.971 & -0.455 & 0.994 & 0.488 &0.509 & 0.966  & -2.029 & -1.000 &-1.512&-2.003&1.830&2.647 & 0.209  & -0.187 & -0.486&\textemdash & \textemdash
				\\
				SE         & 0.414 & 0.047 &  0.295 &  0.294& 0.438 & 0.403  & 0.301  & 0.266 & 0.346& 0.330 &0.083& 0.531 & 0.040  &  0.113  &  0.090& \textemdash &  \textemdash
				\\
				SD         &0.401  &0.048  & 0.311  & 0.294 &0.467 &0.404    &0.308   &0.258   &0.332  &0.312 &0.082&0.504  &0.040  & 0.110   & 0.097 & \textemdash&  \textemdash
				\\
				RMSE    &0.401  &0.065  &  0.311 & 0.294 & 0.467 &0.406   &0.309   &0.257    &0.332 & 0.312 &0.188&0.820  & 0.041 &  0.141 & 0.098 & \textemdash&  \textemdash
				\\
				CP	      & 0.970 & 0.840 &0.938  & 0.954 & 0.926 & 0.946   &0.946 &0.960    & 0.958 &0.960 &0.460& 0.786 & 0.960 &0.884  & 0.944 &  \textemdash&  \textemdash 
				\\
				ECP	     & 0.962 & 0.848 & 0.950 & 0.954 & 0.948 & 0.946  &0.950 & 0.954   & 0.954 & 0.956 &0.442& 0.764 & 0.956 & 0.874 & 0.956 & \textemdash&  \textemdash
				\\
				\cdashline{1-18}
				Scenario 2   & &  &   &   &    &   &  &  &  &  &
				\\
				\cdashline{1-18}
				Est.     &10.009   & -0.465&0.980 &0.481   &0.493&0.950   &-2.147&-1.054   &-1.566&-2.089 &2.024&1.706   & 0.191  &-0.091   &-0.599 & \textemdash& 0.477 
				\\
				SE       &0.414     & 0.048 &  0.297&  0.295& 0.439& 0.404 & 0.332& 0.282  & 0.361 & 0.349 &0.137 &0.668   & 0.042 &  0.120  &  0.133&  \textemdash& 0.163 
				\\
				SD       & 0.400   & 0.048 & 0.309 &  0.295& 0.470 & 0.402 & 0.345 & 0.276 & 0.352 & 0.340 &0.145 & 0.704  & 0.041 & 0.117    & 0.155 &  \textemdash &0.180
				\\
				RMSE  & 0.400   & 0.059 & 0.309 & 0.295 &0.469  &0.404  & 0.375 & 0.281 & 0.358 & 0.351  &0.147 & 0.762  &0.042  & 0.117   & 0.184 & \textemdash  & 0.182 
				\\
				CP	    & 0.968   & 0.890 & 0.944  & 0.950 & 0.920& 0.946  &0.916  &0.946  &0.954  & 0.956 &0.932& 0.914  & 0.946 & 0.958  & 0.864  & \textemdash &0.930  
				\\
				ECP	   & 0.956  & 0.894  & 0.950  & 0.950 & 0.948 & 0.946 & 0.926 & 0.944 & 0.950 & 0.950&0.952& 0.926 & 0.946 & 0.952  & 0.894  & \textemdash &0.930 
				\\
				\cdashline{1-18}
				Scenario 3   & &  &   &   &    &   &  &  &  &  &
				\\
				\cdashline{1-18}
				Est.    &10.001  &-0.466 &0.984  &0.484  & 0.488 & 0.954 &-2.100& -1.032& -1.545&-2.053 &2.057 &1.564   &0.187   &-0.078  &-0.571 & 0.038&  0.495 
				\\
				SE      & 0.416  & 0.049 &  0.298 &  0.296 & 0.440& 0.405 & 0.329 & 0.276 & 0.356 & 0.347 &0.170 & 0.822  & 0.042 &  0.126 &  0.132 & 0.138  & 0.179 
				\\
				SD      & 0.401  & 0.049 &  0.305 & 0.292 & 0.460 &  0.402& 0.323 & 0.262 & 0.341 & 0.339 &0.180 & 0.842  & 0.043 &  0.121 &  0.142& 0.148  &0.190
				\\
				RMSE & 0.401  & 0.059 & 0.305  &0.292  & 0.460  &0.405  & 0.338 &0.264 & 0.343 & 0.343 &0.188 &0.947  & 0.045  & 0.123  & 0.158 & 0.485 & 0.190
				\\
				CP	    & 0.970  & 0.906 & 0.940  & 0.956 & 0.936& 0.952 &0.944  &0.962  &0.958  & 0.958 &0.918 & 0.888 & 0.936 & 0.954  & 0.892 & 0.090 &0.930  
				\\
				ECP	   & 0.962 & 0.902 & 0.948  & 0.952 & 0.948 & 0.952 & 0.940& 0.954 & 0.948  & 0.952 &0.942 & 0.900 & 0.946 & 0.946  & 0.914 & 0.110 & 0.932
				\\
				\cdashline{1-18}
				Scenario 4   & &  &   &   &    &   &  &  &  &  &
				\\
				\cdashline{1-18}
				Est.    &9.984   &-0.501 &1.015     &0.501   & 0.525 & 0.998 &-2.026 & -1.003& -1.518 &-2.013 &2.021 &1.757    &0.196  &-0.090 &-0.511 & 0.525 &  0.475
				\\
				SE      & 0.418   & 0.052 &  0.299 &  0.297 &  0.442& 0.407  & 0.314 & 0.263  & 0.338 & 0.331  &0.150 & 0.745  & 0.042 &  0.125&  0.123 & 0.213 & 0.181 
				\\
				SD      & 0.397   & 0.054 &  0.306 &  0.289& 0.473 &  0.403 & 0.314  & 0.259 & 0.331 & 0.325  &0.152 & 0.730 & 0.042 &  0.120 &  0.136 &0.256  &0.180  
				\\
				RMSE  & 0.397  & 0.054 & 0.306  & 0.288 & 0.474 &0.402   & 0.315  & 0.258 & 0.331  & 0.325  &0.153 & 0.769 & 0.042 & 0.120  & 0.136  & 0.257 & 0.182
				\\
				CP	    & 0.970   & 0.944 & 0.938  & 0.960 & 0.928 & 0.958  &0.948  &0.952   &0.948  & 0.958 &0.950 & 0.936 & 0.950 & 0.954  & 0.932 & 0.942&0.934 
				\\
				ECP	   & 0.962  & 0.954  & 0.942 & 0.956 & 0.950 & 0.956  & 0.948  & 0.950 & 0.940  & 0.956 &0.958 & 0.934 & 0.952 & 0.950  & 0.946 &0.968 &0.934 
				\\
				\hline
		\end{tabular}}
	\label{sim6carcar}
	\end{table}
	\endgroup
	\vspace*{\fill}
\end{landscape}

\begin{landscape}
	\vspace*{\fill}
	\begingroup
	\setlength{\tabcolsep}{6pt} 
	\renewcommand{\arraystretch}{1.18} 
	\begin{table}[H]
		 \renewcommand\thetable{A.4}
	\tbl{\textbf{{\scriptsize Outputs of parameter estimations for \textbf{case 2} (continuous AR1 with $0.5^{|t-10|}$ for $\bm R_{(t_{i})(\bm y_{i})}$) in \textit{scenario 1}: $\bm R_{i}$ is misspecified as $\bm I_{n_{i}},$  \textit{scenario 2}: $\bm R_{(\bm y_{i})}$ is correctly specified but $\bm R_{(t_{i})(\bm y_{i})}$ is misspecified as $\bm0,$ \textit{scenario 3}: $\bm R_{(\bm y_{i})}$ is correctly specified but $\bm R_{(t_{i})(\bm y_{i})}$ is misspecified as an exchangeable structure and \textit{scenario 4}: $\bm R_{i}$ is correctly specified when there are at most 6 measurements.}}}
		{\begin{tabular}{lccccccccccccccccc}
				\toprule
				&$\beta_{10}$&$\beta_{11}$&$\beta_{12}$&$\beta_{13}$&$\beta_{14}$&$\beta_{15}$&$\beta_{21}$&$\beta_{22}$&$\beta_{23}$&$\beta_{24}$&$\sigma$&$D_{11}$&$D_{22}$&$D_{12}$&$\alpha$&$\rho_{ty}$  &$\rho_{y}$
				\\
				& 10      & -0.5     & 1         & 0.5        &  0.5       & 1         &-2        &-1        & -1.5        &-2       &2      & 2       & 0.2       & -0.1          & -0.5         &     0.5        &     0.5  
				\\
				\hline
				Scenario 1 &                   &                   &                    &                   &                   &                   &                    &                    &                     &                   &
				\\
				\cdashline{1-18}
				Est.        &9.954 & -0.464 & 1.003 & 0.494 &0.512  & 0.967   & -2.004 & -0.990 &-1.500&-1.985 &1.790 &2.821   & 0.222  & -0.212 & -0.461& \textemdash & \textemdash
				\\
				SE         & 0.411 & 0.045  &  0.292 &  0.290& 0.434 & 0.399  & 0.289  & 0.259  & 0.338 & 0.321 &0.059 & 0.501   & 0.038 &  0.109  &  0.075 & \textemdash &  \textemdash
				\\
				SD        &0.400  &0.046  & 0.308   & 0.290 &0.459  &0.398   &0.296   &0.253   &0.324 &0.303  &0.057 &0.490   &0.038  & 0.108  & 0.078  & \textemdash &  \textemdash
				\\
				RMSE   & 0.402 & 0.059 & 0.308   & 0.290  & 0.458 &0.399  & 0.296  &0.253   & 0.324 &0.303 &0.217  &0.955  &0.043  & 0.155  & 0.087  &  \textemdash&  \textemdash
				\\
				CP	      & 0.968 & 0.878  &0.944  & 0.950  & 0.926 & 0.950  &0.944  &0.956   & 0.958 &0.968  &0.054 & 0.636  & 0.914 &0.830   & 0.904  &  \textemdash&  \textemdash 
				\\
				ECP	     & 0.960 & 0.886 & 0.954 & 0.950  & 0.948 & 0.950  &0.946  & 0.948  & 0.956 & 0.950 &0.050 & 0.630  &0.908 & 0.824 & 0.922  &  \textemdash&  \textemdash
				\\
				\cdashline{1-18}
				Scenario 2   & &  &   &   &    &   &  &  &  &  &
				\\
				\cdashline{1-18}
				Est.     &9.999   & -0.471&0.986 &0.488   &0.493 &0.949  &-2.135  &-1.050 &-1.562 &-2.081&2.023 &1.717   & 0.193  &-0.089 &-0.585 & \textemdash& 0.507 
				\\
				SE       &0.410   & 0.045 &  0.291&  0.290& 0.432 & 0.398 & 0.322  & 0.277 & 0.356  & 0.344&0.112 &0.605  & 0.040 &  0.115 &  0.118 &  \textemdash& 0.076 
				\\
				SD       & 0.397  & 0.046 & 0.305&  0.288 & 0.458& 0.393 & 0.334 & 0.269  & 0.342 & 0.327 &0.115 & 0.610 & 0.039 & 0.116  & 0.127 &  \textemdash &0.084  
				\\
				RMSE  & 0.396 & 0.054  & 0.305 & 0.288 &0.457 & 0.396 & 0.360 & 0.273  & 0.347 & 0.336 &0.117 & 0.672 & 0.039 & 0.117  & 0.152 & \textemdash  &0.084 
				\\
				CP	    & 0.970  & 0.902 & 0.942 & 0.950 & 0.924 & 0.950  &0.934  &0.952 &0.954  & 0.958&0.928 & 0.920 & 0.958 & 0.948 & 0.882 &  \textemdash &0.968 
				\\
				ECP	   & 0.966 & 0.908 & 0.954 & 0.950 & 0.948 & 0.948  & 0.944 & 0.942 & 0.944 & 0.946&0.934 & 0.922 & 0.956 & 0.948 & 0.898 & \textemdash  & 0.978  
				\\
				\cdashline{1-18}
				Scenario 3   & &  &   &   &    &   &  &  &  &  &
				\\
				\cdashline{1-18}
				Est.    &9.991  &-0.472  &0.988  &0.491    & 0.496 & 0.958 &-2.109  & -1.041 & -1.546&-2.064 &2.030 &1.689   &0.193   &-0.088 &-0.569  & 0.018   &  0.508  
				\\
				SE      & 0.411  & 0.046  &  0.292 &  0.290 &  0.433& 0.399 & 0.327  & 0.276 & 0.356 & 0.346 &0.126  & 0.685  & 0.040 &  0.119 &  0.123  & 0.126    & 0.084 
				\\
				SD      & 0.399 & 0.047  &  0.308 &  0.287 & 0.457 &  0.392 & 0.333 & 0.266 & 0.341 & 0.334  &0.131  & 0.687  & 0.040 &  0.122 &  0.130  &0.138    &0.091 
				\\
				RMSE  &0.398 & 0.055  & 0.308  & 0.287  &0.456  & 0.393  &0.350  & 0.269 & 0.344 & 0.340  &0.134 & 0.753  & 0.041 &  0.123 &   0.147  & 0.502  &  0.091
				\\
				CP	    & 0.974  & 0.898 & 0.938  & 0.954 & 0.924  & 0.956 &0.934  &0.950  &0.952  & 0.960  &0.926 & 0.916   & 0.956 & 0.940  & 0.900  & 0.038   &0.956 
				\\
				ECP	   & 0.960  & 0.912 & 0.954  & 0.952 & 0.950  & 0.952 & 0.942 & 0.946 & 0.940& 0.944  &0.936  & 0.916   & 0.956 & 0.952  & 0.912  & 0.046   & 0.974
				\\
				\cdashline{1-18}
				Scenario 4   & &  &   &   &    &   &  &  &  &  &
				\\
				\cdashline{1-18}
				Est.    &9.987   &-0.502 &1.019     &0.504  & 0.520  & 0.988 &-2.029 & -1.007 & -1.510&-2.006&2.015  &1.787    &0.197   &-0.093&-0.511 & 0.516  &  0.500
				\\
				SE      & 0.412  & 0.048 &  0.294  &  0.292 &  0.435& 0.400  & 0.307  & 0.259 & 0.335 & 0.327 &0.115  & 0.632  & 0.039 &  0.116&  0.112 & 0.186  & 0.080 
				\\
				SD      & 0.393  & 0.051 &  0.303  &  0.285 & 0.460  &  0.391 & 0.301  & 0.260 & 0.324 & 0.311 &0.119  & 0.622 & 0.039 &  0.118 &  0.119 &0.235  &0.088 
				\\
				RMSE  & 0.393 & 0.051  & 0.304  & 0.285  &  0.460 &  0.391 & 0.303  & 0.259 & 0.324 & 0.310 &0.120 & 0.657  & 0.039 & 0.118  &  0.119 & 0.235 &0.088 
				\\
				CP	    & 0.972 & 0.942  & 0.934  & 0.956   & 0.926 & 0.964 &0.954   &0.954  &0.950  & 0.974 &0.938 & 0.952 & 0.956 & 0.942  & 0.952 & 0.922&0.968 
				\\
				ECP	   & 0.964 & 0.954  & 0.942  & 0.950  & 0.952  & 0.954 & 0.954  & 0.954 &0.946 & 0.962 &0.942 & 0.946 & 0.956 & 0.944  & 0.956 &0.958 & 0.982 
				\\
				\hline
		\end{tabular}}
		\label{sim4carcar}
	\end{table}
	\endgroup
	\vspace*{\fill}
\end{landscape}

\section*{Appendix B}
Assuming random intercept and slope for random effect components, there is a set of joint data generated under a general correlation matrix  
\begin{eqnarray*}
	\displaystyle
	\bm R_{i}=\left\{
	\begin{array}{ll}
		\displaystyle
		1& \bm R_{(t_{i})(\bm y_{i})}
		\\
		\displaystyle
		\\
		\bm R_{(\bm y_{i})(t_{i})}& \bm R_{(\bm y_{i})}
	\end{array}
	\right\},  \mbox{  } i=1,...,n.
\end{eqnarray*}  
If  
\begin{eqnarray*}
	\displaystyle
	\bm R_{i}^{*}=\left\{
	\begin{array}{ll}
		\displaystyle
		1& \bm0
		\\
		\displaystyle
		\\
		\bm0& \bm R_{(\bm y_{i})}
	\end{array}
	\right\},
\end{eqnarray*} 
which corresponds to assuming conditional independence  between the survival and longitudinal processes as in scenario 2,  some explicit updating expressions can be derived for the EM algorithm.

In the E-step
\begin{eqnarray*}
	Q(\bm\theta|\bm\theta^{(l)})&=&\sum_{i=1}^{n}\int_{\bm b_{i}}\left\{\delta_{i}\mbox{log}f_{T_{i}^{*},\bm y_{i}}(t_{i},\bm y_{i}|\bm b_{i})+(1-\delta_{i})\mbox{log}f_{T_{i}^{*},\bm y_{i}}(T_{i}^{*}>t_{i},\bm y_{i}|\bm b_{i})+\mbox{log}f_{\bm b_{i}}(\bm b_{i})\right\}\\ \nonumber
	&\times&f_{\bm{b_{i}}}^{\bm R_{i}^{*}}(\bm b_{i}|t_{i},\delta_{i},\bm y_{i};\bm\theta^{(l)})\,d\bm b_{i}\\ \nonumber
	&=&-\frac{N+nq}{2}\mbox{log}2\pi-\frac{n}{2}\mbox{log}|\bm D|-\frac{1}{2}\sum_{i=1}^{n}E_{\bm b_{i}|\bm y_{i},t_{i}}^{\bm R_{i}^{*}(l)}\left(\bm b_{i}^{'}\bm D^{-1}\bm b_{i}\right)-\sum_{i=1}^{n}\frac{n_{i}}{2}\mbox{log}\sigma^{2}\\                                          \nonumber
	&&-\sum_{i=1}^{n}\frac{1}{2\sigma^{2}}E_{\bm b_{i}|\bm y_{i},t_{i}}^{\bm R_{i}^{*}(l)}\left(\left(\bm y_{i}-\bm X_{i1}\bm\beta_{1}-\bm Z_{i1}\bm b_{i}\right)^{'}\bm R_{(\bm y_{i})}^{-1}\left(\bm y_{i}-\bm X_{i1}\bm\beta_{1}-\bm Z_{i1}\bm b_{i}\right)\right)\\                                          \nonumber
	&&+\sum_{i=1}^{n}\delta_{i}E_{\bm b_{i}|\bm y_{i},t_{i}}^{\bm R_{i}^{*}(l)}\left(\mbox{log}f_{T_{i}^{*}}(t_{i}|\bm b_{i})\right)+\sum_{i=1}^{n}(1-\delta_{i})E_{\bm b_{i}|\bm y_{i},t_{i}}^{\bm R_{i}^{*}(l)}\left(\mbox{log}S_{T_{i}^{*}}(t_{i}|\bm b_{i})\right),                                                              \nonumber
\end{eqnarray*}
where $f_{\bm{b_{i}}}^{\bm R_{i}^{*}}(\bm b_{i}|t_{i},\delta_{i},\bm y_{i};\bm\theta^{(l)})$ denotes the distribution of $\bm b_{i}$ conditional on the observed event time and longitudinal measurements when the correlation structure is specified as $\bm R_{i}^{*}$ and $E_{\bm b_{i}|\bm y_{i},t_{i}}^{\bm R_{i}^{*}(l)}\left(h(\bm b_{i})\right)=\int_{\bm b_{i}}h(\bm b_{i})f_{\bm{b_{i}}}^{\bm R_{i}^{*}}(\bm b_{i}|t_{i},\delta_{i},\bm y_{i};\bm\theta^{(l)})\,d\bm b_{i}$ at the $l$th iteration.

In the M-step

\begin{equation}
	\begin{split}
		\widehat{\bm D}^{(l+1)}=&\frac{1}{n}\sum_{i=1}^{n}E_{\bm b_{i}|\bm y_{i},t_{i}}^{\bm R_{i}^{*}(l)}\left(\bm b_{i}\bm b_{i}^{'}\right)\\                                                                                                            
		=&\frac{1}{n}\sum_{i=1}^{n}\left\{E_{\bm b_{i}|\bm y_{i},t_{i}}^{\bm R_{i}^{*}(l)}\left(\left(\bm b_{i}-E_{\bm b_{i}|\bm y_{i},t_{i}}^{\bm R_{i}^{*}(l)}\left(\bm b_{i}\right)\right)^{'}\left(\bm b_{i}-E_{\bm b_{i}|\bm y_{i},t_{i}}^{\bm R_{i}^{*}(l)}\left(\bm b_{i}\right)\right)\right)+E_{\bm b_{i}|\bm y_{i},t_{i}}^{\bm R_{i}^{*}(l)}\left(\bm b_{i}\right)E_{\bm b_{i}|\bm y_{i},t_{i}}^{\bm R_{i}^{*}(l)}\left(\bm b_{i}^{'}\right)\right\}\\                                                                         
		=&\frac{1}{n}\sum_{i=1}^{n}\left\{var_{\bm b_{i}|\bm y_{i},t_{i}}^{\bm R_{i}^{*}(l)}\left(\bm b_{i}\right)+E_{\bm b_{i}|\bm y_{i},t_{i}}^{\bm R_{i}^{*}(l)}\left(\bm b_{i}\right)E_{\bm b_{i}|\bm y_{i},t_{i}}^{\bm R_{i}^{*}(l)}\left(\bm b_{i}^{'}\right)\right\},
	\end{split}	\tag{B.1} \label{emupdateD}
\end{equation}

\begin{equation}
	\begin{split}
		\widehat{\sigma^{2}}^{(l+1)}=&\frac{1}{N}\sum_{i=1}^{n}E_{\bm b_{i}|\bm y_{i},t_{i}}^{\bm R_{i}^{*}(l)}\left(\left(\bm y_{i}-\bm X_{i1}\bm\beta_{1}-\bm Z_{i1}\bm b_{i}\right)^{'}\bm R_{(\bm y_{i})}^{-1}\left(\bm y_{i}-\bm X_{i1}\bm\beta_{1}-\bm Z_{i1}\bm b_{i}\right)\right)\\                                                                                                                                                                                     
		=&\frac{1}{N}\sum_{i=1}^{n}\left\{\left(\bm y_{i}-\bm X_{i1}\bm\beta_{1}\right)^{'}\left(\bm y_{i}-\bm X_{i1}\bm\beta_{1}-2\bm Z_{i1}E_{\bm b_{i}|\bm y_{i},t_{i}}^{\bm R_{i}^{*}(l)}\left(\bm b_{i}\right)\right)+tr\left(Z_{i}^{'}\bm R_{(\bm y_{i})}^{-1}Z_{i}var_{\bm b_{i}|\bm y_{i},t_{i}}^{\bm R_{i}^{*}(l)}\left(\bm b_{i}\right)\right)+\right.\\
		& \left.E_{\bm b_{i}|\bm y_{i},t_{i}}^{\bm R_{i}^{*}(l)}\left(\bm b_{i}^{'}\right)Z_{i}^{'}\bm R_{(\bm y_{i})}^{-1}Z_{i}E_{\bm b_{i}|\bm y_{i},t_{i}}^{\bm R_{i}^{*}(l)}\left(\bm b_{i}\right)\right\},
	\end{split}	\tag{B.2} \label{emupdatesigma}
\end{equation}

\begin{equation}
	\begin{split}
		\widehat{\bm\beta_{1}}^{(l+1)}=&E_{\bm b_{i}|\bm y_{i},t_{i}}^{\bm R_{i}^{*}(l)}\left(\left(\sum_{i=1}^{n}\bm X_{i1}^{'}\bm R_{(\bm y_{i})}^{-1}\bm X_{i1}\right)^{-1}\left(\sum_{i=1}^{n}\bm X_{i1}^{'}\bm R_{(\bm y_{i})}^{-1}\left(\bm y_{i}-\bm Z_{i1}\bm b_{i}\right)\right)\right)\\      
		=&\left(\sum_{i=1}^{n}\bm X_{i1}^{'}\bm R_{(\bm y_{i})}^{-1}\bm X_{i1}\right)^{-1}\left(\sum_{i=1}^{n}\bm X_{i1}^{'}\bm R_{(\bm y_{i})}^{-1}\left(\bm y_{i}-\bm Z_{i1}E_{\bm b_{i}|\bm y_{i},t_{i}}^{\bm R_{i}^{*}(l)}\left(\bm b_{i}\right)\right)\right).
	\end{split}	\tag{B.3} \label{emupdatebeta} 
\end{equation}
Let the baseline hazard function be a piecewise-constant function with $K-1$ internal knots,
\[h_{0}(t)=\sum_{k=1}^{K}\lambda_{k}I(v_{k-1}<t\leq v_{k}),\]
where $0=v_{0}<v_{1}<\cdots<v_{K}=\mbox{max}\{t_{i},\,i=1,...,n\}$ split the time scale into $K$ intervals with a different constant baseline hazard at each interval.
The survival function of subject $i$ with covariate $\bm w_{i}$ is:
\begin{eqnarray*}
	S_{T_{i}^{*}}(t)&=&\mbox{exp}\left\{-\sum_{k=1}^{K}\left[I(t\geq v_{k})\int_{v_{k-1}}^{v_{k}}\lambda_{k}\mbox{exp}\left(\bm w_{i}^{'}\bm\beta_{2}+\alpha_{1} b_{i0}+\alpha_{2} b_{i1}s\right)\,ds\right]\right.\\
	&-&\left.\sum_{k=1}^{K}\left[I(v_{k-1}<t<v_{k})\int_{v_{k-1}}^{t}\lambda_{k}\mbox{exp}\left(\bm w_{i}^{'}\bm\beta_{2}+\alpha_{1} b_{i0}+\alpha_{2} b_{i1}s\right)\,ds\right]\right\}.
\end{eqnarray*}
The corresponding updating expression for $\lambda_{k},\,k=1,...,K$ is:
\newline\newline

\begin{equation}
	\begin{split}
		\lambda_{k}^{(l+1)}=&\sum_{i=1}^{n}I(v_{k-1}<t_{i}\leq v_{k})\delta_{i}\\
		\div&\sum_{i=1}^{n}E_{\bm b_{i}|\bm y_{i},t_{i}}^{\bm R_{i}^{*}(l)}\left\{\frac{\mbox{exp}\left(\bm w_{i}^{'}\bm\beta_{2}+\alpha_{1} b_{i0}\right)}{\alpha_{2} b_{i1}}I(t_{i}\geq v_{k})\left(e^{\alpha_{2} b_{i1}v_{k}}-e^{\alpha_{2} b_{i1}v_{k-1}}\right)\right.\\
		+&\left.\frac{\mbox{exp}\left(\bm w_{i}^{'}\bm\beta_{2}+\alpha_{1} b_{i0}\right)}{\alpha_{2} b_{i1}}I(v_{k-1}<t_{i}< v_{k})\left(e^{\alpha_{2} b_{i1}t_{i}}-e^{\alpha_{2} b_{i1}v_{k-1}}\right)\right\}.
	\end{split}	\tag{B.4} \label{emupdatelambda}
\end{equation}

Asymptotically, the  updating equations (\ref{emupdateD}), (\ref{emupdatesigma}), (\ref{emupdatebeta}) and (\ref{emupdatelambda}) combined with numerical maximisation for the remaining parameters would give us unbiased estimators at convergence if $\bm R_{i}$ were $\bm R_{i}^{*}.$

Under correct specification of the joint model, the score vector of observed data can be arranged into the following form:
\begin{eqnarray*}
	S(\bm\theta)=\sum_{i=1}^{n}\int_{\bm b_{i}} A(\bm\theta,\bm b_{i})f_{\bm b_{i}}^{\bm R_{i}}(\bm b_{i}|t_{i},\delta_{i},\bm y_{i};\bm\theta)d\bm b_{i},
\end{eqnarray*}
where 
\begin{eqnarray*}
	A(\bm\theta,\bm b_{i})&=&A(\bm\theta_{\bm y},\bm b_{i})+A(\bm\theta_{t},\bm b_{i})+A(\bm\theta_{\bm b},\bm b_{i})\\
	&=&\frac{\partial}{\partial\bm\theta}\left\{\mbox{log}f_{T_{i},\delta_{i}}(t_{i},\delta_{i}|\bm b_{i},\bm y_{i};\bm\theta)+\mbox{log}f_{\bm y_{i}}(\bm y_{i}|\bm b_{i};\bm\theta)+\mbox{log}f_{\bm b_{i}}(\bm b_{i};\bm\theta)\right\}\\
	&=&\frac{\partial}{\partial\bm\theta}\left\{\mbox{log}\bm\phi_{n_{i}}\left(\bm Z_{\bm y_{i}|\bm b_{i}};\bm\Sigma_{(\bm y_{i})}\right)+\mbox{log}f_{\bm b_{i}}\left(\bm b_{i};\bm\theta\right)\right\}\\
	&+&\delta_{i}\frac{\partial}{\partial\bm\theta}\left\{\mbox{log}\phi\left(Z_{t_{i}|\bm b_{i}};\mu_{i}^{t_{i}|\bm y_{i},\bm b_{i}},\sigma_{i}^{t_{i}|\bm y_{i},\bm b_{i}}\right)+\mbox{log}f_{T_{i}^{*}}\left(t_{i}|\bm b_{i}\right)-\mbox{log}\phi\left(Z_{t_{i}|\bm b_{i}}\right)\right\}\\
	&+&(1-\delta_{i})\frac{\partial}{\partial\bm\theta}\mbox{log}\Phi\left(-\frac{Z_{t_{i}|\bm b_{i}}-\mu_{i}^{t_{i}|\bm y_{i},\bm b_{i}}}{\sigma_{i}^{t_{i}|\bm y_{i},\bm b_{i}}}\right).
\end{eqnarray*}

For the variance components $\bm\theta_{\bm b}$ of random effects,
\begin{eqnarray*}
	A(\bm\theta_{\bm b},\bm b_{i})=\frac{\partial}{\partial\theta_{\bm b}}\mbox{log}f_{\bm b_{i}}\left(\bm b_{i};\bm\theta\right),
\end{eqnarray*}
which implies the updating equation for $\bm D$ under $\bm R_{i}$ in the EM algorithm has the same formulation as ( \ref{emupdateD}):
\begin{eqnarray*}
	\nonumber	\widehat{\bm D}^{(l+1)}&=&\frac{1}{n}\sum_{i=1}^{n}E_{\bm b_{i}|\bm y_{i},t_{i}}^{\bm R_{i}(l)}\left(\bm b_{i}\bm b_{i}^{'}\right)\\                                                                                                                                                                                    
	&=&\frac{1}{n}\sum_{i=1}^{n}\left\{var_{\bm b_{i}|\bm y_{i},t_{i}}^{\bm R_{i}(l)}\left(\bm b_{i}\right)+E_{\bm b_{i}|\bm y_{i},t_{i}}^{\bm R_{i}(l)}\left(\bm b_{i}\right)E_{\bm b_{i}|\bm y_{i},t_{i}}^{\bm R_{i}(l)}\left(\bm b_{i}^{'}\right)\right\}.
\end{eqnarray*}	
But the difference between $E_{\bm b_{i}|\bm y_{i},t_{i}}^{\bm R_{i}(l)}\left(h(\bm b_{i})\right)$ and $E_{\bm b_{i}|\bm y_{i},t_{i}}^{\bm R_{i}^{*}(l)}\left(h(\bm b_{i})\right)$ indicates (\ref{emupdateD}) is a biased estimator for $\bm D$ at convergence. On the other hand, $\sigma^{2}$ and $\bm\beta_{1}$ do no have closed form updating equations in the EM algorithm.
In fact, for parameters $\bm\theta_{\bm y}$ and $\bm\theta_{t}$ in the longitudinal and survival processes,
\begin{eqnarray*}
	A(\bm\theta_{\bm y},\bm b_{i})&=&\frac{\partial}{\partial\bm\theta_{\bm y}}\left\{\mbox{log}\bm\phi_{n_{i}}\left(\bm Z_{\bm y_{i}|\bm b_{i}};\bm\Sigma_{(\bm y_{i})}\right)+\delta_{i}\mbox{log}\phi\left(Z_{t_{i}|\bm b_{i}};\mu_{i}^{t_{i}|\bm y_{i},\bm b_{i}},\sigma_{i}^{t_{i}|\bm y_{i},\bm b_{i}}\right)\right.\\
	&+&\left.(1-\delta_{i})\mbox{log}\Phi\left(-\frac{Z_{t_{i}|\bm b_{i}}-\mu_{i}^{t_{i}|\bm y_{i},\bm b_{i}}}{\sigma_{i}^{t_{i}|\bm y_{i},\bm b_{i}}}\right)\right\},
\end{eqnarray*}

\begin{figure}[H]
	\centering
	\begin{minipage}{0.42\textwidth}
		\includegraphics[width=\linewidth]{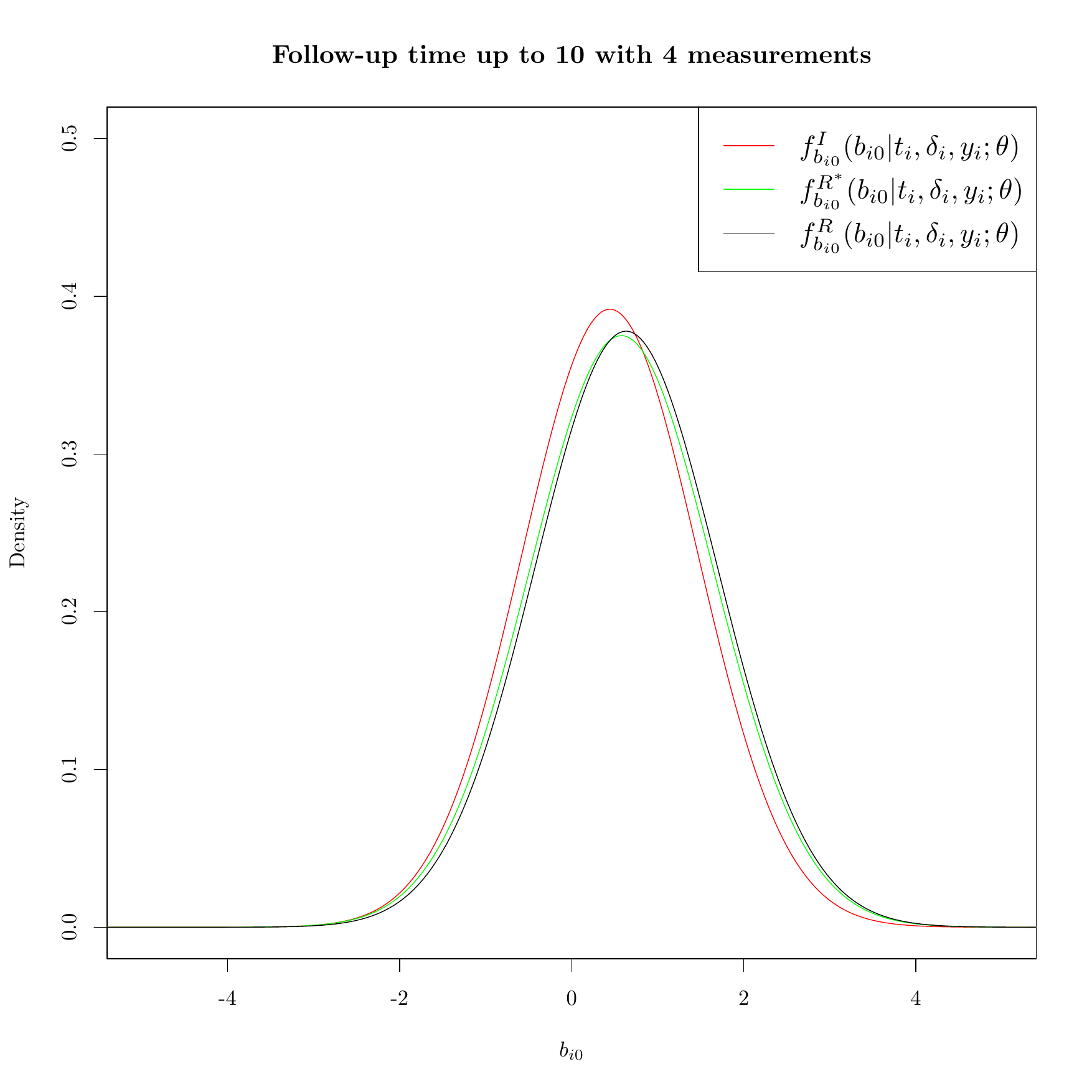}\\
	\end{minipage}
	\begin{minipage}{0.42\textwidth}
		\includegraphics[width=\linewidth]{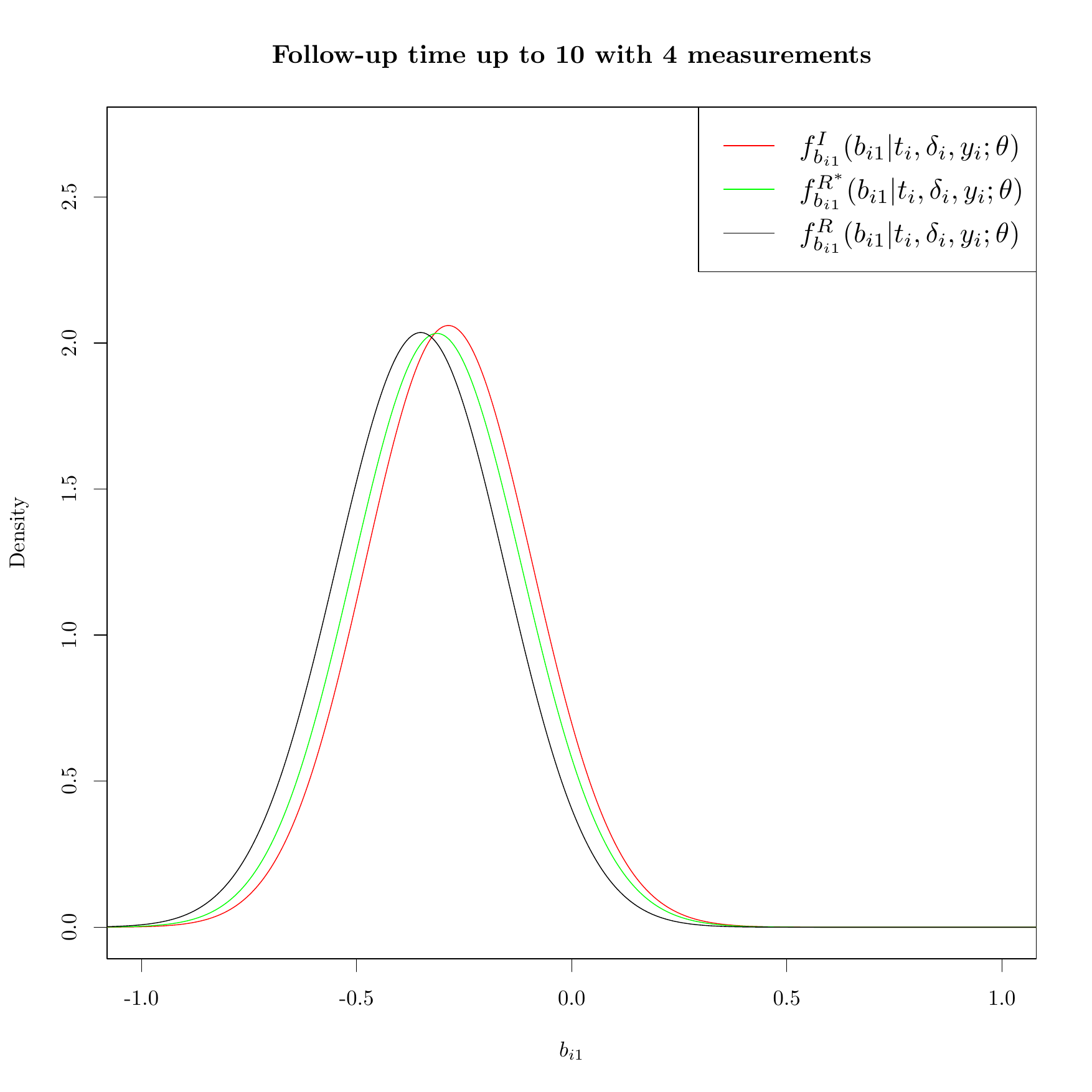}\\
	\end{minipage}
	\begin{minipage}{0.42\textwidth}
		\includegraphics[width=\linewidth]{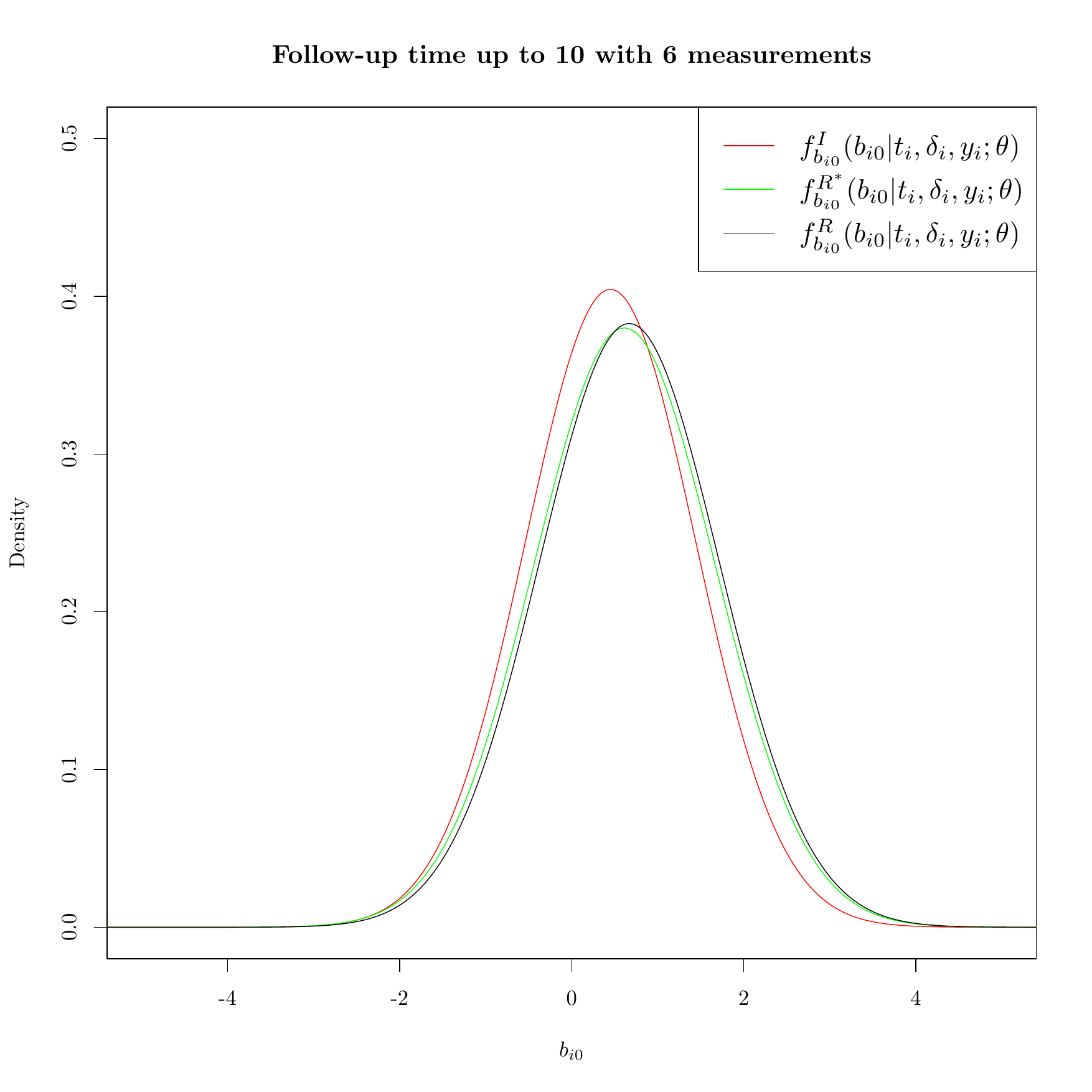}\\
	\end{minipage}
	\begin{minipage}{0.42\textwidth}
		\includegraphics[width=\linewidth]{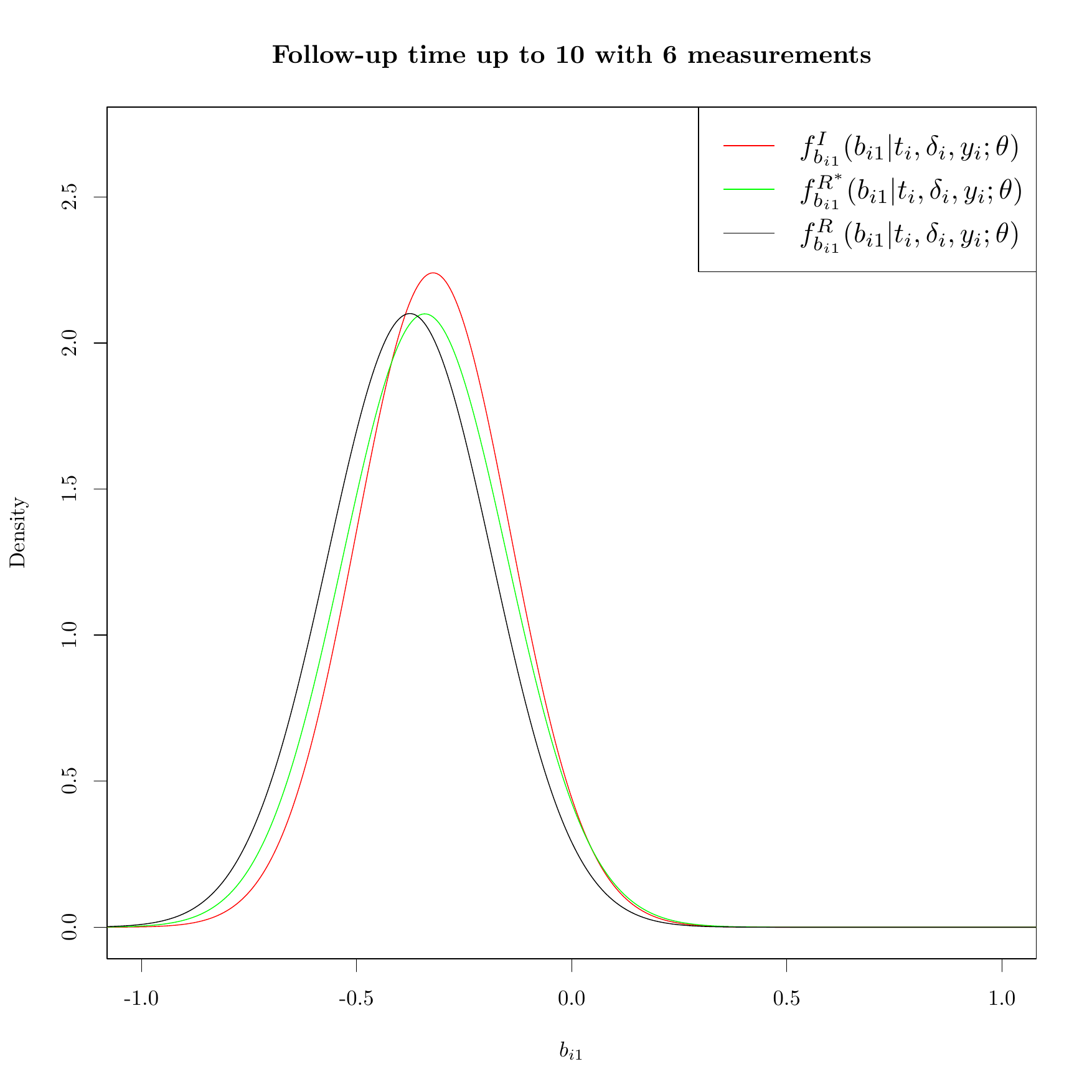}\\
	\end{minipage}
	\begin{minipage}{0.42\textwidth}
		\includegraphics[width=\linewidth]{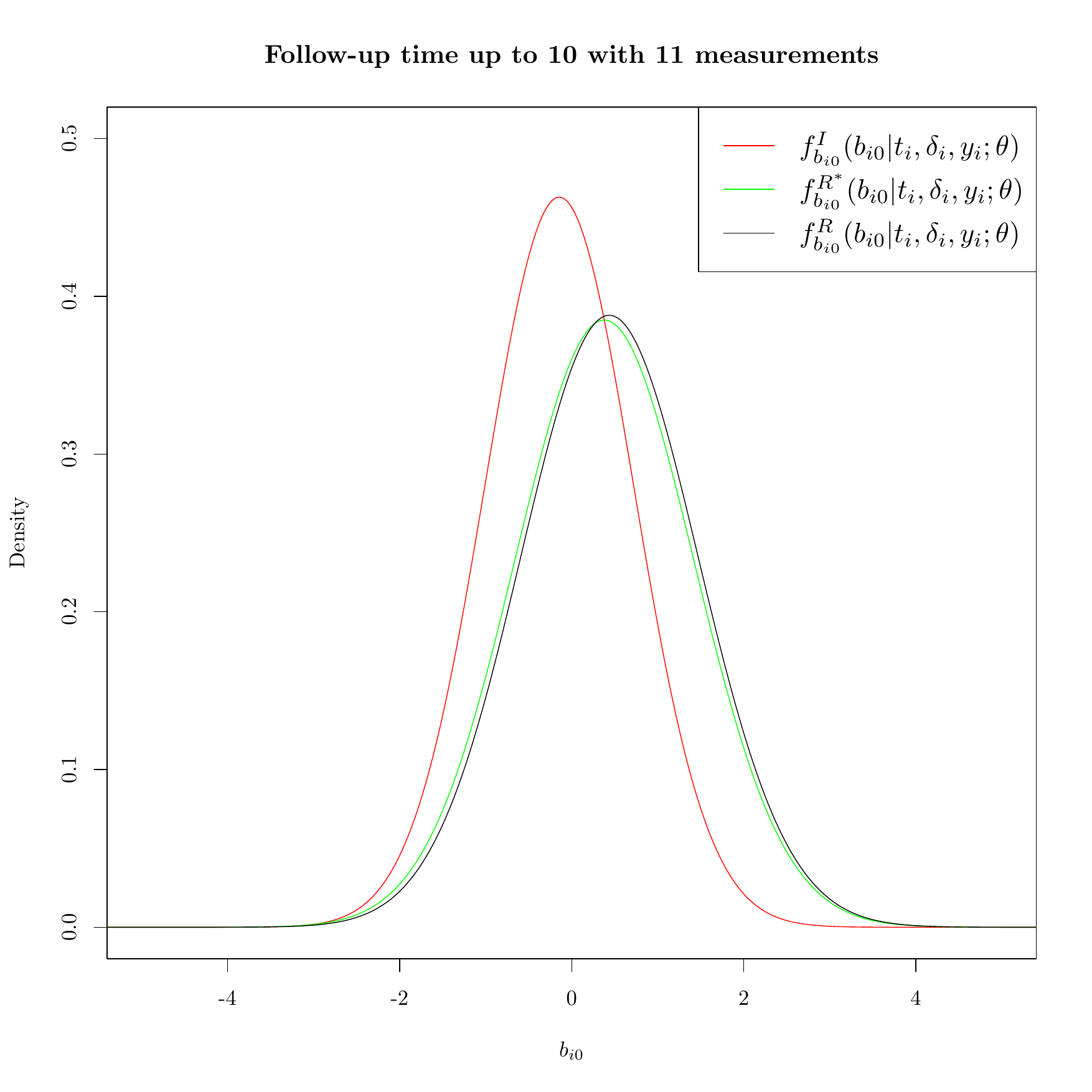}\\
	\end{minipage}
	\begin{minipage}{0.42\textwidth}
		\includegraphics[width=\linewidth]{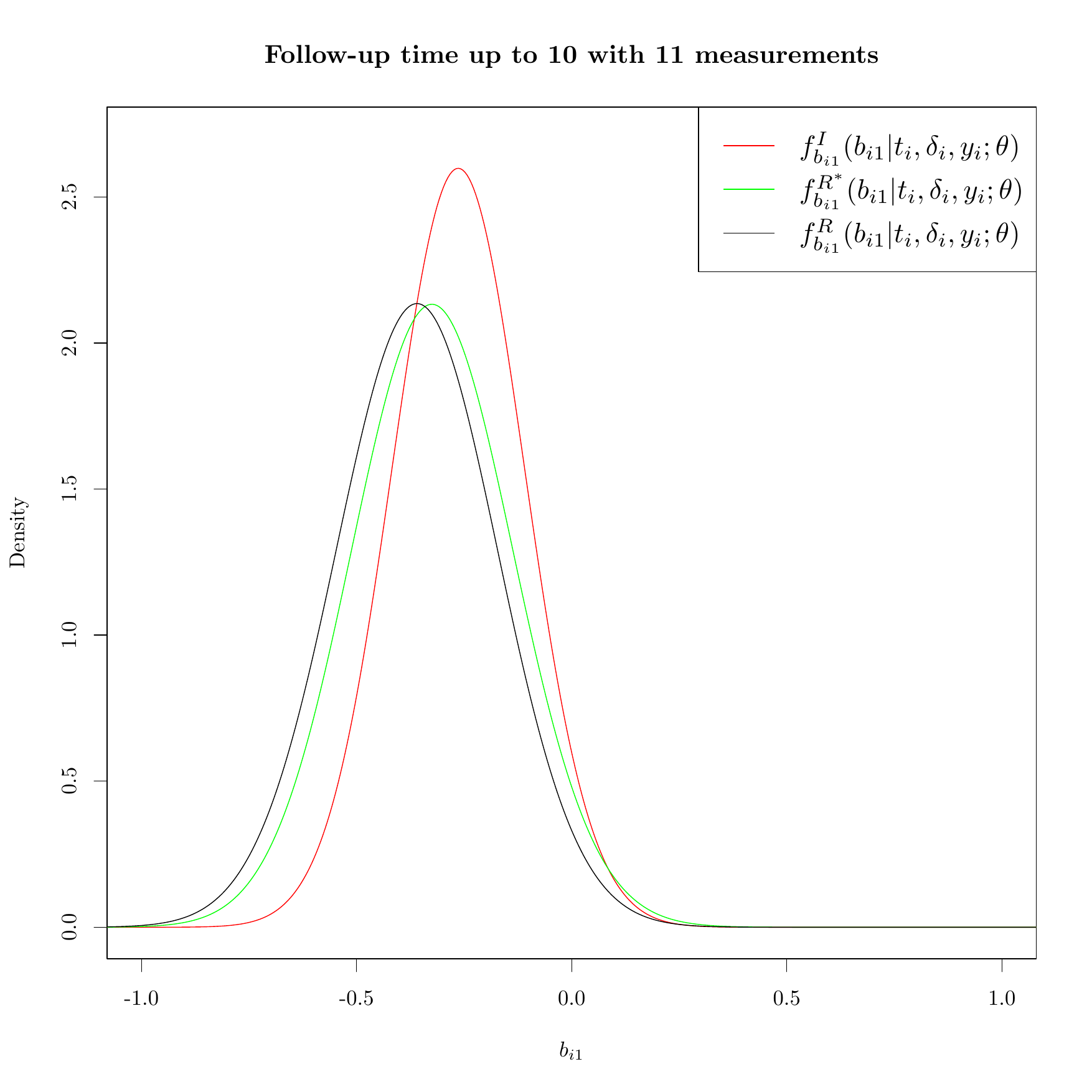}\\
	\end{minipage}
	\caption{\textbf{{\scriptsize Posterior distribution of random effects for a subject from case 2.}}}
\end{figure}

\begin{eqnarray*}
	A(\bm\theta_{t},\bm b_{i})&=&\delta_{i}\frac{\partial}{\partial\bm\theta_{t}}\left\{\mbox{log}\phi\left(Z_{t_{i}|\bm b_{i}};\mu_{i}^{t_{i}|\bm y_{i},\bm b_{i}},\sigma_{i}^{t_{i}|\bm y_{i},\bm b_{i}}\right)+\mbox{log}f_{T_{i}^{*}}\left(t_{i}|\bm b_{i}\right)-\mbox{log}\phi\left(Z_{t_{i}|\bm b_{i}}\right)\right\}\\
	&+&(1-\delta_{i})\frac{\partial}{\partial\bm\theta_{t}}\mbox{log}\Phi\left(-\frac{Z_{t_{i}|\bm b_{i}}-\mu_{i}^{t_{i}|\bm y_{i},\bm b_{i}}}{\sigma_{i}^{t_{i}|\bm y_{i},\bm b_{i}}}\right).
\end{eqnarray*}

Clearly, the estimates of $\bm\theta_{\bm y}$ and $\bm\theta_{t}$ are affected by the misspecification of $\bm R_{i}$ due to its effects on both $f_{\bm b_{i}}^{\bm R_{i}}(\bm b_{i}|t_{i},\delta_{i},\bm y_{i};\bm\theta)$ and $A(\bm\theta,\bm b_{i}).$  Therefore, the misuse of (\ref{emupdatesigma}) and (\ref{emupdatebeta}) as updating equations for $\sigma^{2}$ and $\bm\beta_{1}$ certainly results in biased estimators at convergence, although to what extent is not clear yet.

The posterior distributions of the random effects under scenarios 1, 2 and 4 for the simulated subject of case 2 mentioned in section 3.2 are given in Figure 6. The density $f_{\bm{b_{i}}}^{\bm I_{i}}(\bm b_{i}|t_{i},\delta_{i},\bm y_{i};\bm\theta)$ deviates from $f_{\bm{b_{i}}}^{\bm R_{i}}(\bm b_{i}|t_{i},\delta_{i},\bm y_{i};\bm\theta)$ as the length of longitudinal measurements increases. We note that $f_{\bm{b_{i0}}}^{\bm I_{i}}(\bm b_{i}|t_{i},\delta_{i},\bm y_{i};\bm\theta)$ tends to move to the left of $f_{\bm{b_{i0}}}^{\bm R_{i}}(\bm b_{i}|t_{i},\delta_{i},\bm y_{i};\bm\theta),$ while $f_{\bm{b_{i1}}}^{\bm I_{i}}(\bm b_{i1}|t_{i},\delta_{i},\bm y_{i};\bm\theta)$ tends to move to the right of $f_{\bm{b_{i1}}}^{\bm R_{i}}(\bm b_{i}|t_{i},\delta_{i},\bm y_{i};\bm\theta).$ This partly explains why the estimates of $\bm D$ are  more overestimated with longer longitudinal measurements per subject in scenario 1. The posterior distributions of random effects of other subjects generally follow this similar pattern as $n_{i}$ increases.


\begin{thebibliography}{999}
	
	
	\bibitem{als20}
	Alsefri, M.,  Sudell, M.,
Garc\'{\i}a-Fi\~{n}ana, M. and Kolamunnage-Dona, R. (2020) Bayesian joint modelling of longitudinal and time to event data: a methodological review. \textit{BMC Med Res Methodol} \textbf{20,} 94, https://doi.org/10.1186/s12874-020-00976-2

\bibitem{and18}
Andrinopoulou, E., Eilers, P., Takkenberg, J., and Rizopoulos, D. (2018)  Improved dynamic predictions from joint models of
longitudinal and survival data with time-varying
effects using P-Splines. \textit{Biometrics} \textbf{74}, 685–693
	
	\bibitem{bag17}
	Baghfalaki, T.,  Ganjali, M. and Verbeke, G. (2017) A shared parameter model of longitudinal
measurements and survival time with
heterogeneous random-effects distribution.
	\textit{Journal of Applied Statistics,} VOL.\textbf{ 44}, NO. 15, 2813–2836, http://dx.doi.org/10.1080/02664763.2016.1266309
	
	\bibitem{cox72}
	Cox, D. R. (1972). Regression models and life tables (with discussion). \textit{Journal of the Royal Statistical Society, Series
	B} \textbf{34}, 187–220.

    \bibitem{daf98}
   Dafni, U. G. and Tsiatis, A. A. (1998). Evaluating surrogate markers of clinical outcome measured with error. \textit{Biometrics} \textbf{54}, 1445-1462.
   
   	\bibitem{dai18}
   Dai, H and Pan, J. (2018). Joint Modelling of Survival and Longitudinal Data with Informative Observation Times. \textit{Scandinavian Journal of Statistics}, Vol. \textbf{45}: 571 –589.
   
   
   	\bibitem{den83}
   Dennis, J. E. and Schnabel, R. B. (1983). Numerical Methods for Unconstrained Optimization and Nonlinear Equations. \textit{Prentice-Hall, Englewood Cliffs, NJ.}
   
    \bibitem{dig08}
   Diggle, P.J., Sousa, I., and Chetwynd, A.G. (2008) Joint modelling of repeated measurements and time-to-event outcomes: the fourth Armitage lecture. \textit{Stat Med} 27(16): 2981–2998.
   
   \bibitem{dut21}
   Dutta, S., Molenberghs, G. and  Chakraborty, A. (2021) Joint modelling of longitudinal response and time-to-event data using conditional distributions: a
   Bayesian perspective. 	\textit{Journal of Applied Statistics,} https://doi.org/10.1080/02664763.2021.1897971
   
   \bibitem{emu17}
   Emura, T., Nakatochi, M., Murotani, K., and Rondeau, V. (2017). A joint frailty-copula model between tumour progression and death for meta-analysis. \textit{Statistical Methods in Medical Research}, \textbf{26(6)}, 2649–2666.
   
   \bibitem{emu18}
   Emura, T., Nakatochi, M., Matsui, S., Michimae, H., and Rondeau, V. (2018) Personalized dynamic prediction of death according to tumour progression and high-dimensional genetic factors: Meta-analysis with a joint model.  \textit{Statistical Methods in Medical Research},  \textbf{27(9)}:2842-2858. doi:10.1177/0962280216688032
   
    \bibitem{fau96}
    Faucett,  C. L. and Thomas, D. C. (1996). Simultaneously modelling censored survival data and repeatedly
measured covariates: A Gibbs sampling approach. \textit{Stat Med} \textbf{15,} pp. 1663–1685.
    
    \bibitem{gan15}
    Ganjali, M. and Baghfalaki, T. (2015). A copula approach to joint modeling of longitudinal measurements
    and survival times using Monte Carlo expectation-maximization with application to AIDS studies. \textit{Journal of Biopharmaceutical Statistics} \textbf{25,} 1077-1099.
    
    \bibitem{gol96}
    Goldman, A. I., Carlin, B. P., Crane, L. R., Launer, C., Korvick, J. A., Deyton, L., and Abrams, D. I.
(1996). Response of CD4+ and Clinical Consequences to Treatment Using ddI or ddC in
Patients with Advanced HIV Infection. \textit{Journal of Acquired Immune Deficiency Syndromes
and Human Retrovirology.} \textbf{11}:161–169.
    
    \bibitem{guo04}
    Guo, X. and Carlin, B. P. (2004). Separate and Joint Modelling of Longitudinal
and Event Time Data Using Standard Computer Packages, \textit{The American Statistician,} \textbf{58:1}, 16-24,
DOI: 10.1198/0003130042854

  \bibitem{hen00}
  Henderson, R., Diggle, P. and Dobson, A. (2000). Joint modelling of longitudinal measurements and event time data.
  \textit{Biostatistics} \textbf{4}, 465–480.
    
    \bibitem{hof18}
    Hofert, M., Kojadinovic, I., Machler, M. and Yan, J. (2018). Elements of copula modelling with R. Berlin, Germany: Springer International Publishing AG. https://doi.org/10.1007/978-3-319-89635-9
    
    \bibitem{ibr10}
    Ibrahim, Joseph, G., Chu, H. and Chen, L. M. (2010). Basic concepts and methods for joint models
of longitudinal and survival data, \textit{Journal of Clinical Oncology,} Vol. \textbf{28}, pp. 2796–2801.

   \bibitem{jac05}
   J\"{a}ckel, P. (2005). A note on multivariate Gauss-Hermite quadrature.

    
    \bibitem{jac10}
    Jacqmin-Gadda, H., Proust-Lima, C., Taylor, J. M. and Commenges, D. (2010). Score test for conditional independence between longitudinal outcome and time to
    event given the classes in the joint latent class model. \textit{Biometrics} \textbf{66}, 11–19.
    
    \bibitem{kal02}
    Kalbfleisch, J. and Prentice, R. (2002). The Statistical Analysis of Failure Time Data, 2nd edition. Wiley, New York.
    
    \bibitem{lai82}
    Laird, N. M. and Ware, J. H. (1982). Random-effects models for longitudinal data. \textit{Biometrics} \textbf{38}, 963–974.
    
    \bibitem{li21}
    Li, C., Xiao, L. and Luo, S. (2021). Joint model for survival and multivariate sparse functional
    data with application to a study of Alzheimer’s Disease. \textit{Biometrics.} 	10.1111/biom.13427
    
    \bibitem{li17}
    Li, K. and Luo, S. (2017). Functional joint model for longitudinal
and time-to-event data: an application to
Alzheimer’s disease. \textit{Stat Med} \textbf{36} 3560–3572, DOI: 10.1002/sim.7381
    
    \bibitem{li19}
    Li, K. and Luo, S. (2019). Bayesian functional joint models for multivariate longitudinal
and time-to-event data. \textit{Comput Stat Data Anal.} 129:14–29, https://doi.org/10.1016/j.csda.2018.07.015
    
    \bibitem{lin02}
    Lin, H., Turnbull, B. W., McCulloch, C. E., Slate, E. H. (2002). Latent class models for joint analysis of longitudinal biomarker and event process data: application
    to longitudinal prostate-specific antigen readings and prostate cancer. \textit{J. Amer. Statist. Assoc.} \textbf{97,} 53–65.
   
   \bibitem{liu15}
   Liu Y., Liu L,. and Zhou, J. (2015). Joint latent class model of survival and longitudinal data. \textit{Comput Stat Data Anal.} \textbf{91:} 40‐ 50.
       
    \bibitem{mal15}
    Malehi, A. S., Hajizadehb, E., Ahmadia, K. A., and Mansouric, P. (2015). Joint modelling of longitudinal biomarker
and gap time between recurrent events:
copula-based dependence. \textit{Journal of Applied Statistics,} 2015
Vol. \textbf{42,} No. 9, 1931–1945, http://dx.doi.org/10.1080/02664763.2015.1014889

\bibitem{nel65}
Nelder, J. A. and Mead, R. (1965). A simplex algorithm for function minimization. \textit{Computer Journal,} \textbf{7}, 308--313. 10.1093/comjnl/7.4.308.

    
    \bibitem{pap19}
    Papageorgiou, G., Mauff, K., Tomer, A., and Rizopoulos, D. (2019). An Overview
of Joint Modelling of Time-to-Event and Longitudinal Outcomes. \textit{Annual Review of Statistics and Its Application.}

 \bibitem{phi17}
Philipson P., Sousa I., Diggle P.J., Williamson P., Kolamunnage-Dona R., Henderson R., and Hickey GL. (2017) \verb|joineR|: Joint Modelling of Repeated Measurements and Time-to-event Data. \verb|R| package version 1.2.0. https://CRAN.R-project.org/package=joineR.
    
    \bibitem{pre82}
    Prentice, R. L., (1982). Covariate measurement errors and parameter estimation in a failure time regression model. \textit{Biometrika} \textbf{69}, 331–342.
    
     \bibitem{riz08a}
    Rizopoulos, D., Verbeke, G., and Lesaffre, E. (2008a). A two-part joint model
    for the analysis of survival and longitudinal binary data with excess zeros.
    \textit{Biometrics} \textbf{64,} 611–619.
    
    \bibitem{riz08b}
    Rizopoulos D., Verbeke, G., and Molenberghs, G. (2008b). Shared parameter models under random effects misspecification. \textit{Biometrika} \textbf{95}, 63–74.
    
      \bibitem{riz10}
    Rizopoulos, D. (2010). JM: An R package for the joint modelling of longitudinal and time-to-event data. \textit{Journal of Statistical Software} \textbf{35 (9),} 1-33.
    
    
    \bibitem{riz11}
    Rizopoulos, D. (2011). Dynamic predictions and prospective accuracy in joint models for longitudinal and time-to-event data. \textit{Biometrics} \textbf{67}, 819–829.
    
    \bibitem{riz12a}
    Rizopoulos, D. (2012a). Fast fitting of joint models for longitudinal and event time data using a pseudo-adaptive Gaussian quadrature rule. \textit{Comput Stat Data Anal.}  \textbf{56} (2012) 491–501, 11, doi:10.1016/j.csda.2011.09.007
    
     \bibitem{riz12b}
    Rizopoulos, D. (2012b). Joint Models for Longitudinal and Time-to-Event Data
    with Applications in R. Boca Raton: Chapman \& Hall/CRC.
    

   \bibitem{son02}
  Song, X., Davidian, M., and Tsiatis, A. A. (2002). A semiparametric likelihood approach to joint modelling of longitudinal and time-to-event data. \textit{Biometrics} \textbf{58}, 742–753.

    
   	  \bibitem{sur21a}
   	Suresh, K., Taylor, J. M. G., and Tsodikov, A. (2021a). A Gaussian copula approach for dynamic prediction of survival with a longitudinal biomarker. \textit{Biostatistics}  Volume \textbf{22}, Issue 3, 504-521, doi:10.1093/biostatistics/kxz049
   	
   	\bibitem{sur21b}
   	Suresh, K., Taylor, J. M. G., and Tsodikov, A. (2021b). A copula-based approach for dynamic prediction of survival with a binary time-dependent covariate. \textit{Biostatistics}  Volume \textbf{40}, Issue 23, 4931-4946, https://doi.org/10.1002/sim.9102
   	
   	
    
     \bibitem{tsi04}
     Tsiatis, A. A. and Davidian, M. (2004). Joint modelling of longitudinal and time-to-event data: An overview. \textit{Statistica Sinica} \textbf{14,} 809-834.
     
     \bibitem{ver00}
     Verbeke, G. and Molenberghs, G. (2000). Linear Mixed Models for Longitudinal Data.
     Springer-Verlag, New York.
     
      \bibitem{wan01}
      Wang, Y. and Taylor, J. M. G. (2001). Jointly modelling longitudinal and event time data with application to acquired immunodeficiency syndrome. \textit{J. Amer. Statist. Assoc.} \textbf{96},
      895-905.
         
     \bibitem{wul97}
    Wulfsohn, M. S. and Tsiatis, A. A. (1997). A joint model for survival and longitudinal data measured with error. \textit{Biometrics} \textbf{53,} 330-339.
    
     \bibitem{xu01}
    Xu, J. and Zeger, S. L. (2001). Joint analysis of longitudinal data comprising repeated measures
    and times to events. \textit{Appl. Statist.} \textbf{50}, 375
    
    	\bibitem{zha21}
    Zhang, Z., Charalambous, C., and Foster, P. (2021). Joint modelling of longitudinal measurements and survival times via a multivariate copula approach. \textit{Journal of Applied Statistics,}	https://doi.org/10.1080/02664763.2022.2081965
    
\end{thebibliography}
\end{document}